\documentclass[showpacs,onecolumn,preprintnumbers,amsmath,amsfonts,amssymb,floatfix,aps,superscriptaddress]{revtex4}

\usepackage{graphicx}
\usepackage{epsfig}
\usepackage[hang,nooneline]{subfigure}
\usepackage[normalem]{ulem}
\usepackage{color}
\usepackage{braket}
\usepackage{hyperref}
\usepackage{todonotes}
\usepackage{algorithm}
\usepackage{multirow}
\usepackage{hhline}
\usepackage{shortvrb}
\usepackage{cprotect}
\usepackage[noend]{algpseudocode}

\definecolor{darkblue}{rgb}{0.00,0.00,0.55}
\definecolor{black}{rgb}{0.00,0.00,0.00}
\hypersetup{
    linkcolor = darkblue,
    anchorcolor = darkblue,
    citecolor = darkblue,
    filecolor = darkblue,
    urlcolor = darkblue,
    colorlinks  = true,
}
\definecolor{brightcerulean}{rgb}{0.11, 0.67, 0.84}
\DeclareMathAlphabet\mathbfcal{OMS}{cmsy}{b}{n}

\newcounter{fig}

\begin{document}

\title{Bifurcation analysis of stationary solutions of two-dimensional 
coupled Gross-Pitaevskii equations using deflated continuation}

\author{E. G. Charalampidis}
\email[Email: ]{echarala@calpoly.edu}
\affiliation{Mathematics Department, California Polytechnic State University, 
San Luis Obispo, CA 93407-0403, USA}
\author{N. Boull\'e}
\email[Email: ]{nicolas.boulle@maths.ox.ac.uk}
\affiliation{Mathematical Institute, University of Oxford, Oxford, UK}
\author{P. E. Farrell}
\email[Email: ]{patrick.farrell@maths.ox.ac.uk}
\affiliation{Mathematical Institute, University of Oxford, Oxford, UK}
\author{P. G. Kevrekidis}
\email[Email: ]{kevrekid@math.umass.edu}
\affiliation{Department of Mathematics and Statistics, University of Massachusetts
  Amherst, Amherst, MA 01003-4515, USA}
\affiliation{Mathematical Institute, University of Oxford, Oxford, UK}

\date{\today}

\begin{abstract}
Recently, a novel bifurcation technique known as the deflated continuation 
method (DCM) was applied to the single-component nonlinear Schr\"odinger 
(NLS) equation with a parabolic trap in two spatial dimensions. The bifurcation
analysis carried out by a subset of the present authors shed light on the
configuration space of solutions of this fundamental problem in the physics 
of ultracold atoms. In the present work, we take this a step further by applying 
the DCM to two coupled NLS equations in order to elucidate the
considerably more complex landscape of 
solutions of this system. Upon identifying branches of solutions, we construct 
the relevant bifurcation diagrams and perform spectral stability analysis 
to identify parametric regimes of stability and instability and to
understand the mechanisms by which these branches emerge.
The method reveals a remarkable wealth of solutions: these do not only
include some of the well-known ones including, e.g., from the
Cartesian or polar small amplitude limits of the underlying linear
problem but also a significant number of branches that arise through
(typically pitchfork) bifurcations. In addition to presenting a
``cartography'' of the landscape of solutions, we comment on the 
challenging task of identifying {\it all} solutions of such a
high-dimensional, nonlinear problem. 
\end{abstract}

\maketitle

\section{Introduction} \label{sec:intro}
The study of Bose-Einstein condensates (BECs) has offered a versatile 
playground for the examination of a diverse host of mesoscopic quantum phenomena 
for more than two decades~\cite{pethick,stringari}. Among them, nonlinear 
wave dynamical features are of particular interest in the form of bright~\cite{tomio} 
and dark~\cite{djf} solitons, vortices~\cite{fetter1,fetter2,mplb} and 
vortex lines and vortex rings~\cite{komineas}. As discussed, e.g., in the 
compendium of Kevrekidis et al.~\cite{siambook},
many of these patterns spontaneously emerge 
in the nonlinear dynamics of BEC systems and subsequently play a critical 
role in their dynamical evolution including their density and phase profiles, 
as well as their regular or chaotic/turbulent phenomena.

Although the majority of studies have considered the one component (single 
species) case, multi-component BECs are also of considerable interest; see, 
e.g.~\cite{revip} for a recent review. These may consist of mixtures of, 
for instance, different spin states of a particular atom (pseudo-spinor 
systems)~\cite{Hall1998a,chap01:stamp} or different Zeeman sub-levels of 
the same hyperfine level (so-called spinor condensates)~\cite{Stenger1998a,kawueda,stampueda}.
Within these multi-component generalizations, various coherent structures 
have been realized experimentally~\cite{Becker2008,Middelkamp2011,Hamner2011,Yan2011,Hoefer2011,Yan2012}, 
with one of the most notable arguably being the dark-bright solitons (and 
their dark-dark cousins). Experimental realizations have also been extended 
to spinor BECs~\cite{Stamper-Kurn2001,Chang2004,Chang2005,kawueda,stampueda}, 
where recently also solitonic states have been observed~\cite{Bersano2018}.

In parallel to experimental and theoretical studies, the development 
of numerical methods can also enhance our understanding of single- and 
multi-component BECs. One such example that a subgroup of the present authors 
recently adapted to single-component atomic BECs~\cite{egc_16} is the method 
of deflation and, more specifically, the deflated continuation method 
(DCM)~\cite{farrell2,farrell3}. Given one numerically exact solution of the 
system (computed, in our work, by means of Newton's method), the aim of deflation
is to construct a new problem where applying Newton's method
will no longer converge to the already obtained solution. Hence, if the solver 
is applied again, and upon successful convergence (within user-prescribed 
tolerances), it will discover an additional solution that was previously unknown. 

To set the stage, let $F: U \to V$ be a nonlinear map between Banach 
spaces whose roots are sought. Suppose that $\phi_1$ is an isolated root of 
$F$. Then, one can construct a new operator $G: U \to V$ such that:
\begin{equation}
G(\phi)\doteq\left(\frac{1}{\|\phi - \phi_1\|^2_U} + 1   \right) F(\phi),
\label{eq_deflation}
\end{equation}
where $\|\cdot\|_U$ is the norm on $U$. The essential idea is that $\|\phi - \phi_1\|^{-2}_U$ 
approaches infinity as $\phi \to \phi_1$ faster than $F(\phi)$ approaches
$0$, hence avoiding the convergence to $\phi_1$ of the fixed-point iteration 
when applied to $G$. The addition of unity ensures that the deflated problem 
$G(\phi)=0$ behaves like the original problem $F$ far away from $\phi_1$.

In the present work, we apply the method of deflation and DCM to examine the 
possible solutions of the multi-component atomic BEC system. This will enable
us to obtain significant insights into the pattern formation of this system. 
The present analysis obtains solutions that, to the best of our knowledge, 
were previously unknown, offering a significant tool for understanding the 
landscape of nonlinear waveforms featured by the system. In addition, by computing 
the spectral stability of the resulting solutions one can identify not only 
which of these solutions are potentially stable (and where this is parametrically 
so) but also bifurcations and further solutions arising from the DCM-identified
solutions. 

Our presentation is structured as follows. In section II, we provide an overview
of the existence and stability problems for the two-component system. We also
discuss an important twist on the deflation method to properly account for 
symmetries of the problem (to avoid discovering multiple copies of the same solution
that are e.g.~related by rotation). In section III, we discuss our numerical results
for the different branches of the system. Finally, in section IV, we summarize our
findings and present some conclusions and future directions. The Appendix presents 
some technical details regarding the eigenvalue computations.
While the emphasis of our analysis will be on the existence of the
branches
and their bifurcation diagrams, it is worthwhile to highlight that to
perform the spectral analysis, we
will utilize the state-of-the-art capabilities of FEAST,  an
eigenvalue
solver combining 
highly
desirable features of accuracy, efficiency and
robustness for problems such as the one considered herein.

\section{The model and setup} \label{sec:setup}
In this work, the model of interest is a two-component nonlinear Schr\"odinger
(NLS) system in (2+1)-dimensions (two spatial and one temporal variable) 
given by
\begin{subequations}
\begin{align}
i\frac{\partial\Phi _{-}}{\partial t}&=-\frac{D_{-}}{2}\nabla ^{2}\Phi _{-}+
\left( g_{11}|\Phi _{-}|^{2}+g_{12}|\Phi _{+}|^{2}\right) \Phi
_{-}+V(\mathbf{r})\Phi _{-}, \label{gpe_2D_2comp_m}\\
i\frac{\partial\Phi _{+}}{\partial t}&=-\frac{D_{+}}{2}\nabla ^{2}\Phi _{+}+
\left( g_{12}|\Phi _{-}|^{2}+g_{22}|\Phi _{+}|^{2}\right) \Phi
_{+}+V(\mathbf{r})\Phi _{+}, \label{gpe_2D_2comp_p}
\end{align}
\end{subequations}
where $\nabla^{2}=\partial^{2}/\partial x^{2} + \partial^{2}/\partial y^{2}$ 
is the Laplacian operator, and $D_{\pm }$ and $g_{ij}$, $i,j=\lbrace{1,2\rbrace}$ 
(with $g_{12}=g_{21}$) are the dispersion and interaction coefficients, respectively. 
The function $V(\mathbf{r})$ describes the external harmonic confinement and 
takes the form of 
\begin{equation}
V(\mathbf{r})=\frac{1}{2}\Omega^{2} |{\mathbf{r}}|^{2},
\label{trap}
\end{equation}
with the parameter $\Omega$ capturing its strength and $|\mathbf{r}|^{2}=x^2+y^2$. 
In the realm of atomic BECs, the functions $\Phi_{\pm}(\mathbf{r},t):%
\overline{D\times\mathbb{R}^{+}\cup{\lbrace 0\rbrace}}\mapsto \mathbb{C}$
in Eqs.~\eqref{gpe_2D_2comp_m} and \eqref{gpe_2D_2comp_p} represent the 
macroscopic wave functions with $D\subseteq\mathbb{R}^{2}$ being the 
(two-dimensional) spatial domain. For the reduction of the original 
three-dimensional (in space) BEC problem to the lower-dimensional setting 
of Eqs.~(\ref{gpe_2D_2comp_m})-(\ref{gpe_2D_2comp_p}), see, e.g., the 
discussion of~\cite{siambook}.

Stationary solutions to Eqs.~\eqref{gpe_2D_2comp_m} and~\eqref{gpe_2D_2comp_p} 
with chemical potentials $\mu_{\pm}$ are found by assuming the
standing wave  ansatz
\begin{equation}
\Phi_{\pm}(\mathbf{r},t)=\phi_{\pm}(\mathbf{r})e^{-i\mu_{\pm}t}, \quad
\phi_{\pm}(\mathbf{r}):\overline{D}\mapsto \mathbb{C}.
\label{stat_ansatz}
\end{equation}
Upon inserting Eq.~\eqref{stat_ansatz} into Eqs.~\eqref{gpe_2D_2comp_m} 
and~\eqref{gpe_2D_2comp_p} we arrive at a boundary value problem consisting 
of two coupled (elliptic) nonlinear partial differential equations
\begin{subequations}
\begin{align}
{F_-((\phi_-,\phi_+),\mu_-)\doteq } -\frac{D_{-}}{2}\nabla ^{2} \phi _{-}+%
\left( g_{11}|\phi _{-}|^{2}+g_{12}|\phi _{+}|^{2}\right)%
\phi_{-}+V(\mathbf{r})\phi _{-}-\mu_{-}\phi_{-}&=0, \label{gpe_steady_2D_2comp_m}\\
{F_+((\phi_-,\phi_+),\mu_+)\doteq }-\frac{D_{+}}{2}\nabla ^{2}\phi _{+}+%
\left( g_{12}|\phi _{-}|^{2}+g_{22}|\phi _{+}|^{2}\right) %
\phi_{+}+V(\mathbf{r})\phi _{+}-\mu_{+}\phi_{+}&=0, \label{gpe_steady_2D_2comp_p}
\end{align}
\end{subequations}
together with zero Dirichlet boundary conditions, i.e., $\phi_{\pm}(\mathbf{r})|_{\partial D}=0$. 
The latter conditions are rather inconsequential (meaning that zero Neumann 
or other boundary conditions would function equally well) because we will 
consider wide enough domains that the confining potential of Eq.~(\ref{trap}) 
forces the density to tend to vanish well before we reach the edge of the 
computational domain. Furthermore 
\begin{equation}
F((\phi_-,\phi_+),(\mu_-,\mu_+)):=
\lbrace
F_{-}((\phi_-,\phi_+),\mu_-),F_{+}((\phi_-,\phi_+),\mu_+)
\rbrace
\end{equation}
represents the set of equations considered in the deflated continuation method (DCM).

We apply the DCM to Eqs.~\eqref{gpe_steady_2D_2comp_m} and~\eqref{gpe_steady_2D_2comp_p}
in order to find steady-state solutions $\phi_{\pm}^{0}(\mathbf{r})$ for various 
values of the bifurcation parameter $\mu_+$. This algorithm for performing 
bifurcation analysis is based on Newton's method and the appropriate choice 
of a deflation operator $G$ to compute multiple solutions to a nonlinear partial 
differential equation. It should be noted that the work of~\cite{egc_16} for 
the single-component NLS equation constructed deflated problems via
\begin{equation}
G(\phi) = \left(\frac{1}{\||\phi|^2-|\phi_1|^{2}\|_U^2}+1\right)F(\phi)
\end{equation}
in order to deflate the group orbit $\{e^{i\theta}\phi_1\mid \theta\in[0,2\pi)\}$. 
That is, steady-state solutions to the (single-component) NLS equation are not 
isolated: if $\phi_1$ is solution then so is $e^{i\theta}\phi_1$ for any 
$\theta\in [0,2\pi)$. The deflation operator 
given by Eq.~\eqref{eq_deflation} is not appropriate, as the solutions are not 
isolated, due to the Lie group of symmetries.

In this work, we overcome this problem by further extending the deflation
operator to eliminate rotations of solutions to the NLS equation. That is, 
we wish to deflate the group orbit 
\begin{equation}
\{\phi_1^{\theta}: (x, y) \mapsto \phi_1(x\cos\theta-y\sin\theta,x\sin\theta+y\cos\theta)\mid\theta\in[0,2\pi)\}.
\end{equation}
To do so, we define the rotationally invariant transformation of a function
$u$ denoted by $\hat{u}$ as 
\begin{equation}
\hat{u}(x, y)\doteq \frac{1}{2\pi}\int_0^{2\pi}u(x\cos\theta-y\sin\theta,x\sin\theta+y\cos\theta)\,d\theta. 
\label{eq_rotation_transf}
\end{equation}
Let $(\phi_{1-},\phi_{1+})$ be a solution to Eqs.~\eqref{gpe_steady_2D_2comp_m}
and~\eqref{gpe_steady_2D_2comp_p}, computed by Newton's method. We construct 
the deflated problem for finding the steady-state solutions to the two-component 
NLS system as follows:
\begin{equation}
G((\phi_-,\phi_+),(\mu_-,\mu_+)) = %
\left(\frac{1}{\left\|\left(\widehat{|\phi_-|^2}-\widehat{|\phi_{1-}|^2}\right)%
\left(\widehat{|\phi_+|^2}-\widehat{|\phi_{1+}|^2}\right)\right\|_U^2}+1\right)%
F((\phi_-,\phi_+),(\mu_-,\mu_+)), 
\label{eq_deflated_problem}
\end{equation}
where the norm for $U$ is the $L^2(D;\mathbb{R})$ norm. Since the amplitude is 
invariant under phase shift and the transformation defined by Eq.~\eqref{eq_rotation_transf} 
is unchanged by rotations, the modified problem defined by Eq.~\eqref{eq_deflated_problem}
ensures nonconvergence to a solution in the group orbit of $(\phi_-,\phi_+)$.

Having identified the steady-state solutions $\phi_{\pm}^{0}(\mathbf{r})$, 
we perform a stability analysis by assuming the perturbation ansatz
\begin{subequations}
\begin{eqnarray}
\widetilde{\Phi}_{-}(\mathbf{r},t)&=&e^{-i\mu _{-}t}\left[\phi _{-}^{0}+\varepsilon
\left( a(\mathbf{r})e^{i\omega t}+b^{\ast }(\mathbf{r})e^{-i\omega ^{\ast }t}\right) \right],\\
\widetilde{\Phi}_{+}(\mathbf{r},t)&=&e^{-i\mu _{+}t}\left[ \phi _{+}^{0}+\varepsilon
\left( c(\mathbf{r})e^{i\omega t}+d^{\ast }(\mathbf{r})e^{-i\omega ^{\ast }t}\right) \right],
\end{eqnarray}
\label{lin_ansatz}%
\end{subequations}
where $\omega\in\mathbb{C}$ is the eigenfrequency, $\varepsilon\ll 1$ is a 
(formal) small parameter, and $\ast$ indicates complex conjugation. 
Upon inserting Eqs.~(\ref{lin_ansatz}) into Eqs.~\eqref{gpe_2D_2comp_m} 
and~\eqref{gpe_2D_2comp_m}, we obtain at order $\mathcal{O}(\varepsilon )$
an eigenvalue problem written in matrix form as
\begin{equation}
\rho
\begin{pmatrix}
a \\
b \\
c \\
d%
\end{pmatrix}
=%
\begin{pmatrix}
A_{11} & A_{12} & A_{13} & A_{14} \\
-A_{12}^{\ast } & -A_{11} & -A_{14}^{\ast } & -A_{13}^{\ast } \\
A_{13}^{\ast } & A_{14} & A_{33} & A_{34} \\
-A_{14}^{\ast } & -A_{13} & -A_{34}^{\ast } & -A_{33}%
\end{pmatrix}
\begin{pmatrix}
a \\
b \\
c \\
d%
\end{pmatrix}%
,  \label{eig_prob}
\end{equation}%
with eigenvalue $\rho =-\omega $, eigenvector $\mathcal{V}=\left[a\,b \,c \,d\right]^{T}$, 
and matrix elements given by
\begin{subequations}
\begin{eqnarray}
A_{11} &=&-\frac{D_{-}}{2}\nabla^{2}+%
\left(2g_{11}|\phi _{-}^{0}|^{2}+g_{12}|\phi _{+}^{0}|^{2}\right) +V-\mu _{-},
\label{A11} \\
A_{12} &=&g_{11}\,\left( \phi _{-}^{0}\right) ^{2}, \\
A_{13} &=&g_{12}\,\phi _{-}^{0}\left( \phi _{+}^{0}\right) ^{\ast },
\\
A_{14} &=&g_{12}\,\phi _{-}^{0}\phi _{+}^{0}, \\
A_{33} &=&-\frac{D_{+}}{2}\nabla^{2}+\left(
g_{12}|\phi _{-}^{0}|^{2}+2g_{22}|\phi _{+}^{0}|^{2}\right) +V-\mu _{+},
\label{A33} \\
A_{34} &=&g_{22}\,\left( \phi _{+}^{0}\right) ^{2}.
\end{eqnarray}
\end{subequations}
If we decompose the eigenfrequencies $\omega$ into their real and imaginary 
parts according to $\omega=\omega_{r}+i\,\omega _{i}$, then the following 
cases regarding the classification of steady-states $\phi^{0}_{\pm}$ in terms
of their stability can be distinguished. If $\omega_{i}=0$ holds for all 
$\omega$, then the steady-state $\phi^{0}_{\pm}$ is classified as spectrally 
stable. For practical considerations, in our numerical results presented in 
the next section, we assume that $\phi^{0}_{\pm}$ is stable if $\omega_{i}<10^{-3}$ 
holds for all eigenfrequencies; this case scenario is depicted by solid 
blue lines in the bifurcation diagrams presented therein. On the contrary, 
if $\omega_{i}\neq 0$, then two types of instabilities can be identified: 
\begin{itemize}
\item If $\omega_{r}=0$ holds, then $\phi^{0}_{\pm}$ is classified as 
\textit{exponentially unstable} and characterized by a pair of 
imaginary eigenfrequencies. \item If $\omega_{r}\neq 0$, then $\phi^{0}_{\pm}$ 
is classified as \textit{oscillatorily unstable} and characterized by a 
complex eigenfrequency quartet.
\end{itemize}
These two scenarios are depicted by dashed-dotted red and green lines 
respectively in the bifurcation diagrams that follow to highlight 
the nature of the dominant unstable mode. If a change in the dominant 
instability type happens, e.g., from an exponentially unstable to
an oscillatory unstable steady-state solution, that change will be 
highlighted by a change between the respective colors. The eigenvalue 
problem of Eq.~\eqref{eig_prob} is very challenging to solve efficiently 
and accurately; technical details of how to solve these eigenproblems,
using, e.g., FEAST, are described in Appendix~\ref{feast}.

For our numerical computations presented below, we consider the spatial 
domain $D = (-12, 12)^2$. Hereafter, we fix $D_{-}=D_{+}\equiv 1$ and 
set $\Omega=0.2$, $\mu_{-}=1$, $g_{11}=1.03$, $g_{22}=0.97$ and $g_{12}=1$, 
motivated by relevant values of coefficients previously used in the case 
of $^{87}$Rb~\cite{siambook}. The case of Rb is one where the two components
are normally immiscible~\cite{pethick,stringari,siambook}, that is to say
it is energetically favored for them not to occupy the same location in 
space, a key driving force in the pattern formation that we will observe 
below. Slight deviations of the interaction coefficients from the
above values will not change the essence of our results, provided that
one stays within this weakly immiscible regime.

Fixing the above parameters, we perform a natural parameter continuation 
in the value of $\mu_{+}$, up to $\mu_{+}=1.355$ (or $\mu_{+}=1.4$ in some 
cases). A useful notion in this context is that of the Thomas-Fermi
limit, obtained from
Eqs.~(\ref{gpe_steady_2D_2comp_m}) and (\ref{gpe_steady_2D_2comp_p})
when eliminating the Laplacian terms~\cite{pethick,stringari,siambook}.
It should be noted in passing that the Thomas-Fermi solution
of the ``$+$'' component is given (in general) by
\begin{equation}
|\phi_{+}|^{2}=\frac{1}{g_{22}}\left[\mu_{+}-V(\mathbf{r})-g_{12}|\phi_{-}|^{2}\right],
\end{equation}
at those points on the plane where $\mu_{+}>V(\mathbf{r})-g_{12}|\phi_{-}|^{2}$ 
holds whereas the solution is zero elsewhere. However, as $\mu_{+}$ becomes large 
enough during the continuation process, and due to the nonlinear coupling between
the two components, the ``$-$'' component gradually becomes smaller in its amplitude
and eventually vanishes giving (at this limit) a Thomas-Fermi solution exactly the 
same as the single-component case with a Thomas-Fermi radius of $R_{\textrm{TF}}\approx 8.23$ 
(or $\approx 8.37$). On the other hand, and for the linear limit of the ``$+$'' 
component, i.e., the values of $\mu_{+}$ where a bound pair forms but with $|\phi_{+}|\ll 1$, 
the associated Thomas-Fermi solution and radius of the ``$-$'' component again could 
be reduced to the single-component case with a radius in particular of $R_{\textrm{TF}}\approx 7.07$ 
(note that $\mu_{-}=1$ and is kept fixed). To put it otherwise, in the weakly immiscible 
regime considered herein, there exist two distinct Thomas-Fermi configurations, one dominated 
by the ``$+$'' component with the ``$-$'' component being absent and vice-versa. In either of 
these limits nevertheless, the coherent structures sit comfortably inside the chosen 
computational domain.

Two spatial discretizations were used in this work. First, the DCM was applied 
to a finite element discretization of Eqs.~\eqref{gpe_steady_2D_2comp_m} and 
\eqref{gpe_steady_2D_2comp_p} using FEniCS \cite{logg2011} on a relatively coarse mesh (as this requires 
the solution of many nonlinear systems with Newton's method). The resulting 
solutions were then used as initial guesses for a second-order accurate centered 
finite difference scheme on a uniform two-dimensional 
grid of equidistant points with lattice spacing $\Delta x\equiv \Delta y=0.08$.
The underlying linear system arising at each Newton step was solved by 
using the induced dimension reduction IDR(s) algorithm~\cite{idrs_1} (see 
also~\cite{kody_pgk} for its applicability on a similar computation). The solutions
obtained by the latter numerical scheme were subsequently used for calculating their
spectra.

In the bifurcation diagrams presented in the following section, we use the total number 
of atoms (or the sum of the (squared) $L^{2}$-norms of the respective fields) as our 
diagnostic defined via
\begin{equation}
N_{t}=N_{+}+N_{-}, \quad N_{\pm}=\int_{{D}}|\phi_{\pm}|^{2}\,dxdy,
\label{totn}
\end{equation}
as well as the absolute total-number-of-atoms difference
\begin{equation}
\Delta N_{t}=\Big|N_{t}^{(\textrm{a})}-N_{t}^{\textrm{(b)}}\Big|,
\label{totnd}
\end{equation}
between the total number of atoms of branches
$\textrm{(a)}$ and $\textrm{(b)}$.

\section{Numerical results}

\begin{figure}[htp]
\vskip -0.5cm
\begin{center}
\includegraphics[height=.25\textheight, angle =0]{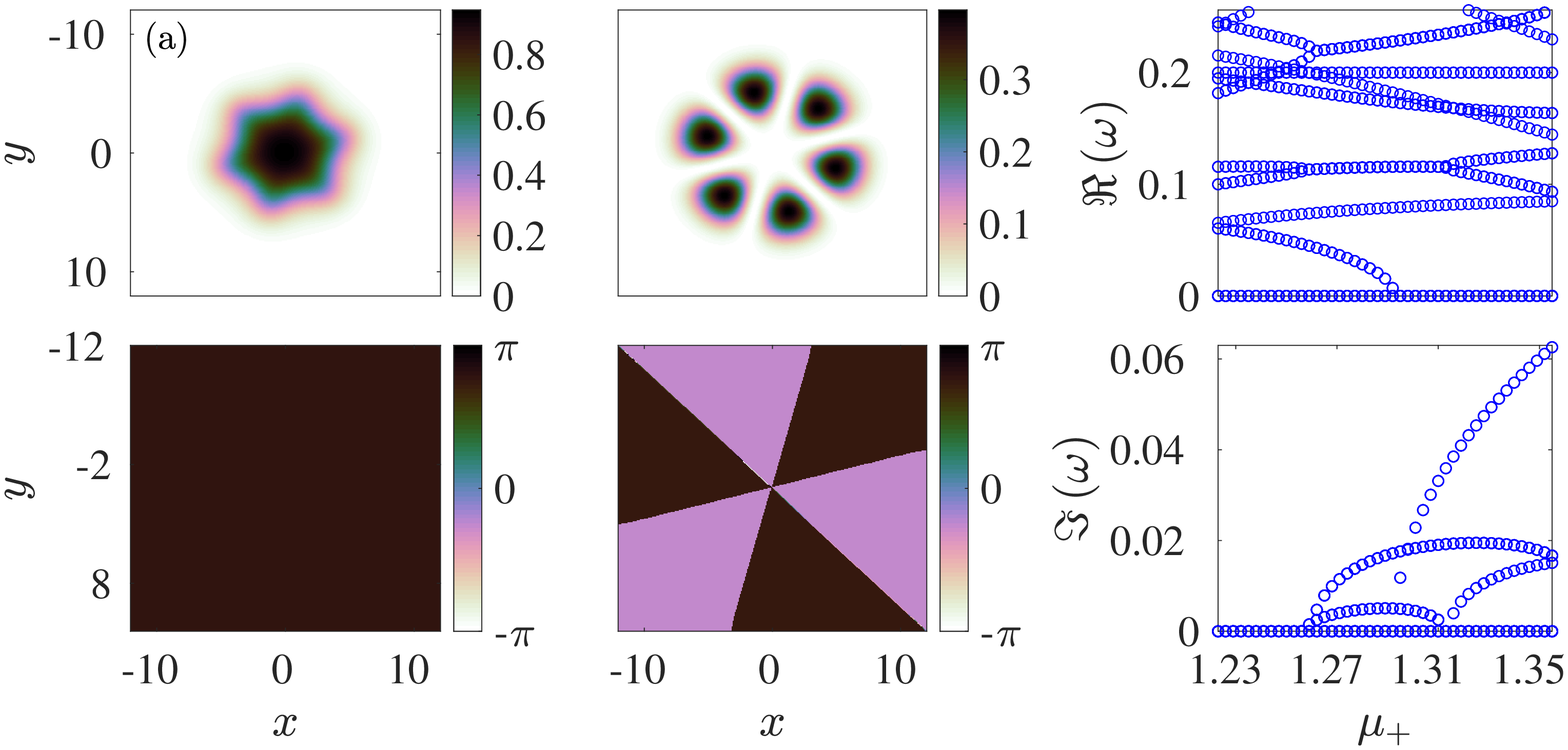}
\includegraphics[height=.25\textheight, angle =0]{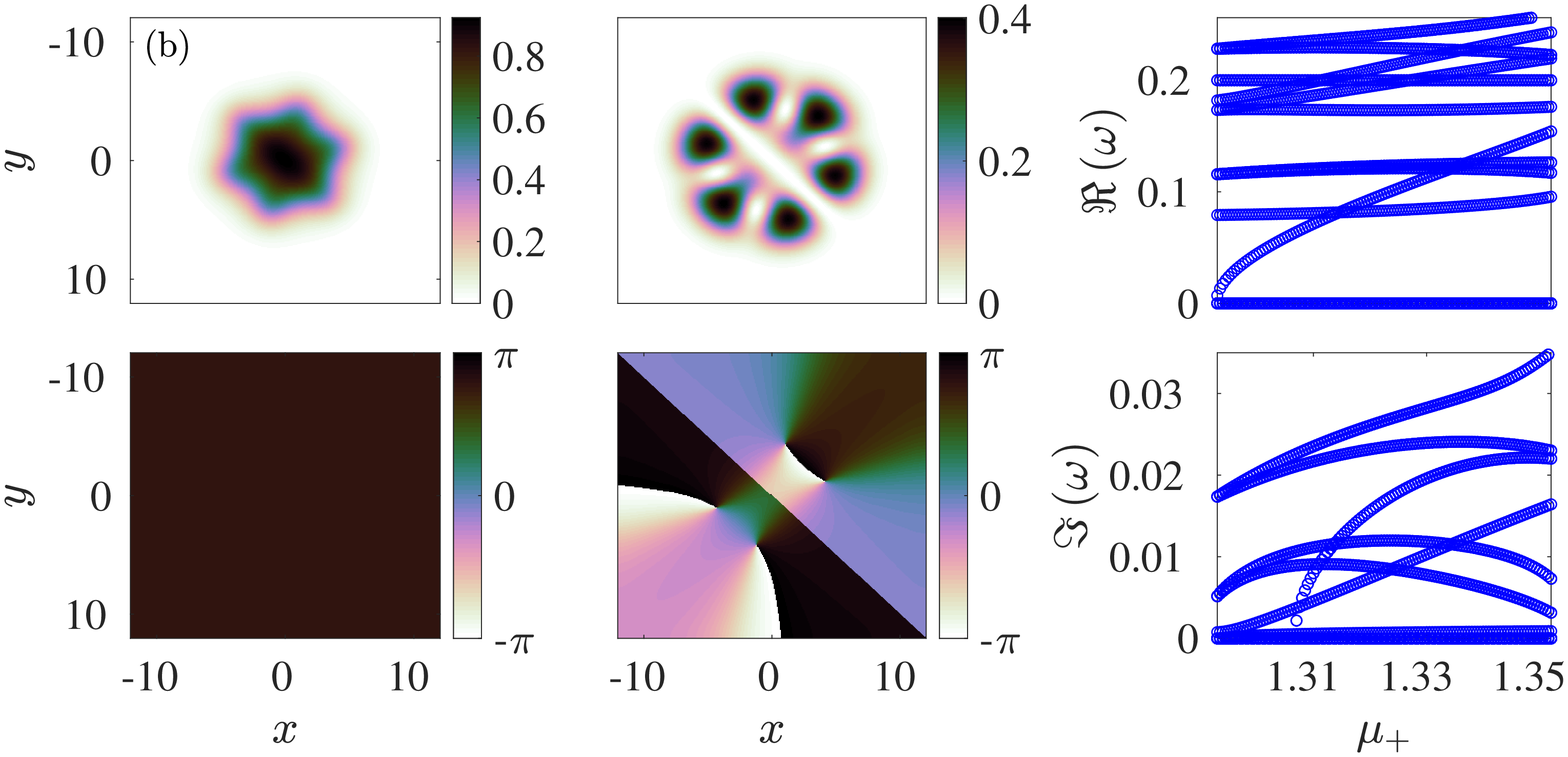}
\includegraphics[height=.17\textheight, angle =0]{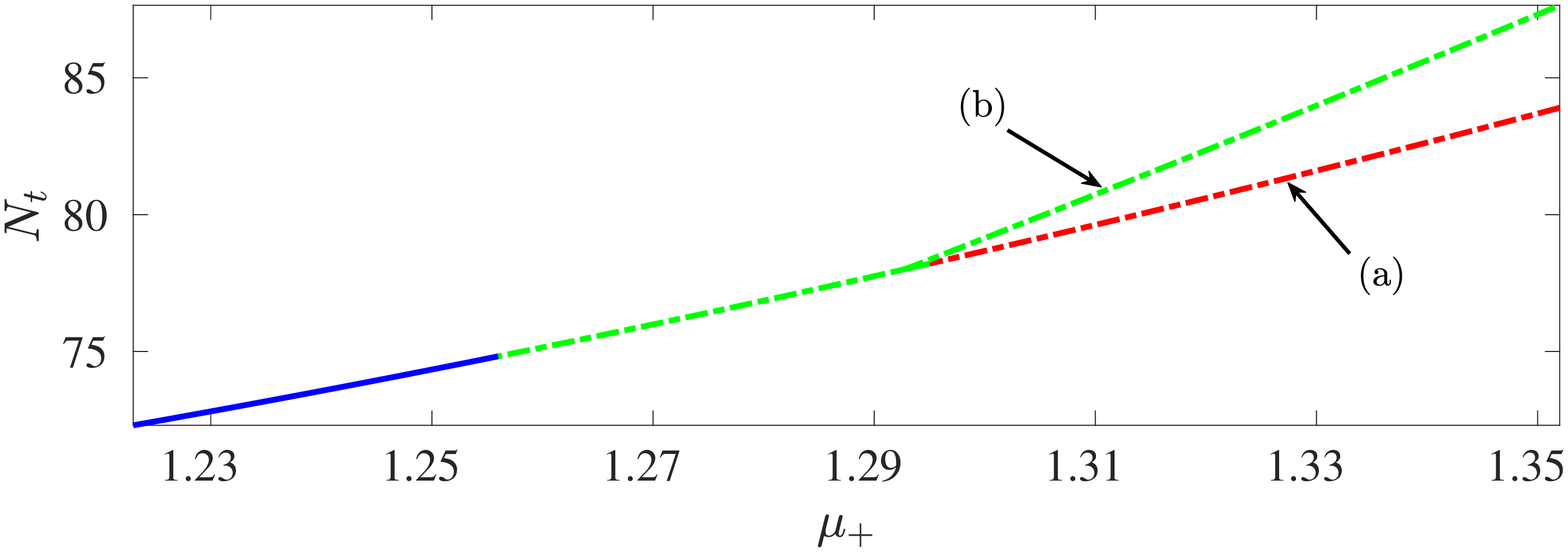}
\includegraphics[height=.17\textheight, angle =0]{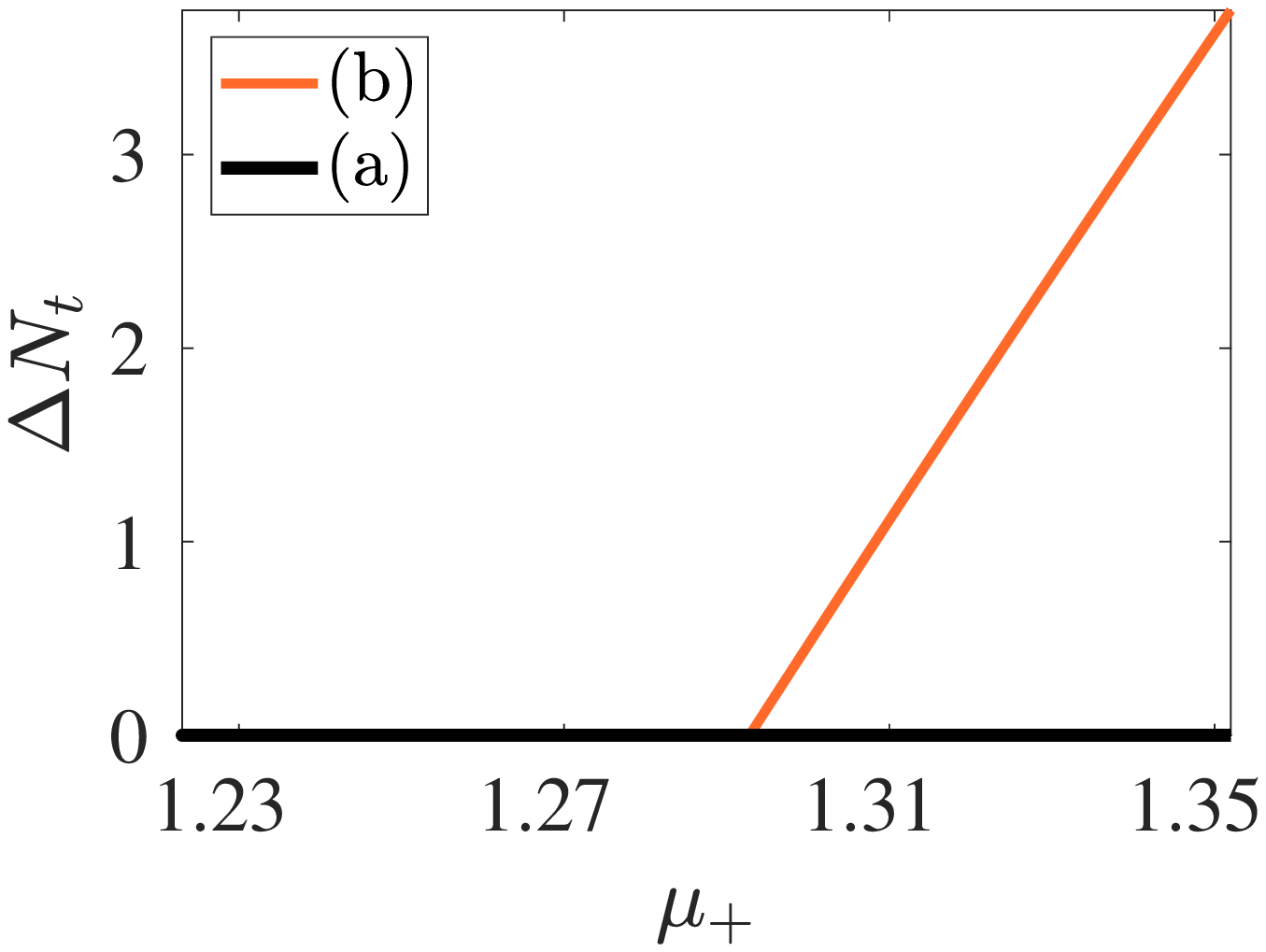}
\end{center}
\caption{
A bifurcation diagram of a bright-soliton-necklace state. In particular, 
the clustered panels (a) present the $|\phi_{-}|^{2}$ (top left) and 
$|\phi_{+}|^{2}$ (top middle), together with the respective phases 
(bottom left and middle panels therein). In addition, the associated 
spectra are presented where the real (top right) and imaginary (bottom
right) parts of the relevant eigenfrequencies $\omega$ are depicted (for 
the real part of these eigenfrequencies only the lowest ones are shown). 
Similarly, the clustered panels (b) correspond to the bifurcated state 
of a stripe and two vortex dipoles in the second component emerging at 
$\mu_{+}\approx 1.293$. Bottom left and right panels correspond to the 
total number of atoms $N_{t}$ (left) and atom number difference $\Delta N_{t}$ 
(right) as functions of $\mu_{+}$ (see text for its definition). The 
vanishing of the latter is used to signal the bifurcation point. Recall 
that solid blue lines denote stability, while red and green dash-dotted 
ones exponential and oscillatory instability, respectively, in the bifurcation 
diagram here and in what follows.
}
\label{fig0}
\end{figure}

We begin the presentation of our results with Fig.~\ref{fig0}. Specifically, 
each panel shows the densities (top row) of $\phi_{-}$ (top left) and 
$\phi_{+}$ (top middle) together with the associated phases (bottom left and
middle) as well as the real (top right) and imaginary parts (bottom right) 
of the eigenfrequencies computed using the highly accurate FEAST eigenvalue solver 
(see Appendix~\ref{feast}). The state in Fig.~\ref{fig0}(a) corresponds to a
bound mode consisting of a ground state in the $``-"$ component and a soliton 
necklace in the $``+"$ component (hereafter, we will refer to the $``-"$ 
and $``+"$ components as first and second components, respectively). The 
latter state was identified in the single-component NLS equation (see~\cite{egc_16}
and references therein) and is generally unstable. It is relevant to note 
that this hexagonal state bears alternating phases between its constituent
``blobs'' (as is customary generally in such dipolar, quadrupolar etc. states) 
and this hexagonal symmetry bears an imprint also on the spatial profile of 
the first component due to the weak immiscibility between the components for 
the case of $^{87}$Rb, as explained above.

\begin{figure}[htp!]
\vskip -0.5cm
\begin{center}
\includegraphics[height=.25\textheight, angle =0]{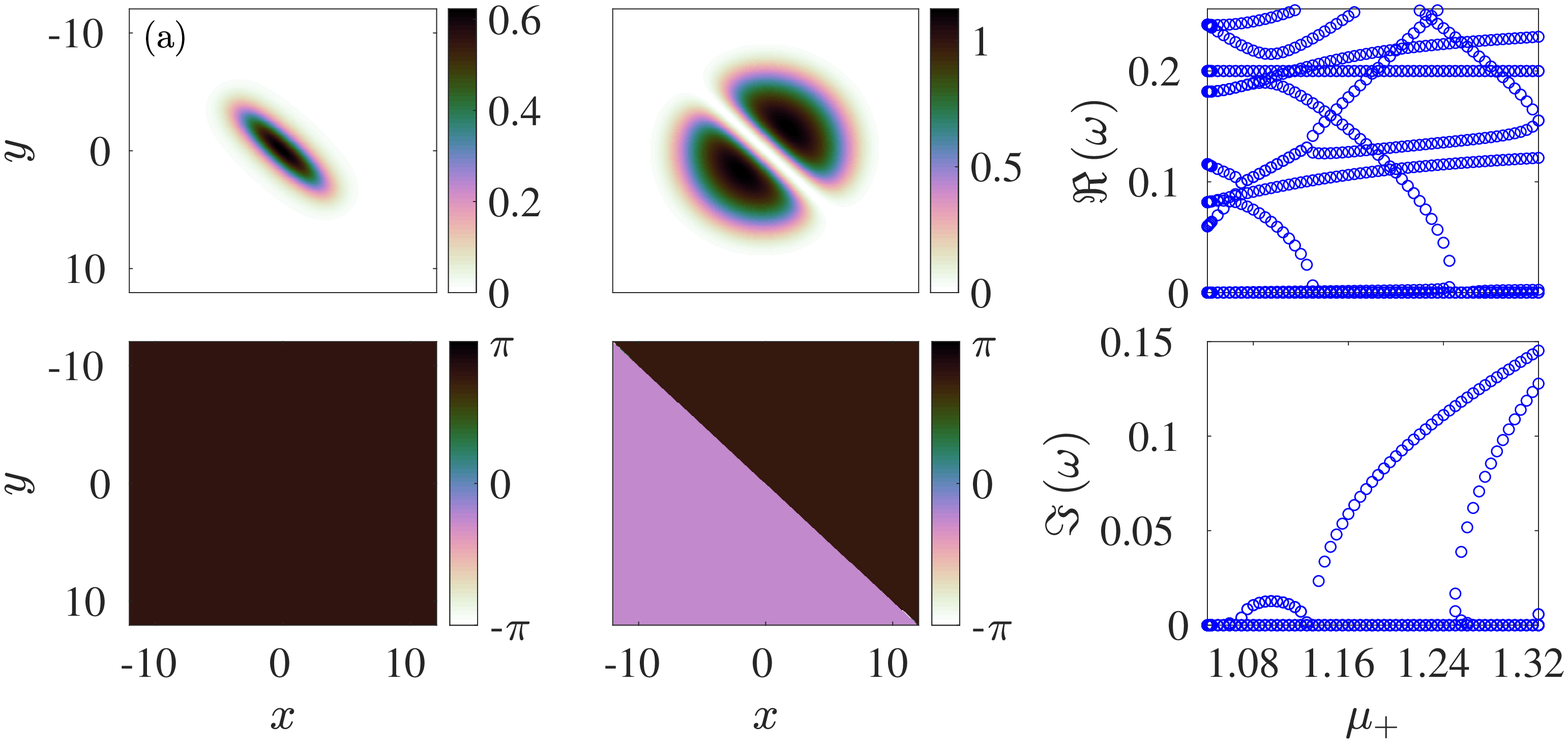}
\includegraphics[height=.25\textheight, angle =0]{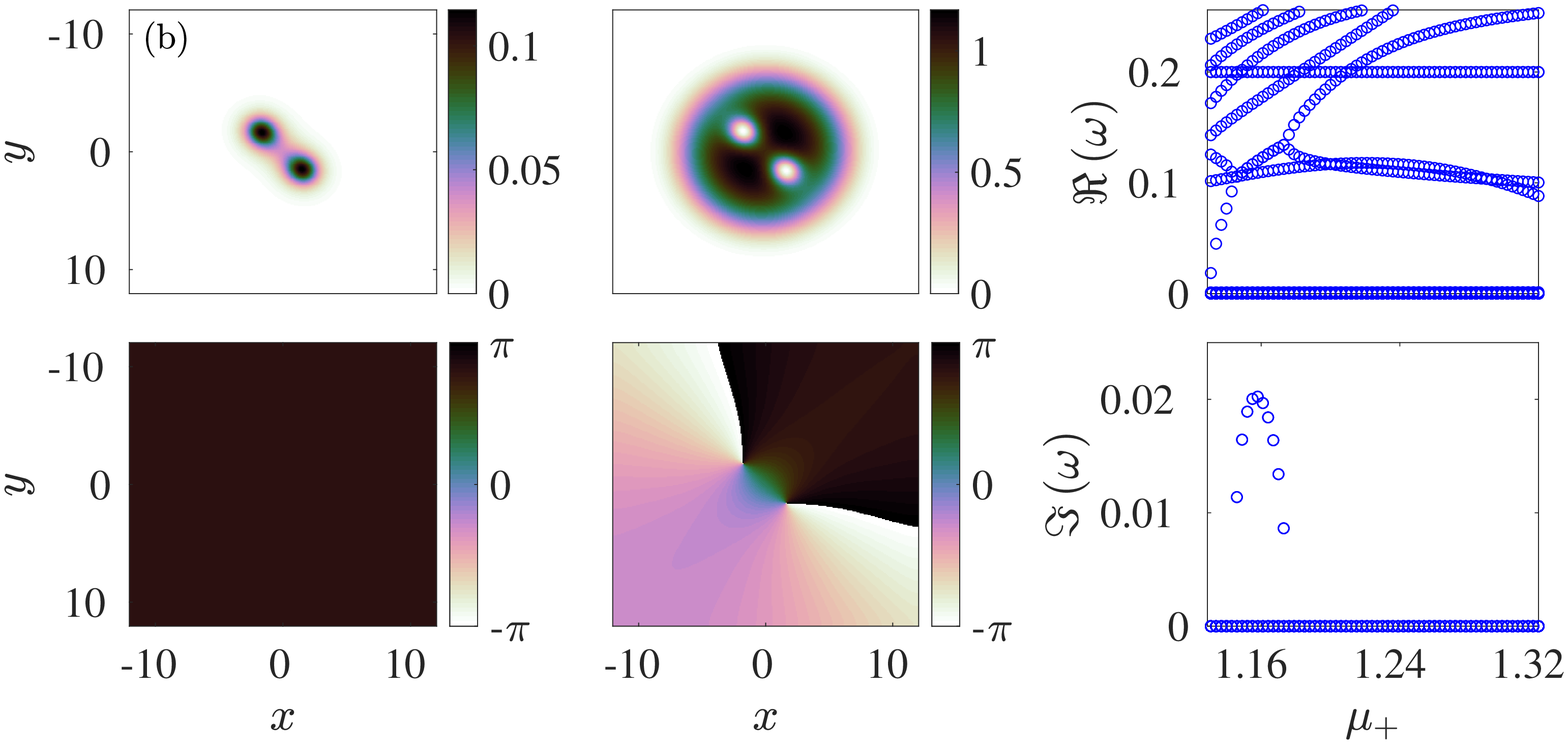}
\includegraphics[height=.25\textheight, angle =0]{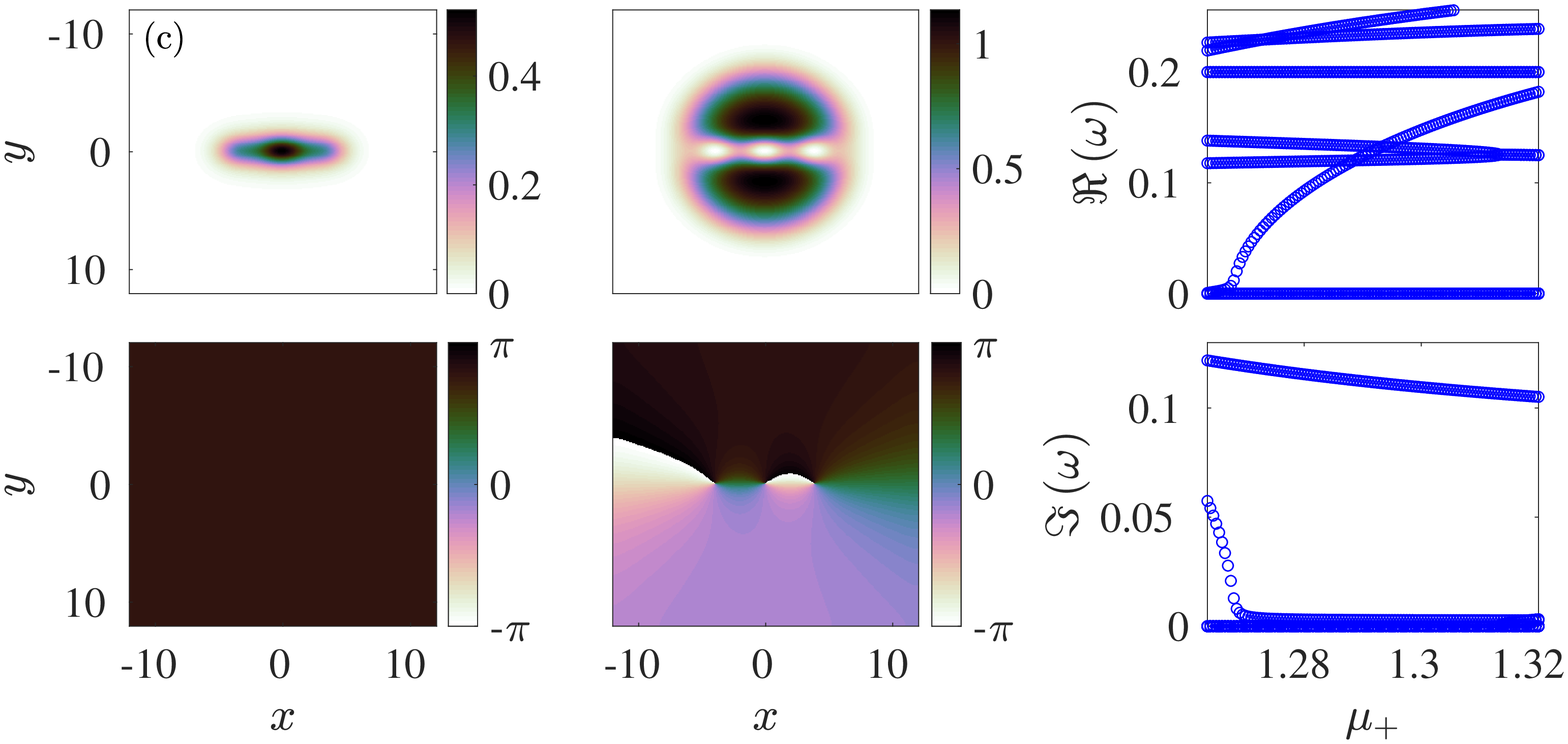}
\includegraphics[height=.17\textheight, angle =0]{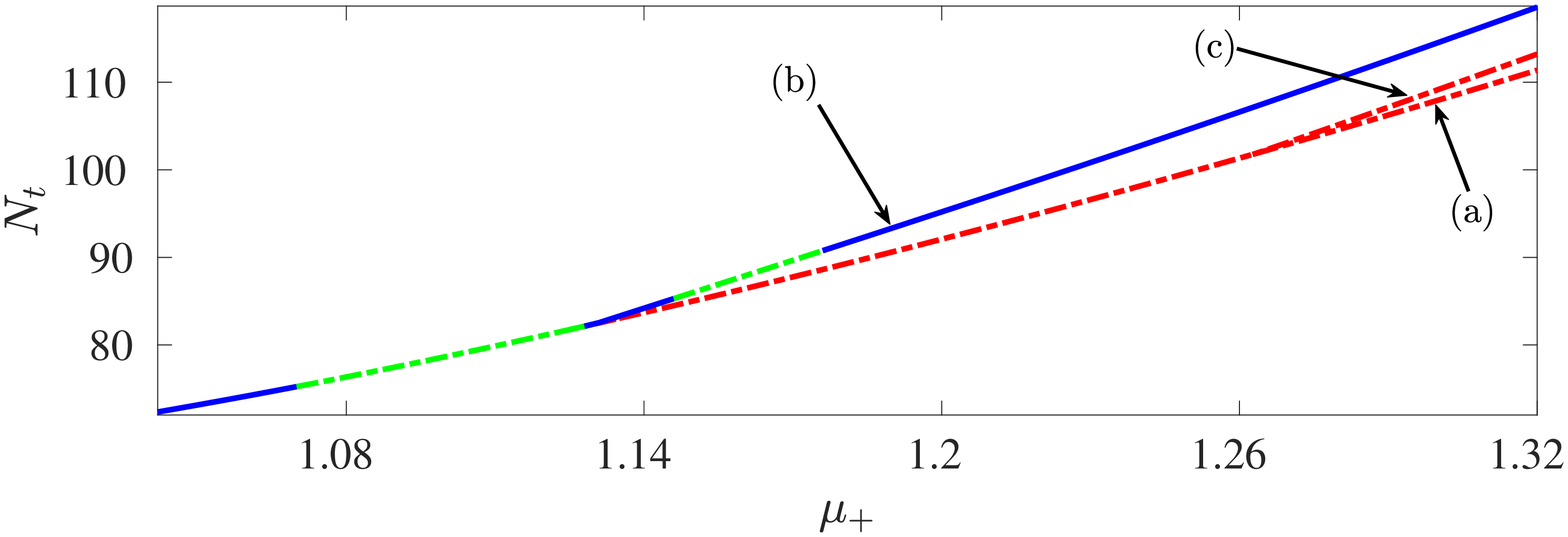}
\includegraphics[height=.17\textheight, angle =0]{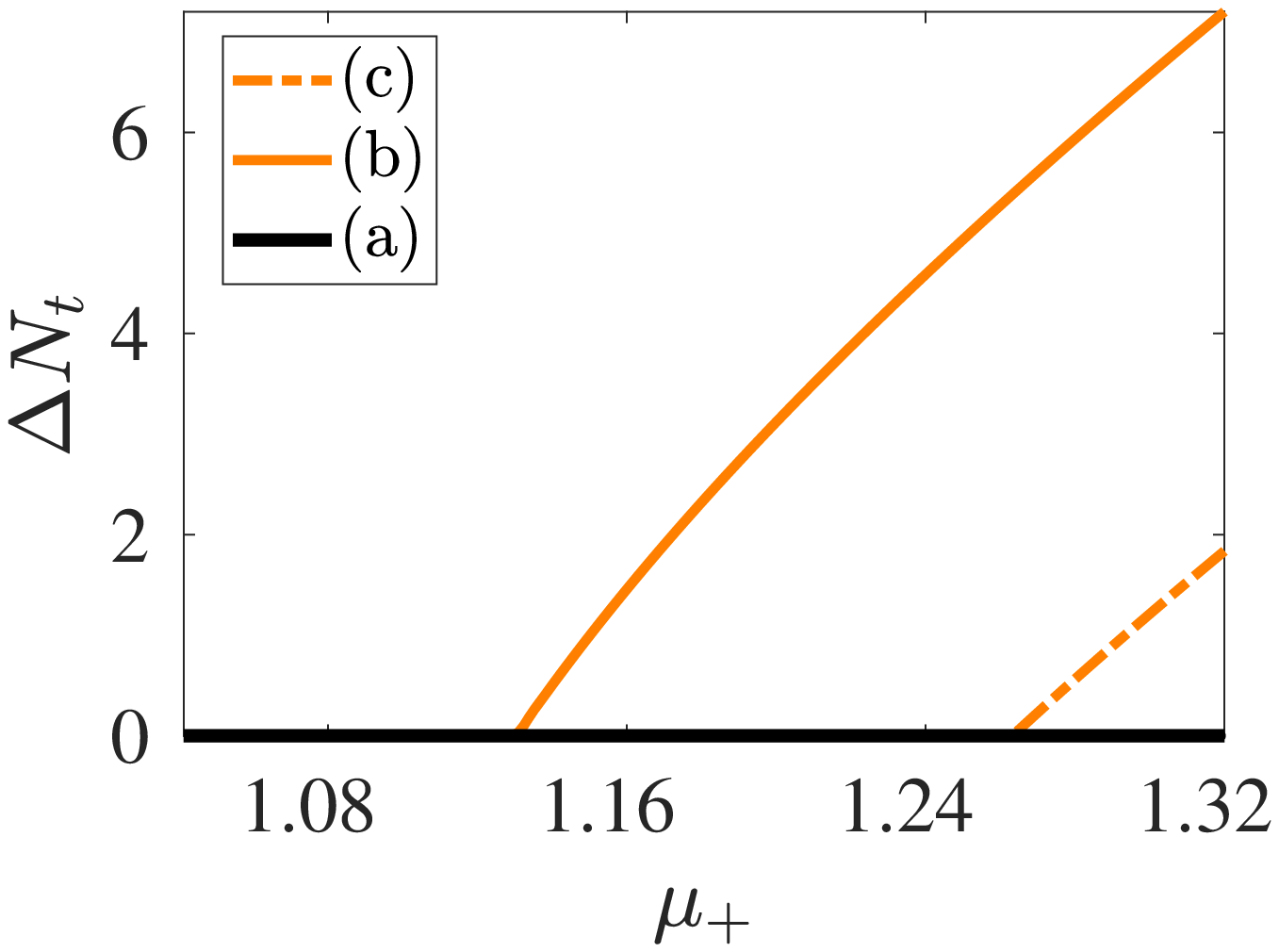}
\end{center}
\caption{
Same as Fig.~\ref{fig0} but for the dark-bright (DB) soliton stripe 
branch. All densities and respective phases presented in panels (a)-(c) 
are shown for values of $\mu_{+}$ of $\mu_{+}=1.32$. Note that 
bifurcations happen at $\mu_{+}\approx 1.13$ (b) and $\mu_{+}\approx 1.263$ (c), 
respectively. The emerging branches from the bifurcations of the
DB stripe of the top panel are the VB dipole of the middle panel 
(b) and the VB soliton tripole of the lower panel (c).
}
\label{fig1}
\end{figure}
\begin{figure}[htp]
\vskip -0.5cm
\begin{center}
\includegraphics[height=.25\textheight, angle =0]{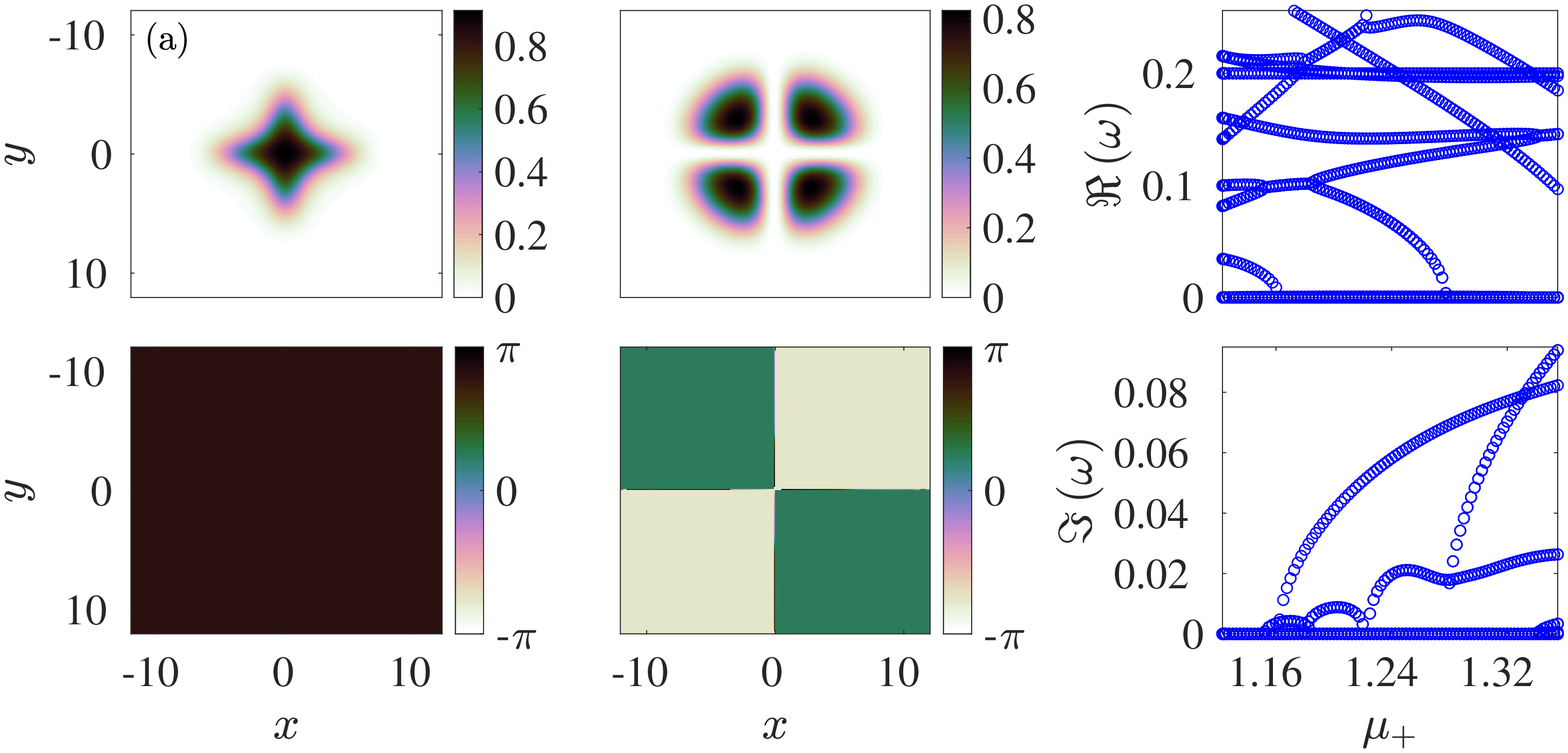}
\includegraphics[height=.25\textheight, angle =0]{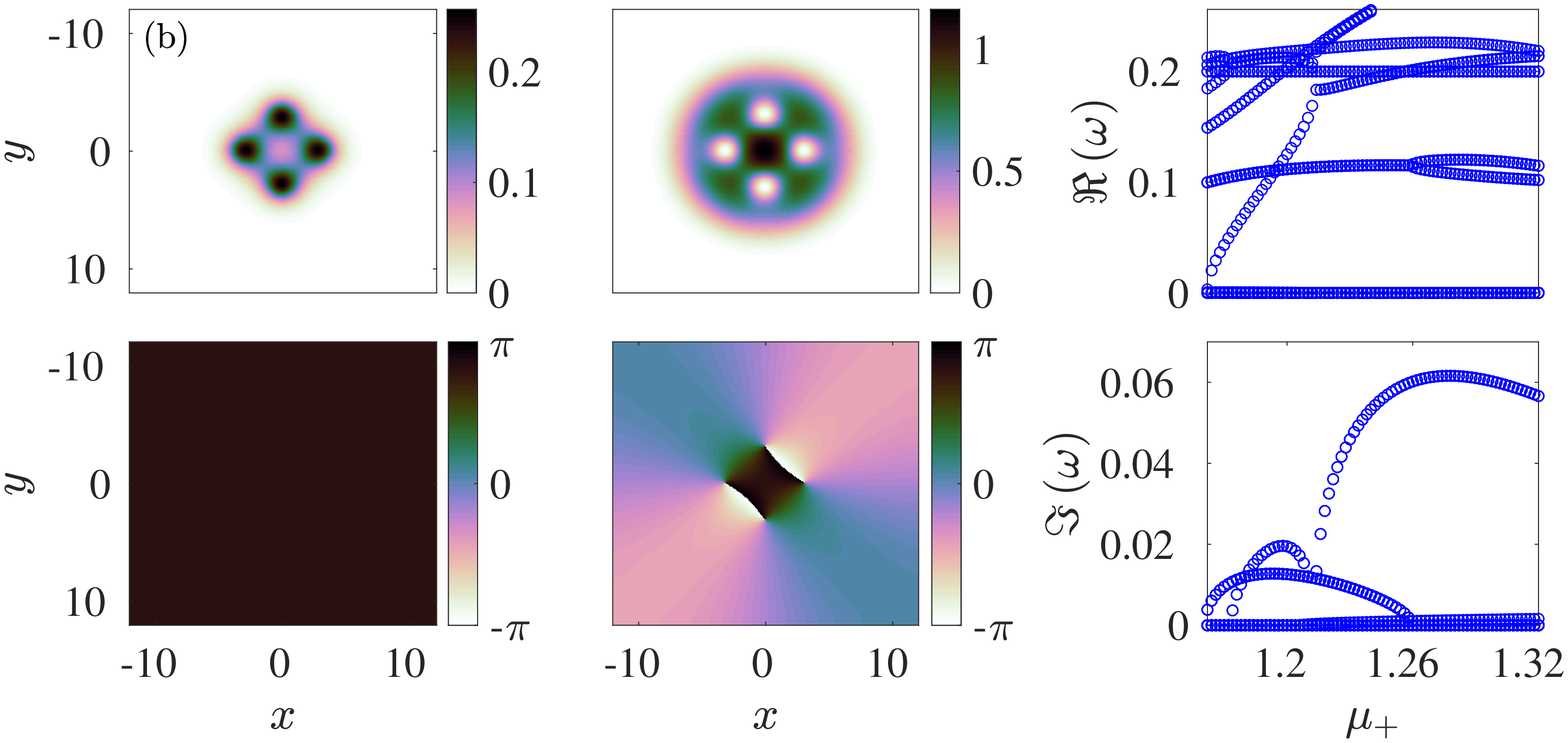}
\includegraphics[height=.25\textheight, angle =0]{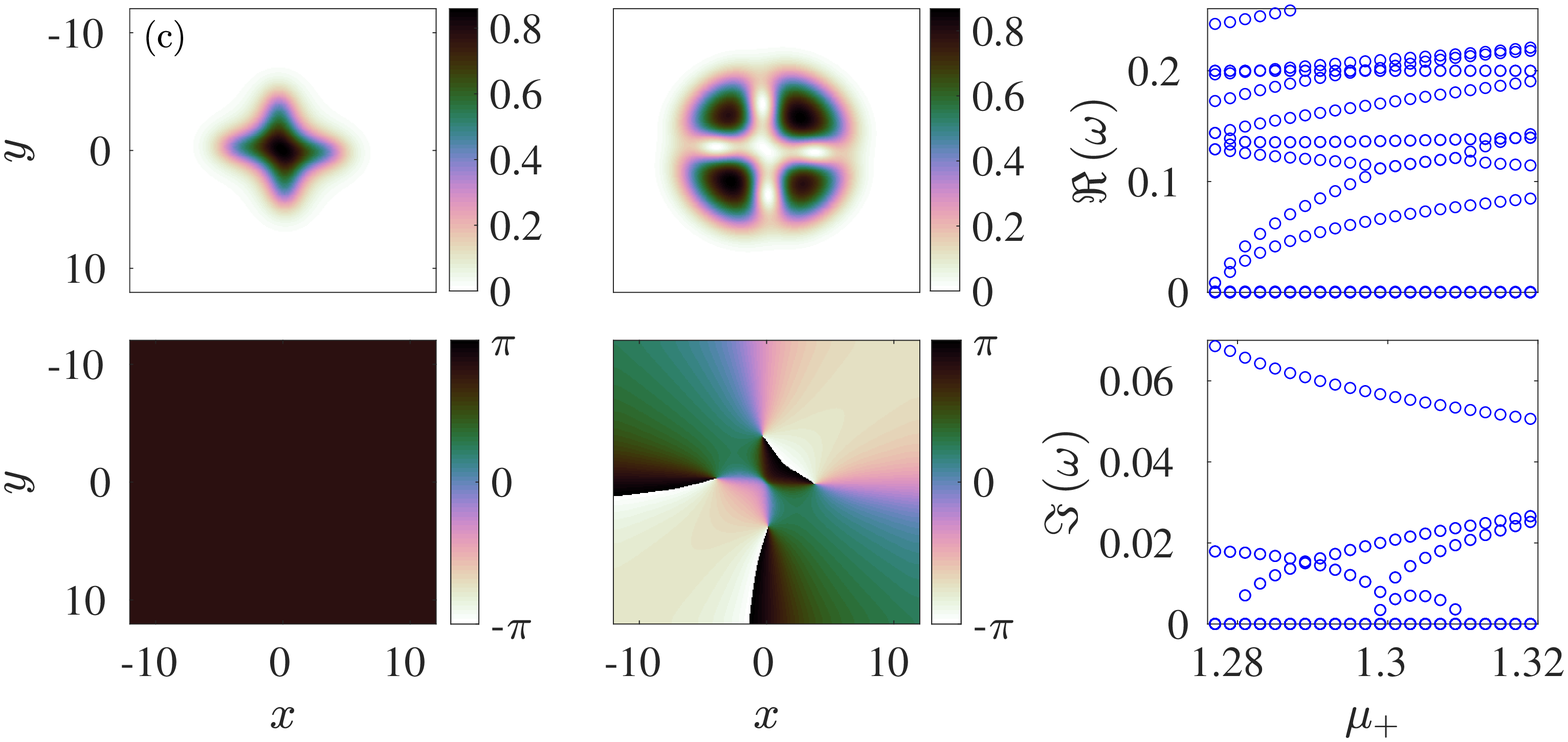}
\includegraphics[height=.17\textheight, angle =0]{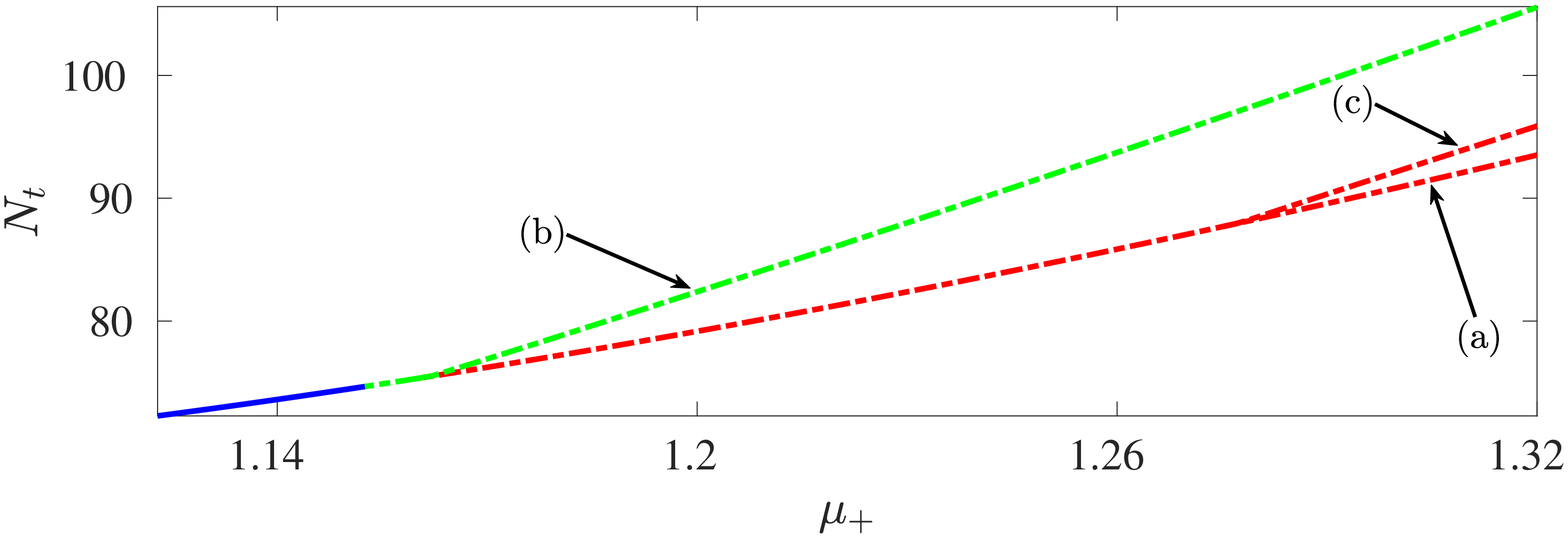}
\includegraphics[height=.17\textheight, angle =0]{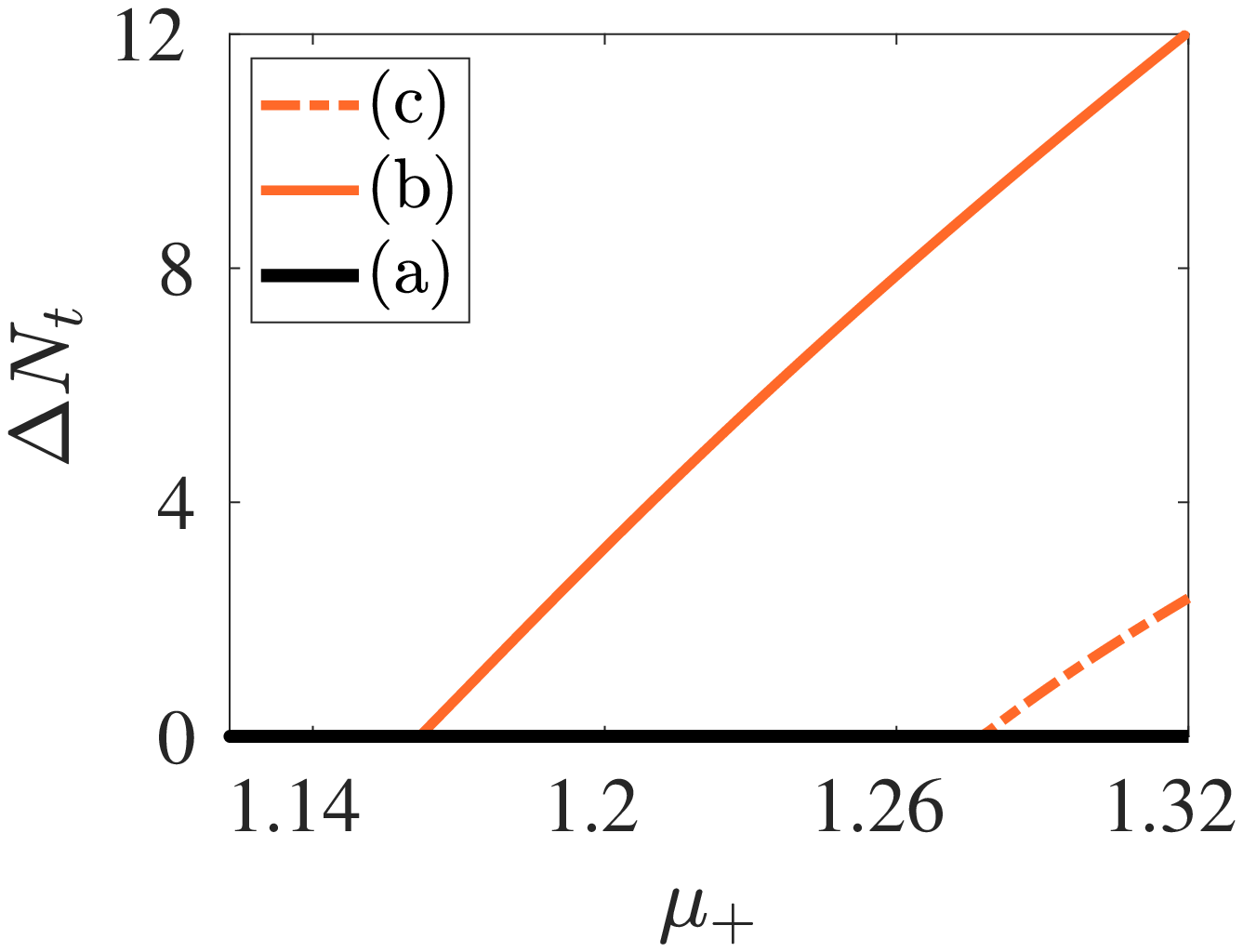}
\end{center}
\caption{
Same as Fig.~\ref{fig0} but for the crossed dark-bright (DB) soliton
branch. Densities and phases shown in panels (a) and (b) correspond 
to a value of $\mu_{+}$ of $\mu_{+}=1.32$ whereas the ones presented
in (c) correspond to $\mu_{+}=1.319$. Note that branch (b) emerges 
at $\mu_{+}\approx 1.162$ and branch (c) at $\mu_{+}\approx 1.279$.
Branch (b) corresponds to a vortex-bright (VB) quadrupole, while in 
(c) a structure with additional vorticity at the center bifurcates 
from branch (a).
}
\label{fig2}
\end{figure}
\begin{figure}[hpt]
\vskip -0.5cm
\begin{center}
\includegraphics[height=.25\textheight, angle =0]{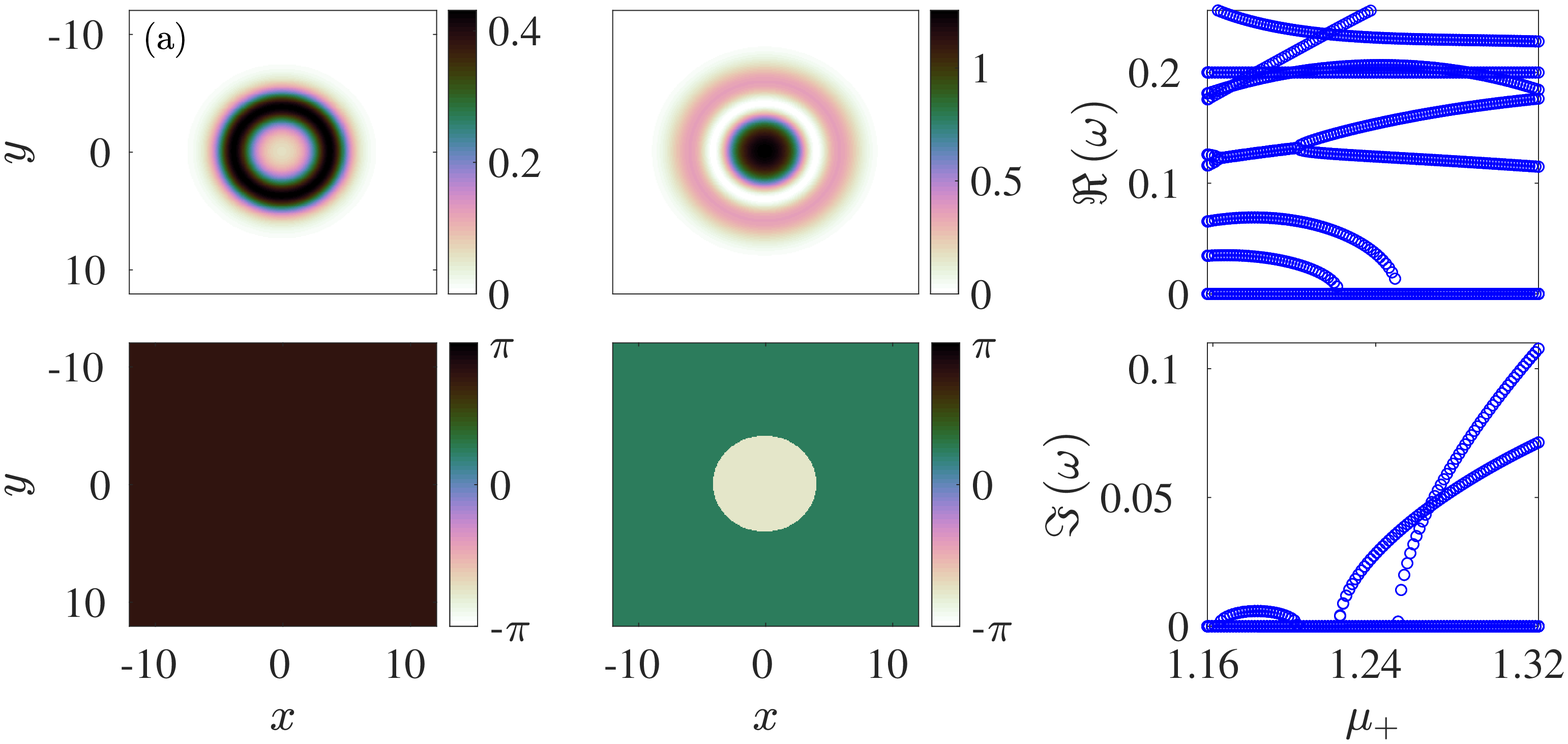}
\includegraphics[height=.25\textheight, angle =0]{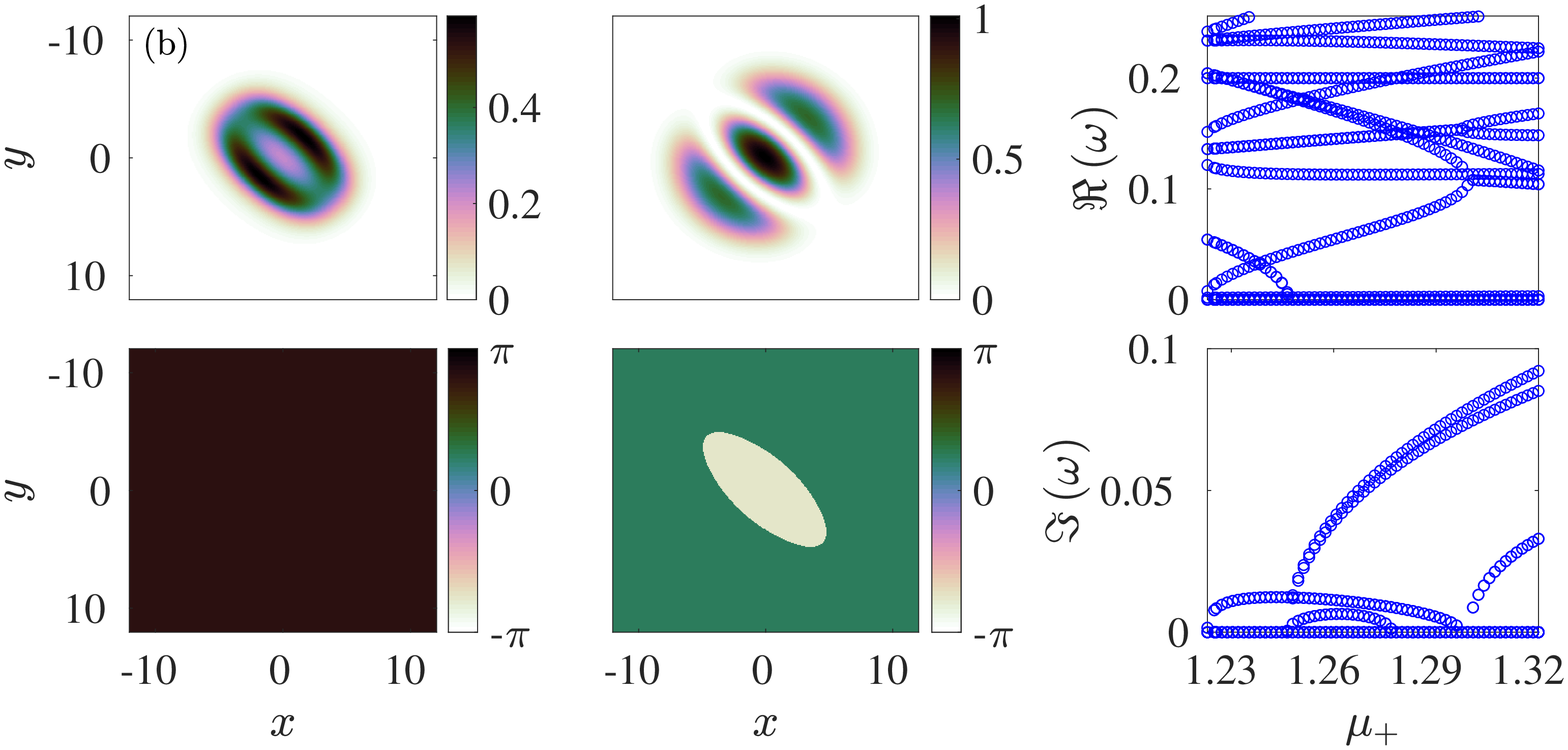}
\includegraphics[height=.25\textheight, angle =0]{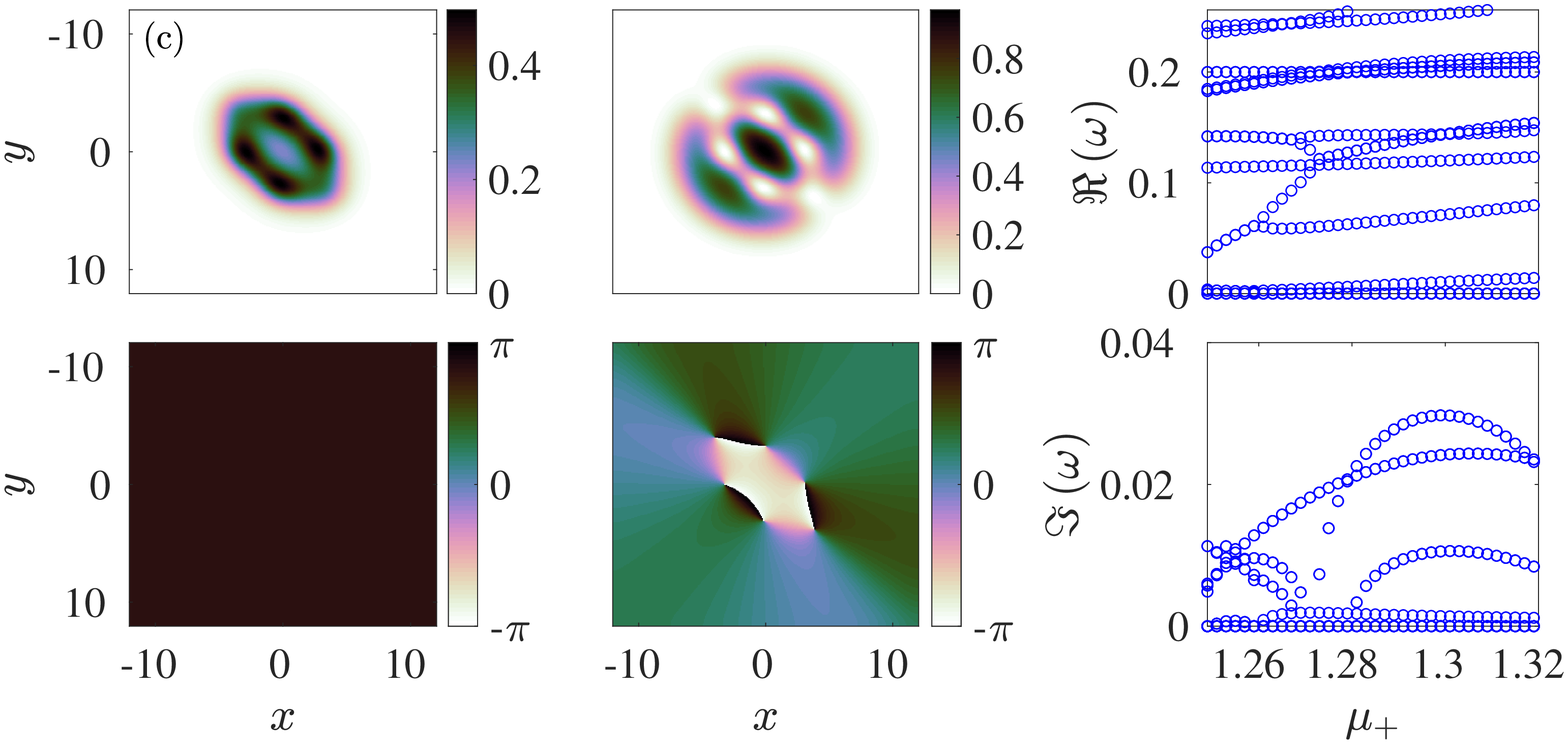}
\includegraphics[height=.25\textheight, angle =0]{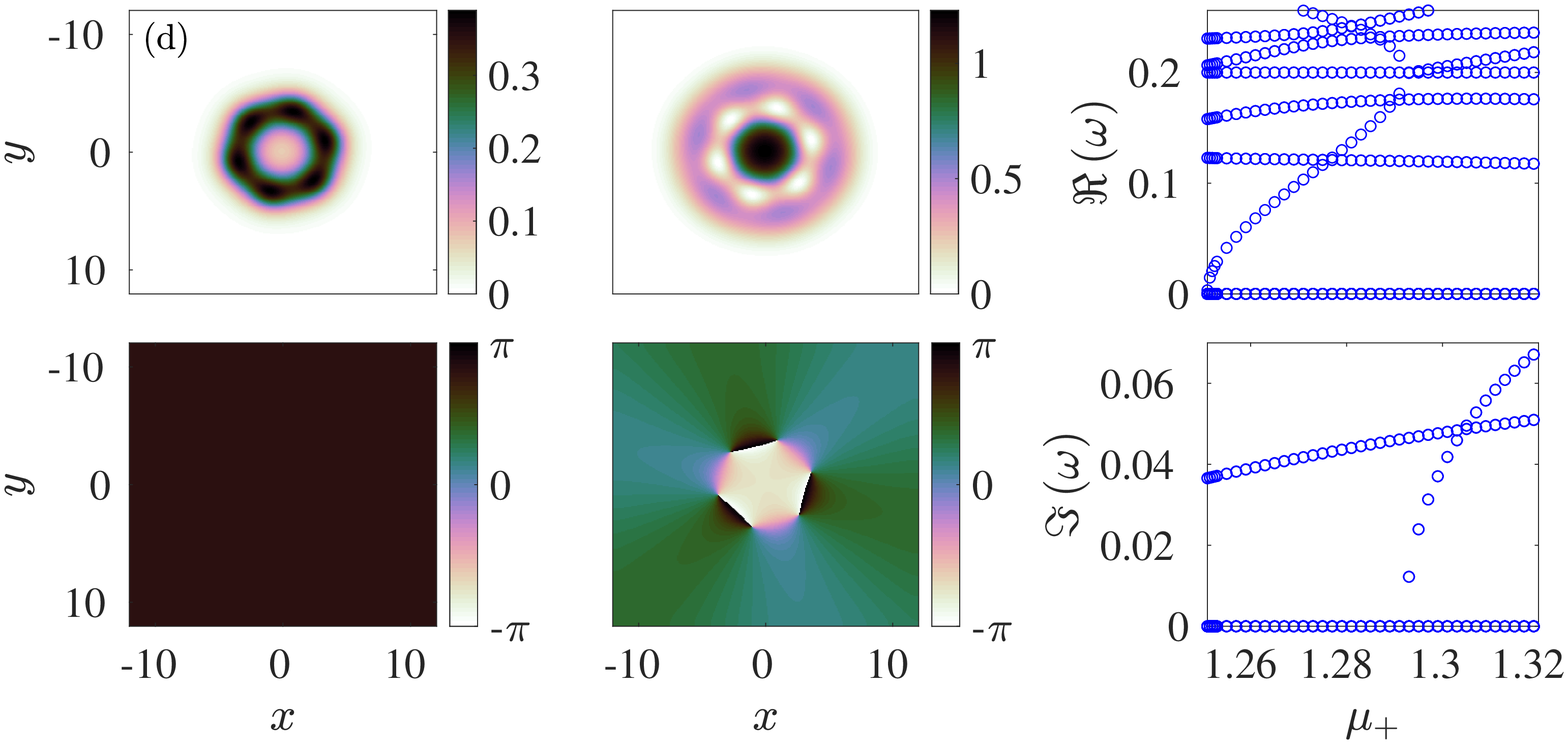}
\end{center}
\caption{
Same as Fig.~\ref{fig0} but for the dark-bright (DB) ring soliton branch. 
Densities and phases shown in panels (a), (b) and (c) correspond to a value 
of $\mu_{+}$ of $\mu_{+}=1.32$ whereas the one presented in (d) corresponds 
to $\mu_{+}=1.319$. Note that branch (b) emerges at $\mu_{+}\approx 1.222$, 
branch (c) at $\mu_{+}\approx 1.247$, and (d) at $\mu_{+}\approx 1.251$. The 
bifurcation diagram and total atom number difference (as a function of $\mu_{+}$) 
are shown in Fig.~\ref{fig3_supp}.
}
\label{fig3}
\end{figure}

\begin{figure}[htp]
\vskip -0.5cm
\begin{center}
\includegraphics[height=.17\textheight, angle =0]{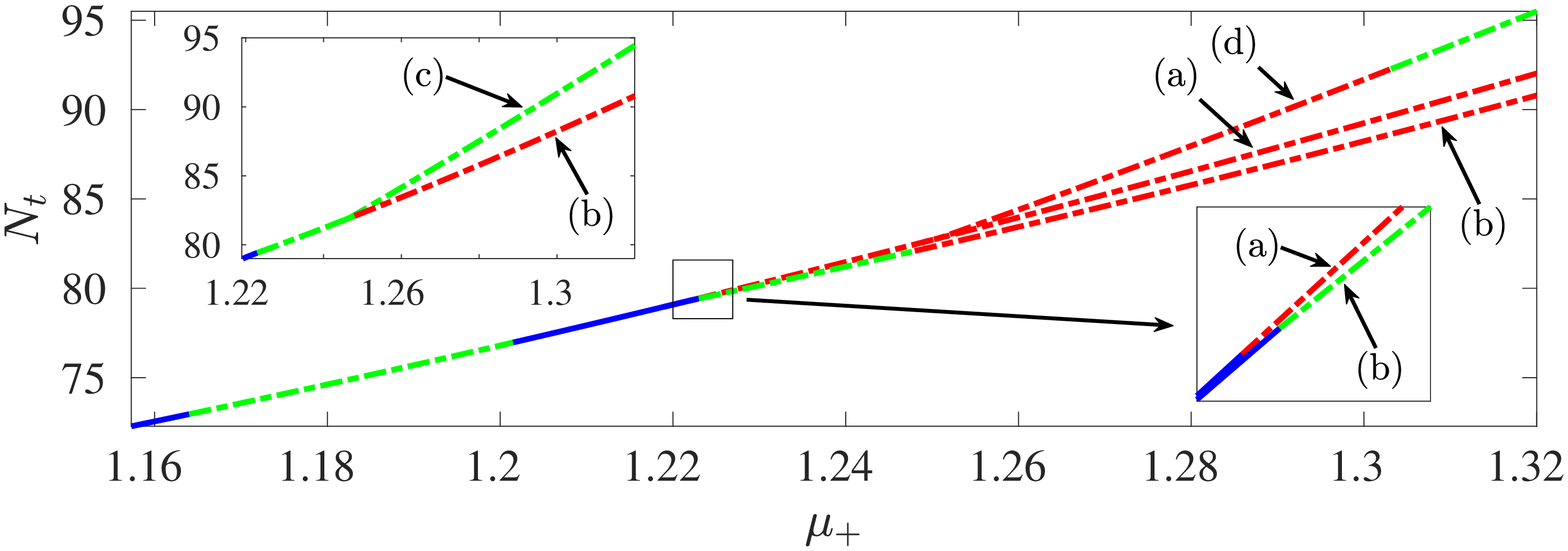}
\includegraphics[height=.17\textheight, angle =0]{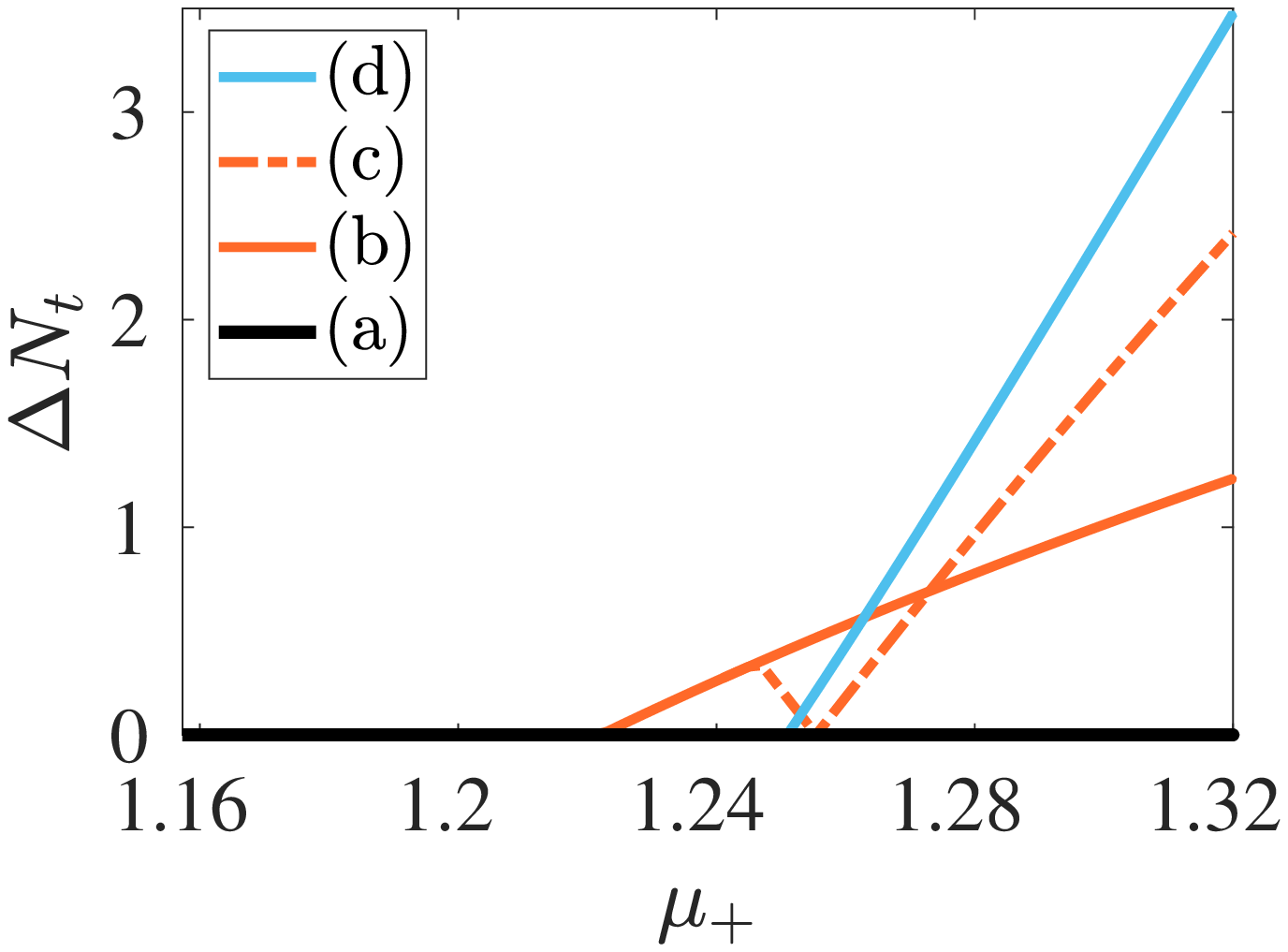}
\end{center}
\caption{Continuation of Fig.~\ref{fig3}. The left and right 
panels present the total number of atoms and total number of 
atoms difference both as functions of $\mu_{+}$. Notice the 
``spike'' shown in the right panel as well as the fact that 
the $\Delta N_{t}$ of branches (b) and (c) from (a) become equal, 
i.e., equidistant from (a), at $\mu_{+}\approx 1.273$ corresponding 
to the intersection of their respective curves (see text for details).
}
\label{fig3_supp}
\end{figure}
According to the bifurcation diagram shown in the bottom left panel of 
Fig.~\ref{fig0}, this bound state is stable from $\mu_{+}\approx 1.223$ 
(where the second component is formed, i.e., it is nontrivial) to 
$\mu_{+}\approx1.256$. Given the instability of the soliton necklace in 
the single-component case of~\cite{egc_16}, we infer that this stabilization
is due to its coupling with the ground state of the first component. The 
state becomes oscillatorily unstable past the latter value of $\mu_{+}$. In addition, 
and as per the eigenvalue continuations in Fig.~\ref{fig0}(a) a real eigenvalue
--or equivalently an imaginary eigenfrequency-- (see the top right panel
therein) passes through the origin at $\mu_{+}\approx 1.293$, thus giving 
birth (through a pitchfork bifurcation~\cite{yang_2012}) to a new state 
having a stripe and two vortex dipoles in the second component one on each 
side of the stripe as is shown in Fig.~\ref{fig0}(b) (see also the bottom 
right panel showcasing the atom difference of the bifurcated state from the
parent branch). Past the bifurcation point ($\mu_{+}\approx 1.293$), the parent 
branch becomes exponentially unstable due to the relevant (purely) imaginary 
eigenfrequency, while the daughter branch bearing vorticity in this quadrupolar 
structure inherits the oscillatory instability that the parent branch featured 
before the bifurcation point (see the bifurcation diagram in the bottom left 
panel of Fig.~\ref{fig0}).

Next, let us examine the results presented in Fig.~\ref{fig1}. Specifically,
Fig.~\ref{fig1}(a) corresponds to the dark-bright (DB) soliton stripe branch. 
This state and its spectrum have been studied in~\cite{pola_pra_2012}. More 
recently the theoretical analysis of such a dark-bright stripe and its transverse 
instability has also been considered in~\cite{wenlong18}. According to the 
bifurcation diagram shown in the bottom left panel of the figure, this state 
is stable emanating from the linear limit (i.e., when the second component is 
almost absent) but it becomes oscillatory unstable for $\mu_{+}\approx [1.07,1.13]$.
However, a sequence of pitchfork bifurcations happens at $\mu_{+}\approx 1.13$
and $\mu_{+}\approx 1.263$ giving birth to a vortex-bright (VB) soliton dipole branch 
of Fig.~\ref{fig1}(b) which has been studied in \cite{pola_pra_2012} (see also~\cite{wenlong18}
for further analysis of this transverse instability) as well as the tripole branch 
(of alternating vortices) of Fig.~\ref{fig1}(c) in the second component (see also
the bottom right panel of the figure). This progression of first the dipole (of 
VB solitons), then the tripole, then an aligned quadrupole etc. is strongly reminiscent 
of the corresponding process of breakup of a stripe into multi-vortex states due to 
transverse instability in single-component BECs, as discussed, e.g., in~\cite{middel10}. 
The dipole branch itself appears dynamically stable except for a narrow interval of 
oscillatory instability, once again analogous to the one-component case~\cite{middel10}.
On the other hand, the tripole branch is classified as exponentially 
unstable, inheriting the exponential instability of its parent DB stripe branch. While 
we are not aware of an experimental manifestation of states bearing multiple vortex-bright 
solitary waves, the ability to tune the experimentally realized DB solitons~\cite{Becker2008,Middelkamp2011,Hamner2011,Yan2011,Hoefer2011,Yan2012}
should, in principle, enable the realization of such observations, including, e.g., 
through the transverse instability of a dark-bright solitonic stripe. It is relevant 
to also mention that a single VB solitary wave has been previously realized experimentally 
in the work of~\cite{brian_old}.

A similar bifurcation pattern, i.e., a cascade of pitchfork bifurcations appears in 
Fig.~\ref{fig2}. In particular, Fig.~\ref{fig2}(a) corresponds to the crossed DB 
soliton~\cite{stockhofe_jpb_2011}, i.e., a fundamental state in the first component 
and the $|1,1\rangle_{(\textrm{c})} \propto x y e^{-\Omega (x^2+y^2)/2}$ quadrupolar 
state in the second component as per its Cartesian representation at the linear limit. 
See, e.g.,~\cite{egc_16} for this representation; here, we remind the reader for 
completeness that Cartesian eigenstates of the linear limit can be denoted as
\begin{eqnarray}
\ket{m,n}_{(\textrm{c})}\doteq \phi_{m,n}
\sim H_m(\sqrt{\Omega} x) H_n(\sqrt{\Omega} y) e^{-\Omega r^2/2},
\label{extra1}
\end{eqnarray}
where $H_{m,n}$ stands for the Hermite polynomials with $m,n>0$, the quantum numbers of 
the harmonic oscillator. The associated energy of such a state (i.e., its eigenvalue) is
$E_{m,n}\doteq (m + n +1)\Omega$ [cf.~\cite{egc_16}]. In the polar representation, 
we also have
\begin{eqnarray}
\ket{k,l}_{(\textrm{p})}\doteq\phi_{k,l} = q_{k,l}(r) e^{i l \theta}
\label{extra2}
\end{eqnarray}
with eigenvalues $E_{k,l}\doteq(1 + |l| + 2 k) \Omega$. Here, $l$ and $k$ stand for the 
eigenvalue of the ($z$-component of the) angular momentum operator and the number of radial 
zeros of the corresponding eigenfunction {$q_{k,l}$}, respectively. The eigenfunction's 
radial part can be denoted by
\begin{eqnarray}
q_{k,l} \sim r^l L_k^l(\Omega r^2) e^{-\Omega r^2/2},
\label{extra3}
\end{eqnarray}
where $L_k^l$ are the associated Laguerre polynomials. Interestingly once again, in the 
pattern of Fig.~\ref{fig2}(a) note the astroid pattern of the first component induced by 
its phase immiscibility with the second component. The relevant state bifurcates from the 
linear limit (i.e., in the absence of the second component) and is dynamically stable for 
$\mu_{+}\approx [1.123,1.153]$. However, this branch becomes oscillatorily unstable for 
$\mu_{+}\approx[1.153,1.163]$ and past the value of $\mu_{+}\approx 1.163$, it becomes 
exponentially unstable (notice that this progression of instability is highlighted by the
change of colors, i.e., from green to red in Fig.~\ref{fig2}). In particular, the branch 
of Fig.~\ref{fig2}(a) gives birth to the VB quadrupole cluster as was discussed in~\cite{stockhofe_jpb_2011} 
at $\mu_{+}\approx 1.162$ shown in Fig.~\ref{fig2}(b). The latter waveform is oscillatorily 
unstable over the interval of $\mu_{+}$ we consider herein. A subsequent pitchfork bifurcation 
happens later at $\mu_{+}\approx 1.279$ (see also the bottom right panel of the figure) giving 
birth to the state of Fig.~\ref{fig2}(c). This state is exponentially unstable (due to the 
instability of its parent  branch of Fig.~\ref{fig2}(a)) and bears one vortex of charge 2 in the 
middle surrounded by another four vortices in the second component. 
\begin{figure}[htp]
\vskip -0.5cm
\begin{center}
\includegraphics[height=.25\textheight, angle =0]{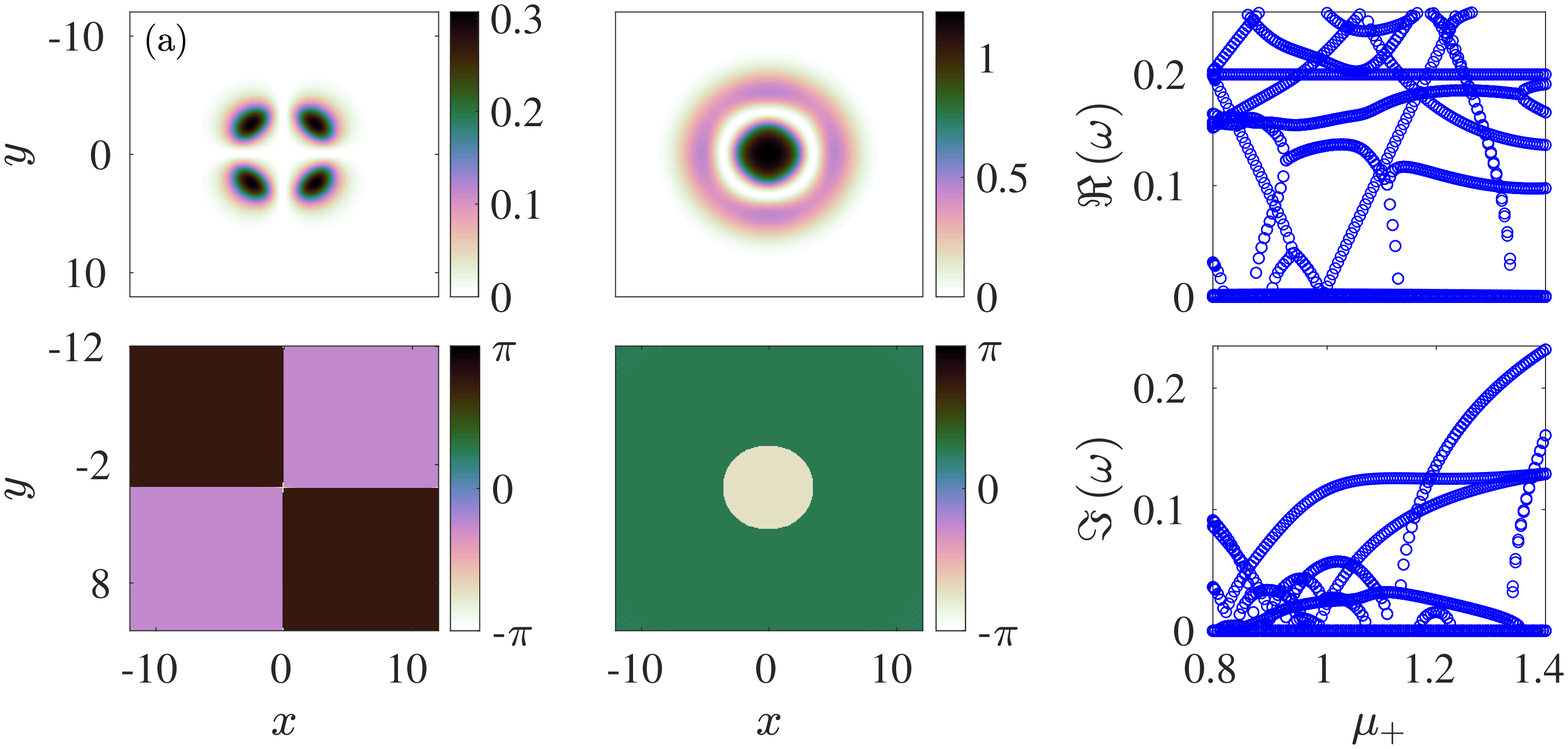}
\includegraphics[height=.25\textheight, angle =0]{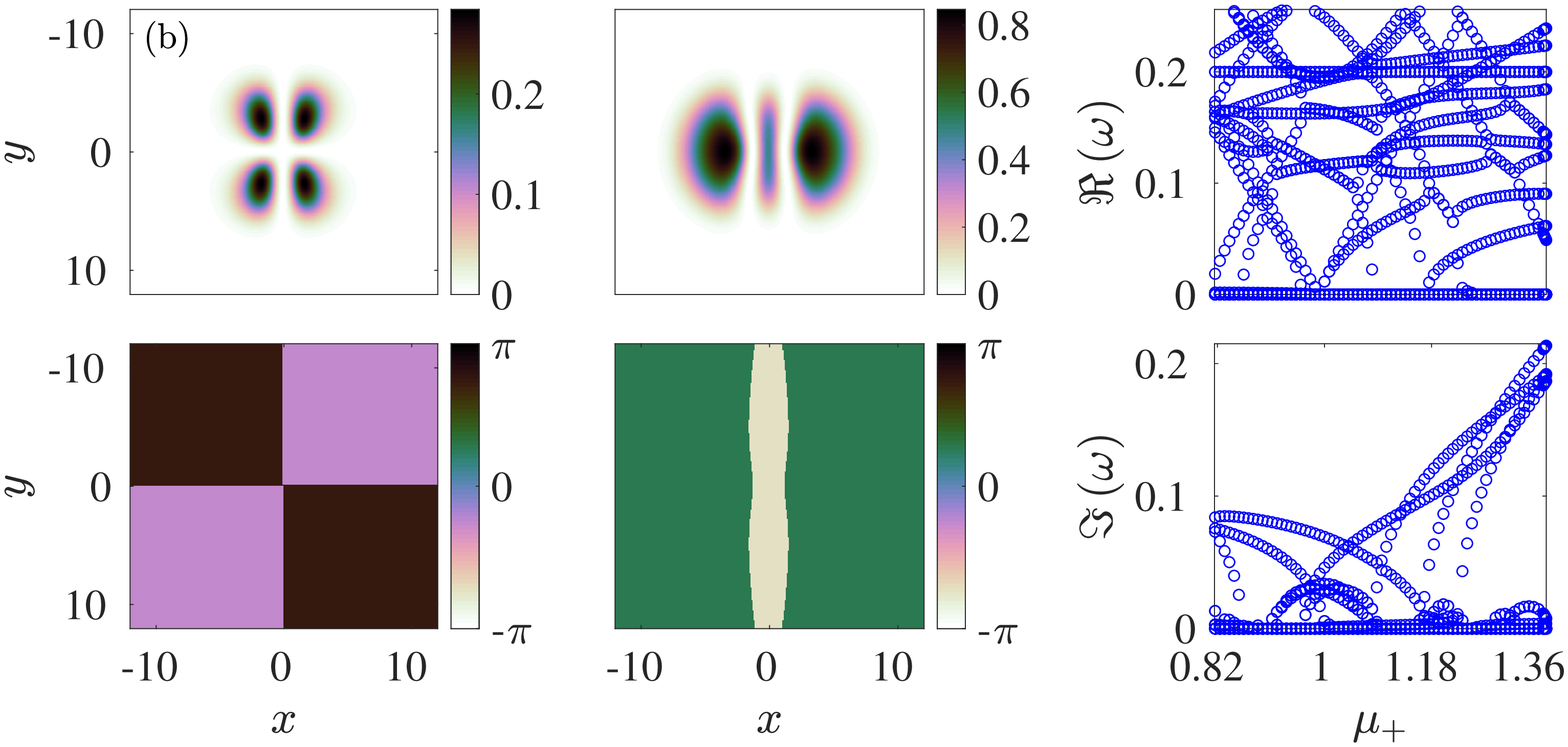}
\includegraphics[height=.25\textheight, angle =0]{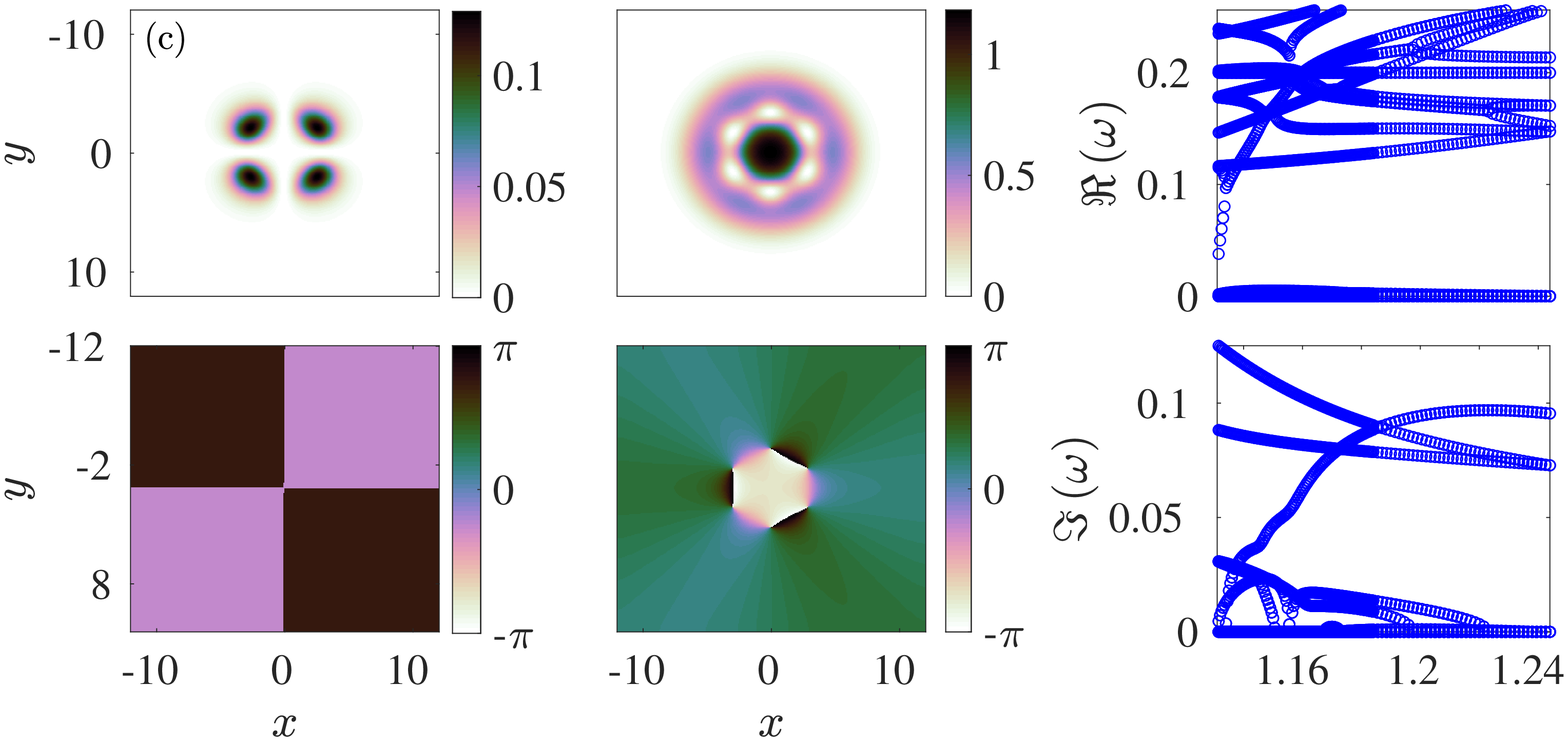}
\includegraphics[height=.25\textheight, angle =0]{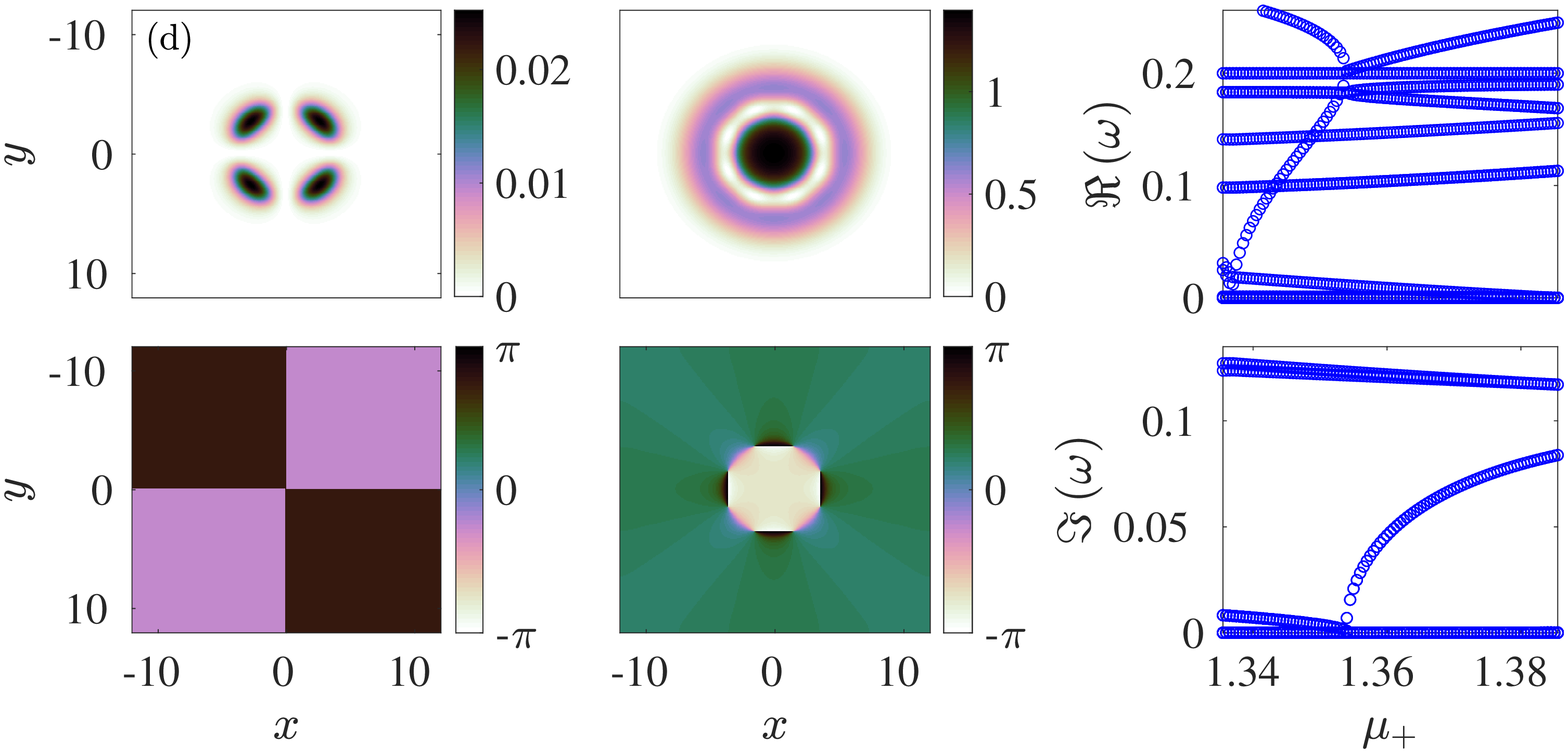}
\end{center}
\caption{
Same as Fig.~\ref{fig0} but for the quadrupolar ring-dark soliton (RDS) 
branch. The densities and respective profiles are shown for $\mu_{+}=1.2$
(a), $\mu_{+}=1.16$ (b), $\mu_{+}=1.2$ (c), and $\mu_{+}=1.377$ (d), 
respectively. The branches of (b), (c), and (d) emerge at values of 
$\mu_{+}$ of $\mu_{+}\approx 0.815$, $\mu_{+}\approx 1.135$ and 
$\mu_{+}\approx 1.336$, respectively.
}
\label{fig4}
\end{figure}

\begin{figure}[htp]
\vskip -0.5cm
\begin{center}
\includegraphics[height=.17\textheight, angle =0]{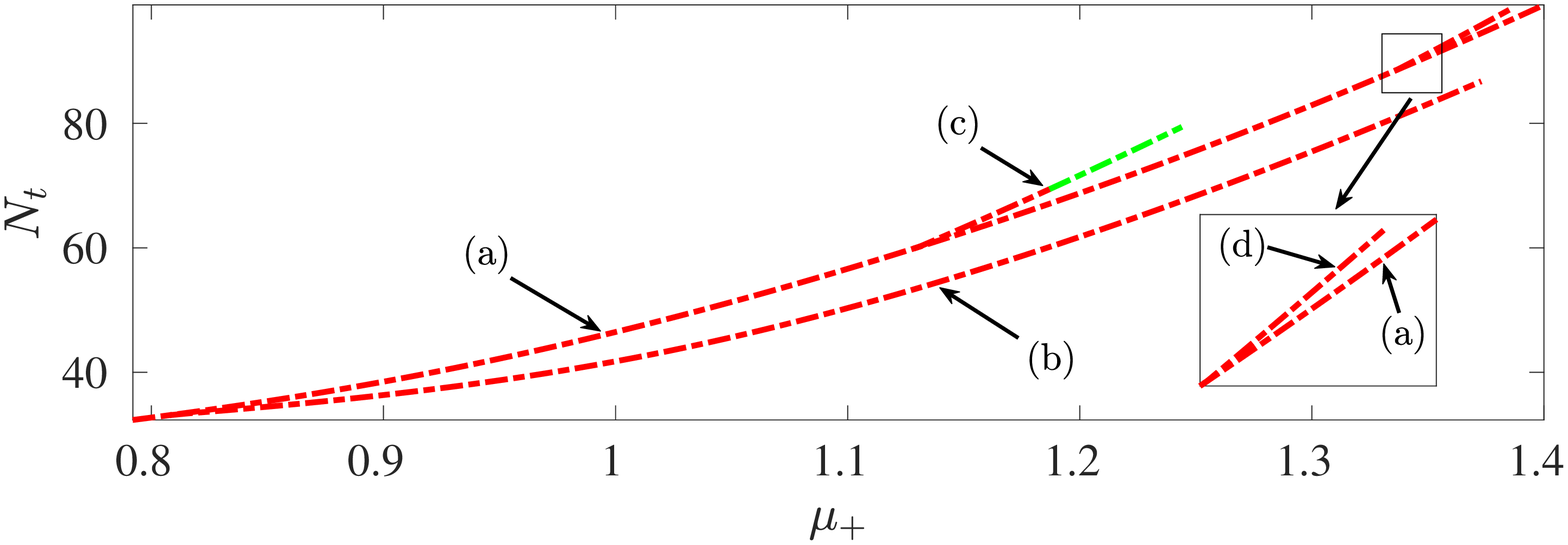}
\includegraphics[height=.17\textheight, angle =0]{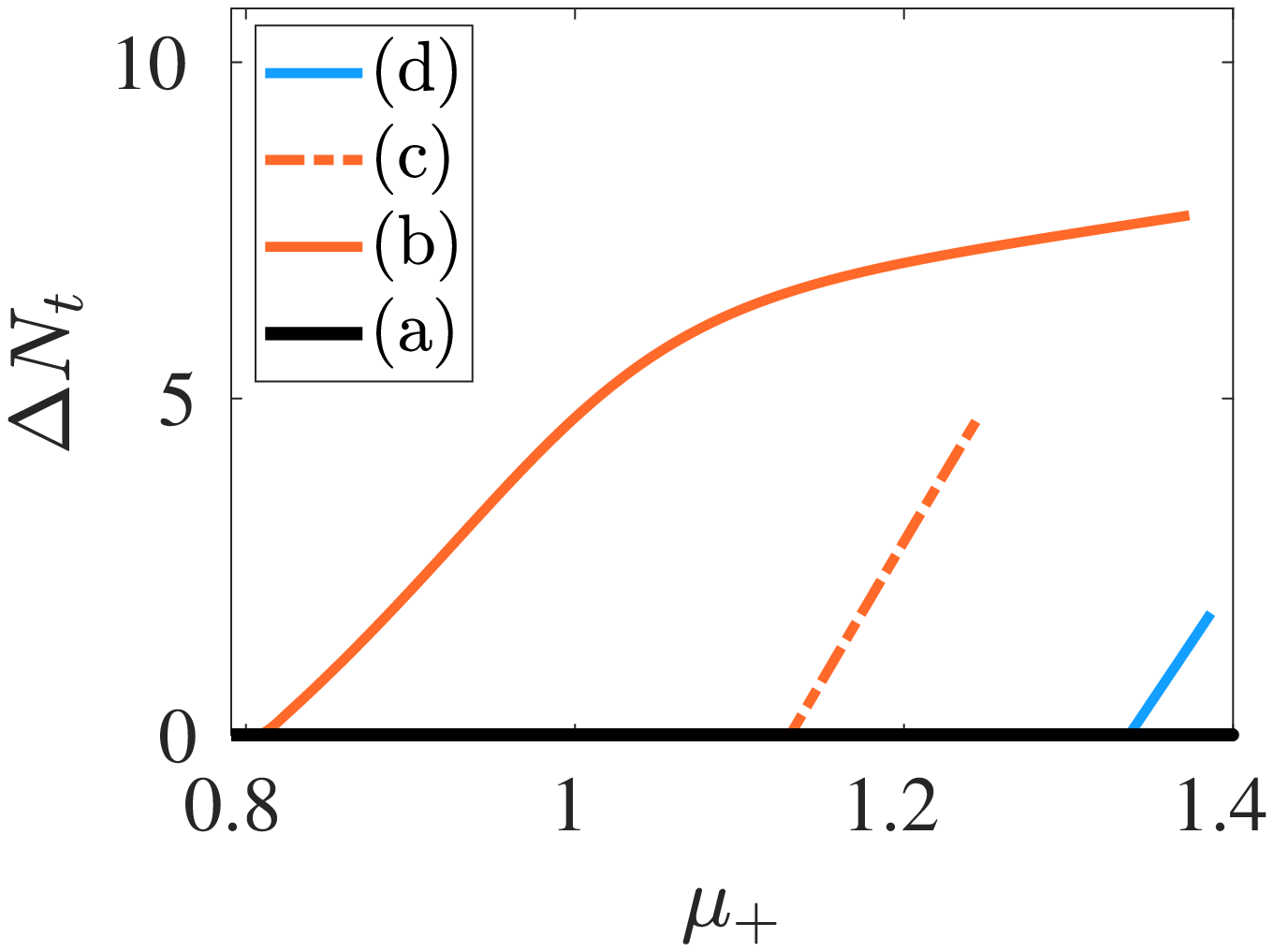}
\end{center}
\caption{Continuation of Fig.~\ref{fig4}. The left and right 
panels present the total number of atoms and total number of 
atoms difference both as functions of $\mu_{+}$. Note that the
first component of the branch (b) vanishes past $\mu_{+}\approx 1.373$
as well as (c) does past $\mu_{+}\approx 1.244$ and (d) past 
$\mu_{+}=1.386$, respectively.
}
\label{fig4_supp}
\end{figure}
Subsequently, we focus on the DB ring soliton state of Fig.~\ref{fig3}(a) and 
its bifurcations. This state (and in particular, with the first and second 
components reversed) has been identified and studied in~\cite{stockhofe_jpb_2011} 
together with its stability. The state of Fig.~\ref{fig3}(a) is generically 
unstable except for parametric intervals of stability $\mu_{+}\approx[1.157,1.164]$ 
and $\mu_{+}\approx[1.202,1.222]$. In particular, the DB ring soliton becomes 
oscillatory unstable past $\mu_{+}\approx 1.164$, then restabilizes itself at 
$\mu_{+}\approx 1.202$, and finally becomes exponentially unstable past 
$\mu_{+}\approx 1.222$. The first pitchfork bifurcation that takes place at 
$\mu_{+}\approx 1.222$ (a double imaginary eigenfrequency emanates from 
this collision) results in the emergence of the two DB soliton stripes shown 
in Fig.~\ref{fig3}(b). This daughter branch inherits the stability of the
parent branch and maintains its stability for a very short parametric interval 
(see the bottom right inset of the left panel of Fig.~\ref{fig3_supp} showcasing 
the exchange of stability). Then, the two DB soliton stripes branch becomes 
oscillatorily unstable past $\mu_{+}\approx 1.223$, and then exponentially 
unstable past $\mu_{+}\approx 1.248$. It is worth pointing out about the additional 
co-existence of a smaller (in its imaginary part) oscillatorily unstable mode 
for $\mu_{+}\approx[1.2464,1.2784]$ being responsible for the emergence of a ``bubble'' 
as is shown in the imaginary part of the eigenfrequency spectrum. However, and slightly
after the appearance of this ``bubble'', a real eigenvalue crosses the origin at a 
value of $\mu_{+}$ of $\mu_{+}\approx 1.247$ giving birth to a secondary bifurcating branch 
shown in Fig.~\ref{fig3}(c). This branch has been identified in~\cite{stockhofe_jpb_2011} 
and is known as six vortex state with four of them being filled (i.e., VBs). This state 
inherits the oscillatory instability of the parent branch of Fig.~\ref{fig3}(b) as well as
gains further oscillatory unstable modes shortly after the bifurcation point $\mu_{+}\approx 1.247$
(see also the upper left inset of the left panel of Fig.~\ref{fig3_supp}). Finally, the 
parent branch of Fig.~\ref{fig3}(a) undergoes one more bifurcation at $\mu_{+}\approx 1.251$
giving birth to the VB hexagon state~\cite{stockhofe_jpb_2011} (again, the emerging unstable 
mode corresponds to a double imaginary eigenfrequency). This branch is shown in Fig.~\ref{fig3}(d) 
and classified as exponentially unstable although past $\mu_{+}=1.303$, it also becomes oscillatorily 
unstable. All the above bifurcations are summarized in Fig.~\ref{fig3_supp}. Specifically, it should 
be noted that the branch of Fig.~\ref{fig3}(c) emerging from the one of Fig.~\ref{fig3}(b) and depicted
with dashed-dotted orange line in the right panel of Fig.~\ref{fig3_supp} has a ``spike'' at 
$\mu_{+}\approx 1.255$. The explanation of why this 
happens is offered in the following. Based on the definition of $\Delta N_{t}$ [cf. Eq.~\eqref{totnd}], 
we measure the distance of $N_{t}$ [cf. Eq.~\eqref{totn}] of a bifurcating branch from a \textit{reference} 
branch. Indeed, although the $N_{t}$ for branch (b) in Fig.~\ref{fig3} is smaller than the one for the 
reference branch (a) over the entire interval in $\mu_{+}$ considered therein, this difference appears 
\textit{positive} in the right panel of Fig.~\ref{fig3_supp}, based on the definition of the relevant 
diagnostic. Furthermore, the ``spike'' of the curve therein corresponding to the difference of $N_{t}$ between 
branches (c) and (a) happens because the $N_{t}$ of the former branch (as this emerges from 
branch (b)) is smaller than the $N_{t}$ of 
branch (a) (and larger than the $N_{t}$ of 
branch (b)) until its curve (as a function of $\mu_{+}$), intersects the 
corresponding one of branch 
(a) at $\mu_{+}\approx 1.255$. Then, and past that value, the
$N_{t}$ of branch (c) 
is larger than the one of
branch (a).

Hereafter, the configurations become more complex. The quadrupolar type solution in the first component described 
as $|1,1\rangle_{(\textrm{c})}$ in its Cartesian classification trapping a ring-dark soliton (RDS) in the second
component is shown in Fig.~\ref{fig4}(a). This branch emerges, i.e., bearing a non-trivial solution in its second
component at $\mu_{+}\approx 0.791$ and is classified as exponentially unstable over the parametric interval of 
$\mu_{+}\approx[0.791, 1.4]$. Similarly, the branch of Fig.~\ref{fig4}(a) undergoes a pitchfork bifurcation at 
$\mu_{+}\approx 0.815$ giving birth to the daughter branch of Fig.~\ref{fig4}(b). This branch bears a two-dark 
soliton stripes waveform (or $|2,0\rangle_{(\textrm{c})}$ as per its Cartesian classification) in the second component
whose total number 
of atoms is less than its parent branch (see the left panel of Fig.~\ref{fig4_supp}). In addition, this state is 
classified as exponentially unstable throughout its interval of existence where the first component vanishes eventually 
at $\mu_{+}\approx 1.373$. A subsequent pitchfork bifurcation of the parent branch [cf. Fig~\ref{fig4}(a)] occurs at 
$\mu_{+}\approx 1.135$ where the branch of Fig.~\ref{fig4}(c) bearing a hexapolar mode in the second component~\cite{egc_16} 
is born. This branch, i.e., the quadrupolar-vortex-hexagon branch is classified as exponentially unstable over 
$\mu_{+}\approx[1.135,1.187]$ and past the value of $\mu_{+}\approx 1.187$ possesses a dominant oscillatory unstable 
mode (see the left panel of Fig.~\ref{fig4_supp}), thus mimicking the spectrum in the single-component case~\cite{egc_16}. 
Finally, and as the value of $\mu_{+}$ increases, the parent branch of Fig.~\ref{fig4}(a) undergoes one more pitchfork 
bifurcation giving birth to the quadrupolar vortex-octagons state of Fig.~\ref{fig4}(d). Such a pattern, i.e., the subsequent
bifurcations of the RDS state was briefly mentioned in the one-component case~\cite{egc_16}. In our case, the branch of 
Fig.~\ref{fig4}(d) has a narrow interval of existence of $\mu_{+}\approx [1.336,1.386]$ in which it is classified as 
exponentially unstable according to our spectral stability analysis (see the left panel of Fig.~\ref{fig4_supp}). Finally, 
it should be noted that the total number of atoms of branch (b) is less than the one of the parent branch of (a). Similarly 
to the case of the branches of Figs.~\ref{fig3}(a) and~\ref{fig3}(b), this difference is shown in the right panel of 
Fig.~\ref{fig4_supp} and depicted with a solid orange line therein.

%
\begin{figure}[htp]
\vskip -0.5cm
\begin{center}
\includegraphics[height=.25\textheight, angle =0]{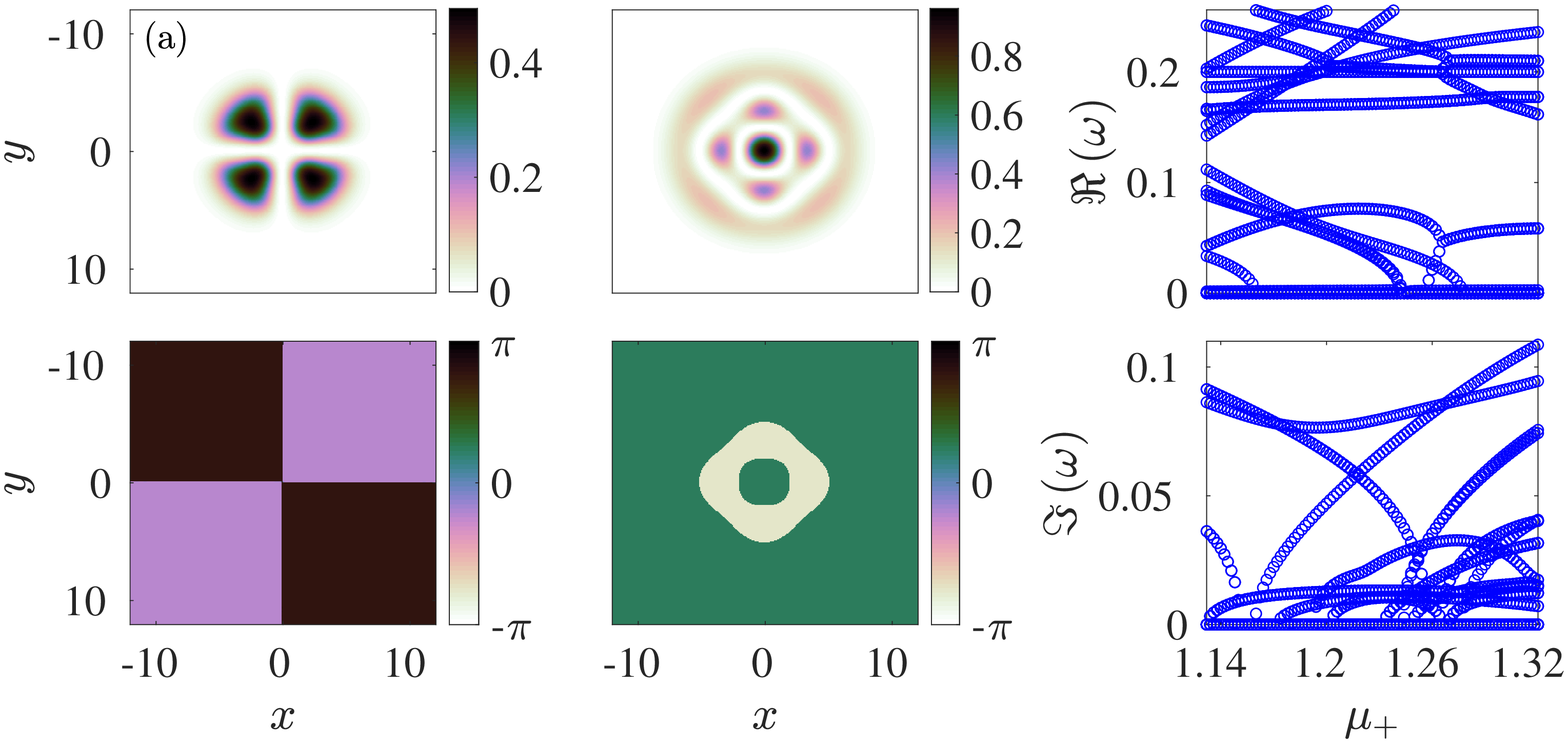}
\includegraphics[height=.25\textheight, angle =0]{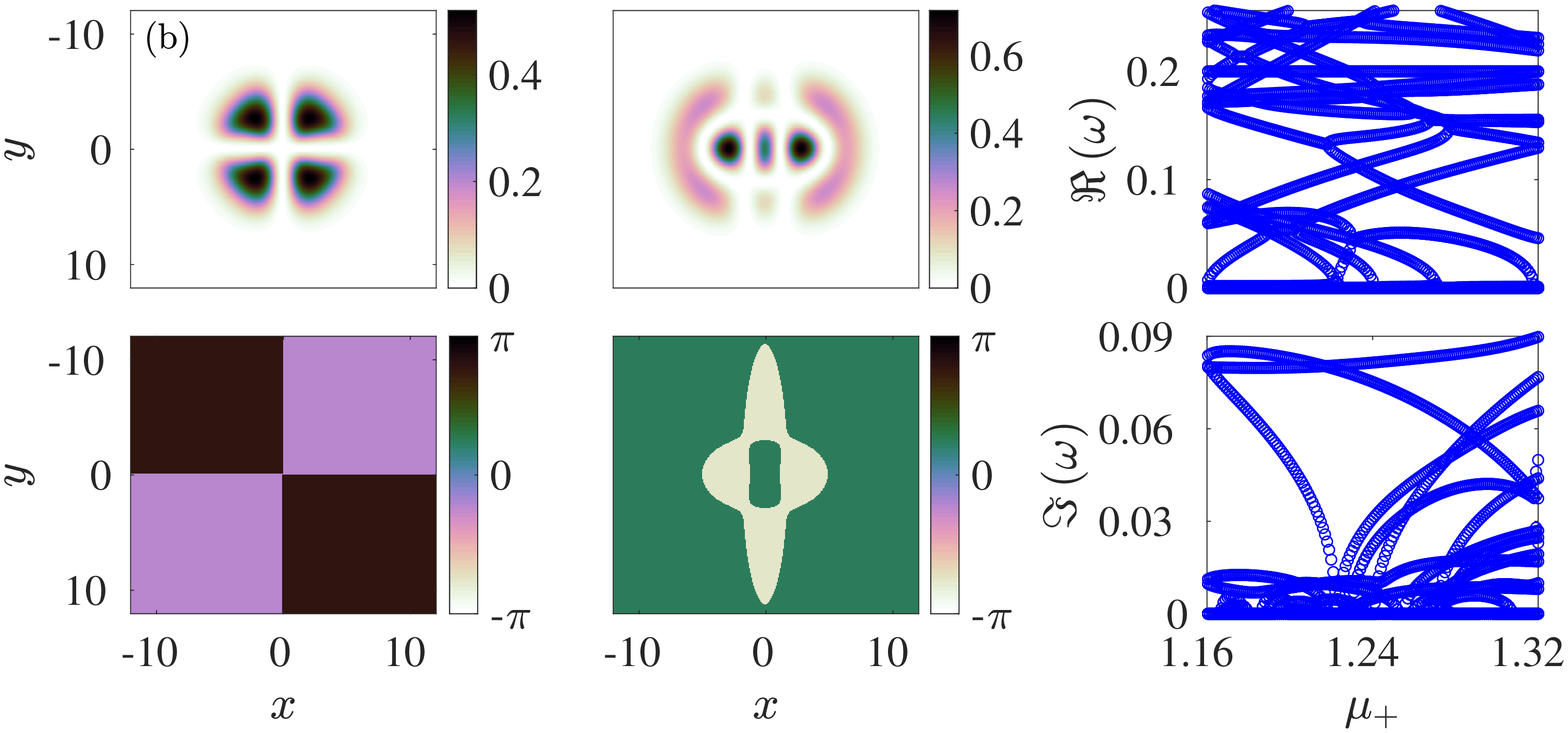}
\includegraphics[height=.25\textheight, angle =0]{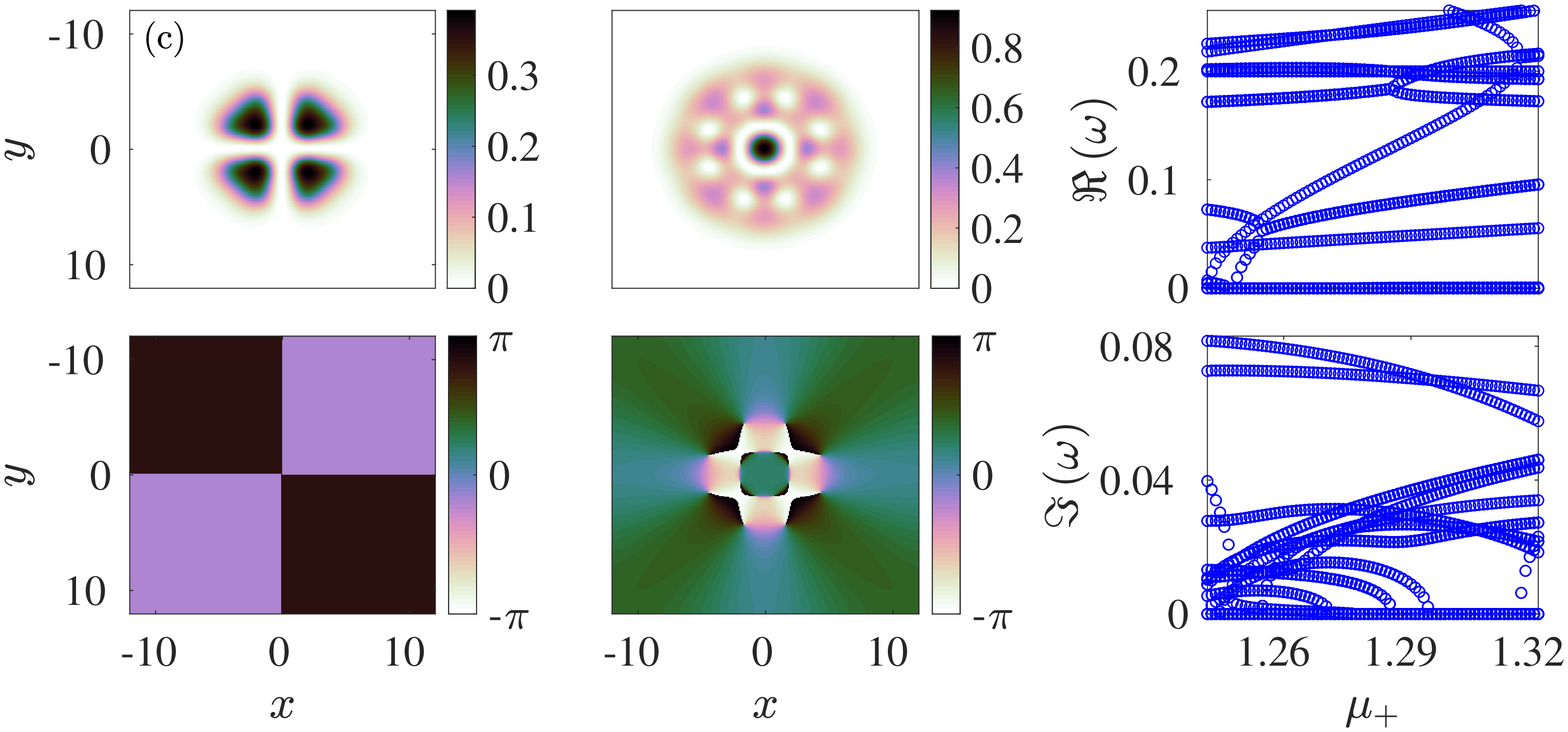}
\includegraphics[height=.25\textheight, angle =0]{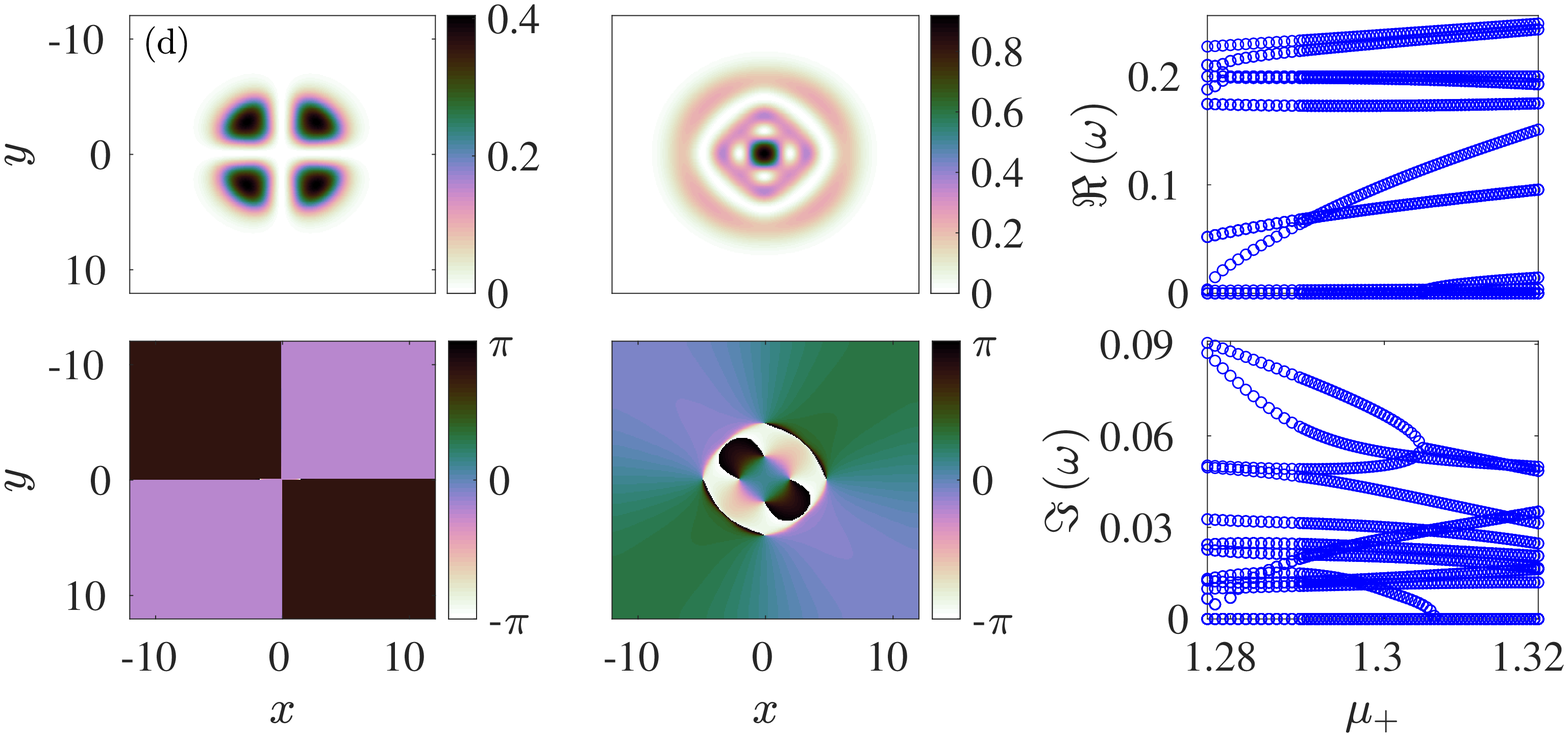}
\end{center}
\caption{
Same as Fig.~\ref{fig0} but for the quadrupolar--two dark 
rings branch. The branches (a)-(d) are presented for a value of $\mu_{+}$ 
of $\mu_{+}=1.32$. The branch of panel (a) undergoes a cascade of 
pitchfork bifurcations occurring at $\mu_{+}\approx 1.16$, $\mu_{+}\approx 1.242$
and $\mu_{+}\approx 1.277$, thus giving birth to the branches of 
panels (b), (c) and (d), respectively.
}
\label{fig5}
\end{figure}

\begin{figure}[htp]
\vskip -0.5cm
\begin{center}
\includegraphics[height=.17\textheight, angle =0]{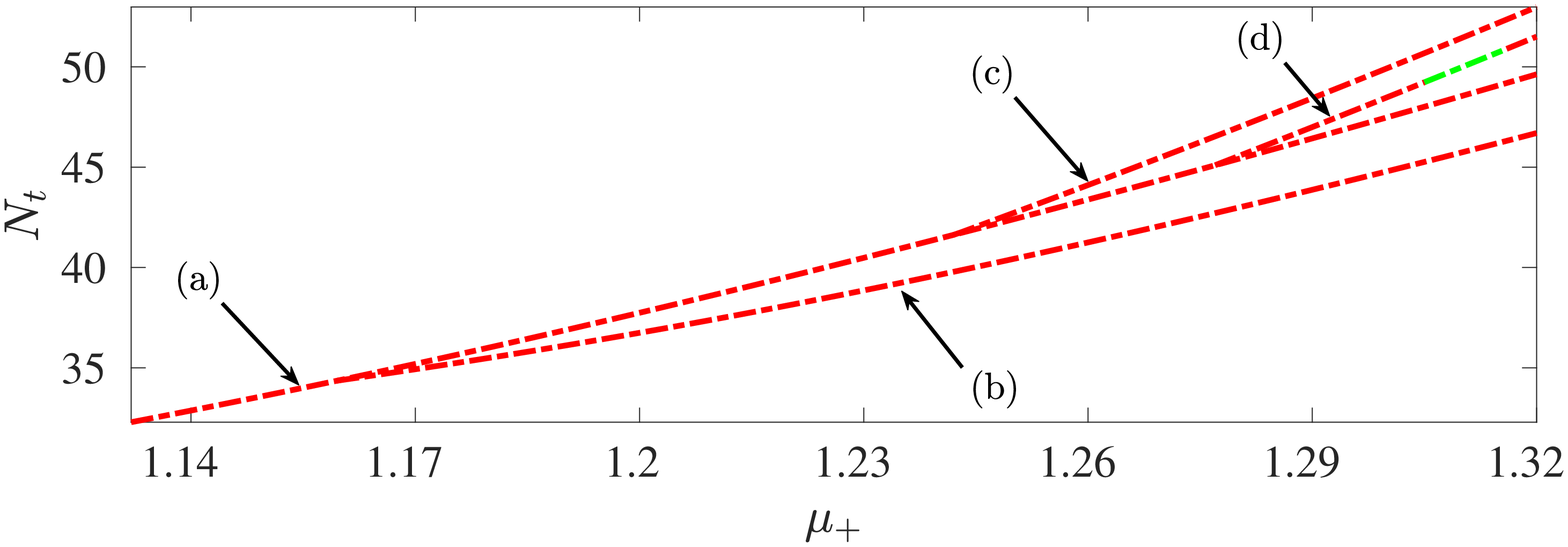}
\includegraphics[height=.17\textheight, angle =0]{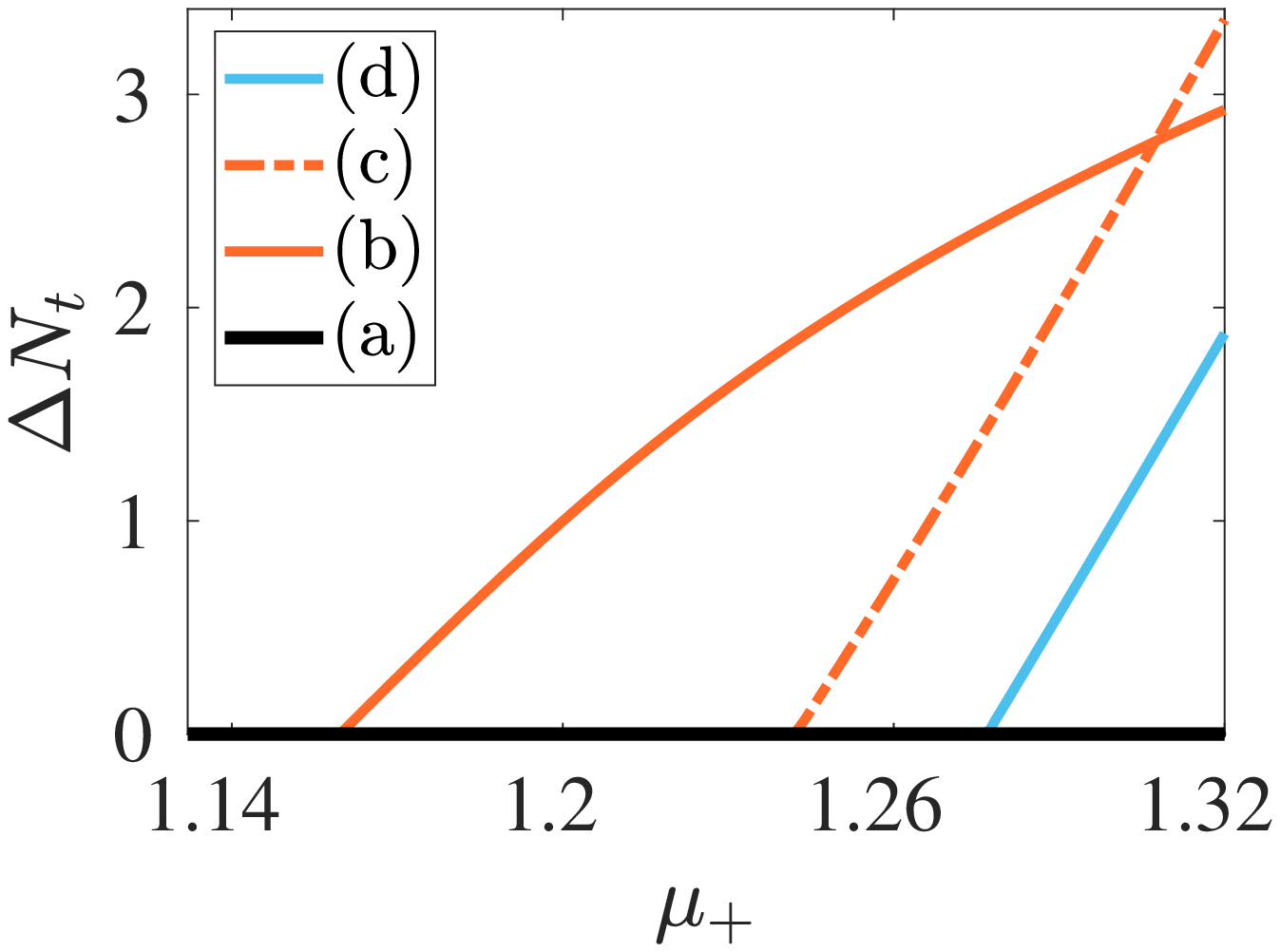}
\end{center}
\caption{Continuation of Fig.~\ref{fig5}. The left and right 
panels present the total number of atoms and total number of 
atoms difference both as functions of $\mu_{+}$. According to
the left panel, the $N_{t}$ of branch (b) is less than the one 
of its parent branch (a). Similarly to Fig.~\ref{fig3_supp} the 
$\Delta N_{t}$ of branches (b) and (c) from (a) become equal, 
i.e., equidistant from (a), at $\mu_{+}\approx 1.3075$ corresponding 
to the intersection of their respective curves (see text for details).
}
\label{fig5_supp}
\end{figure}

Next, we turn to the case of Fig.~\ref{fig5}. In particular, Fig.~\ref{fig5}(a) presents a 
quadrupolar waveform in the first component which, in turn, is coupled to a state bearing 
two dark rings in the second component. This state has lost its radial symmetry, though, 
due to the immiscibility (and the effects of nonlinear coupling) between the two components. 
This bound mode is exponentially unstable over the parametric interval in $\mu_{+}$ considered
herein (see the left panel of Fig.~\ref{fig5_supp}). However, there is again a cascade 
of pitchfork bifurcations, the first of which occurs at $\mu_{+}\approx 1.16$ as is also
highlighted in the right panel of Fig.~\ref{fig5_supp}. In particular, the first bifurcating 
branch is shown in Fig.~\ref{fig5}(b) where the solution in the second component can be thought 
of as a symmetry-broken (between the axes) variant of panel (a) that can be represented as a 
$|4,0\rangle_{(\textrm{c})}+|2,2\rangle_{(\textrm{c})}$ Cartesian state at the linear limit.
In the present two-component case, the branch of Fig.~\ref{fig5}(b) is exponentially unstable 
as well (see the left panel of Fig.~\ref{fig5_supp}). The next pitchfork bifurcation occurs 
at $\mu_{+}\approx1.242$ where Fig.~\ref{fig5}(a) gives birth to the bound mode shown in 
Fig.~\ref{fig5}(c) with the second component featuring an intriguing
combination of a ring and a vortex necklace.
Although this solution has been 
identified as oscillatorily unstable in~\cite{egc_16} in the single-component NLS, the daughter 
branch shown in Fig.~\ref{fig5}(c) is classified to be exponentially unstable (see the 
left panel of Fig.~\ref{fig5_supp}). It is interesting to highlight here that although in some 
cases (such as in Fig.~\ref{fig0}(a) as we saw before) the presence of a second component plays 
a stabilizing role, in others such as the one of Fig.~\ref{fig5}(c), it adds further unstable 
eigendirections to a particular waveform. Finally, the parent branch of Fig.~\ref{fig5}(a) 
undergoes one further pitchfork bifurcation at $\mu_{+}\approx 1.277$ giving birth to the daughter 
branch of Fig.~\ref{fig5}(d). This branch forms four vortices in its inner ring and is generally 
exponentially unstable (over the interval we considered herein) except for a narrow interval of 
$\mu_{+}\approx [1.305,1.316]$ where the dominant instability appears to be an oscillatory one 
(co-existing with a number of weaker exponentially unstable eigendirections). Similarly as in 
Figs.~\ref{fig3_supp} and~\ref{fig4_supp}, the left panel of Fig.~\ref{fig5_supp} suggests that 
although the total number of atoms $N_{t}$ of the bifurcating branch of Fig.~\ref{fig5}(b) is 
less than the one of its parent branch [cf. Fig.~\ref{fig5}(a)], the respective total-number-of-atoms 
difference between those two branches is positive as is 
shown in the right panel of Fig.~\ref{fig5_supp} per the definition of Eq.~(\ref{totnd}).
\begin{figure}[htp]
\vskip -0.5cm
\begin{center}
\includegraphics[height=.25\textheight, angle =0]{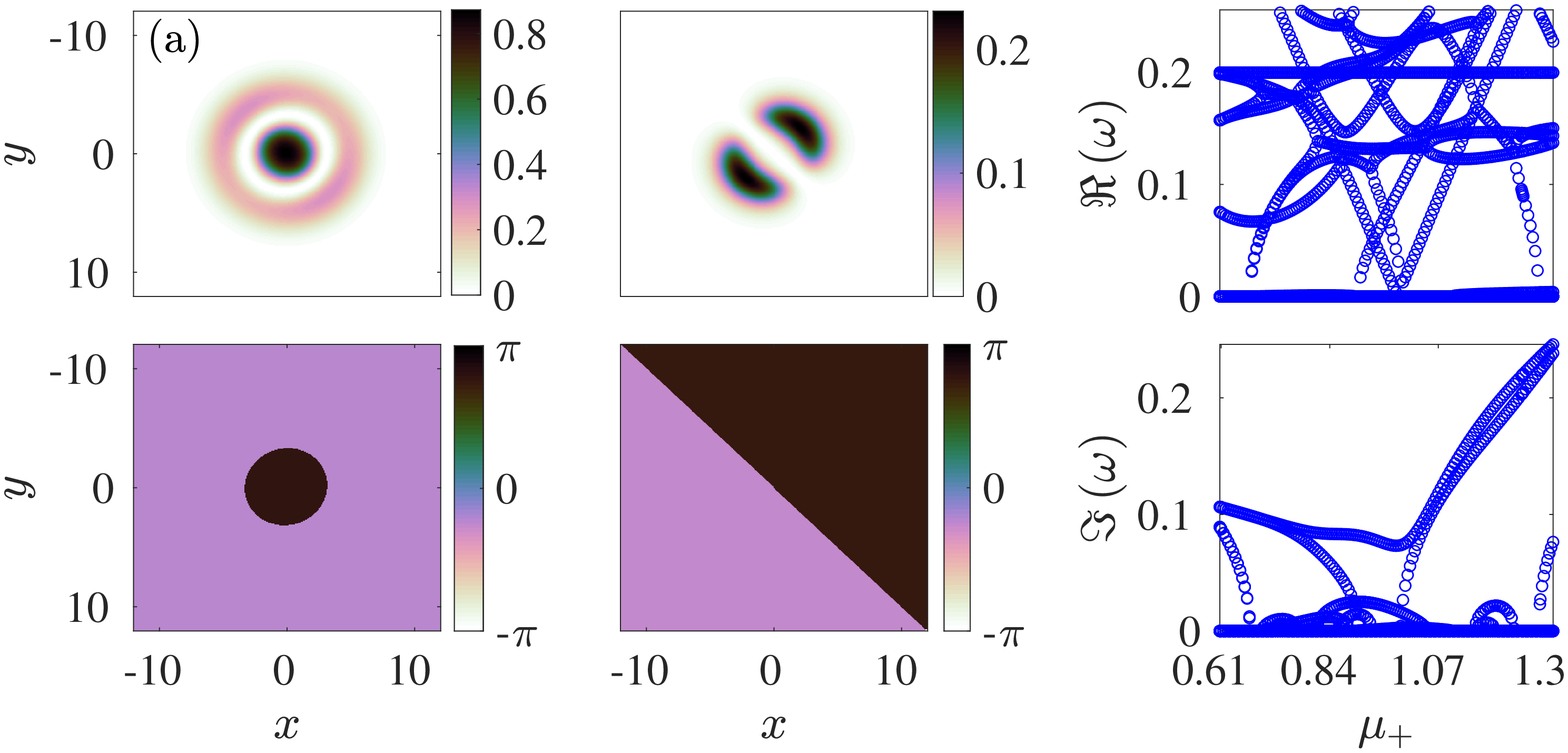}
\includegraphics[height=.25\textheight, angle =0]{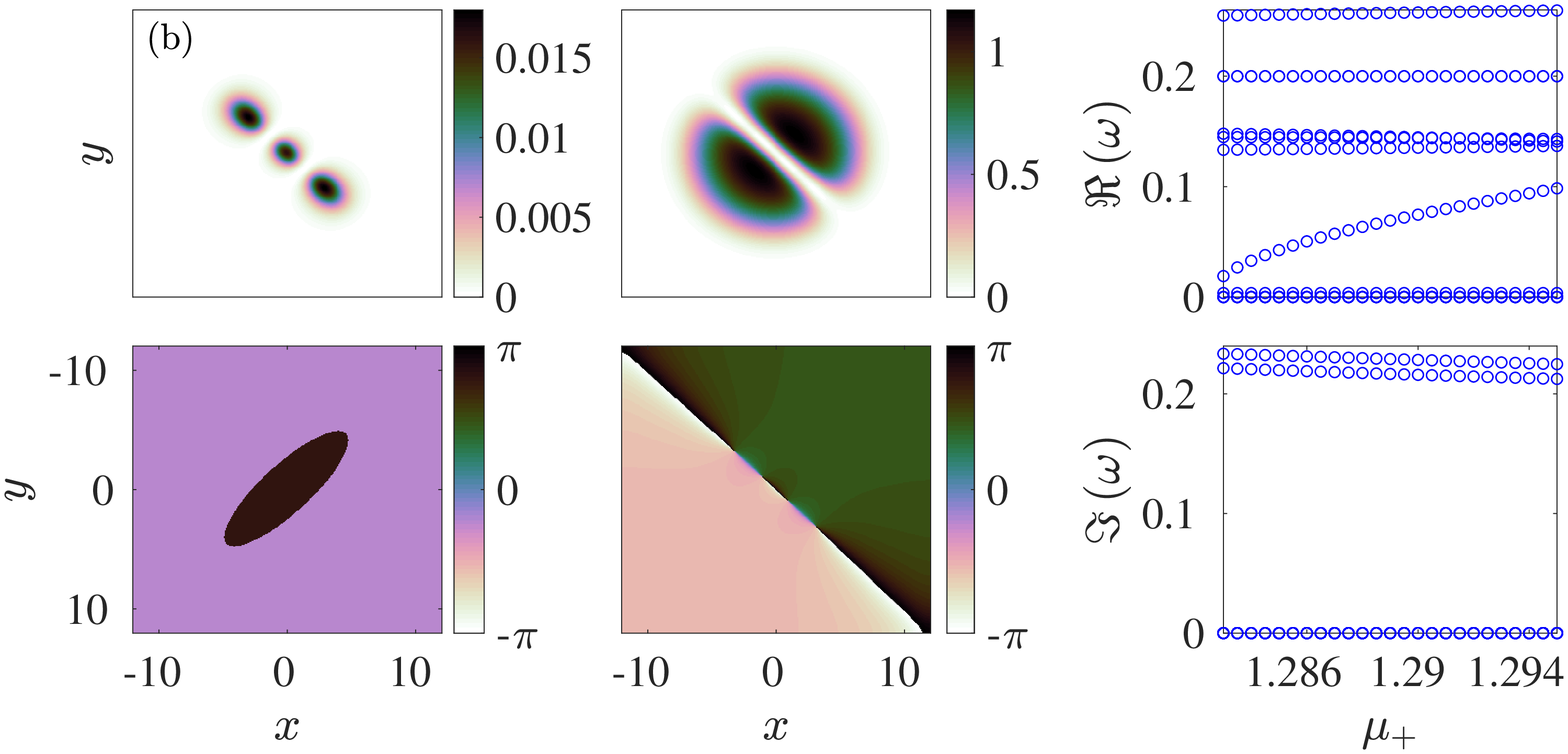}
\includegraphics[height=.17\textheight, angle =0]{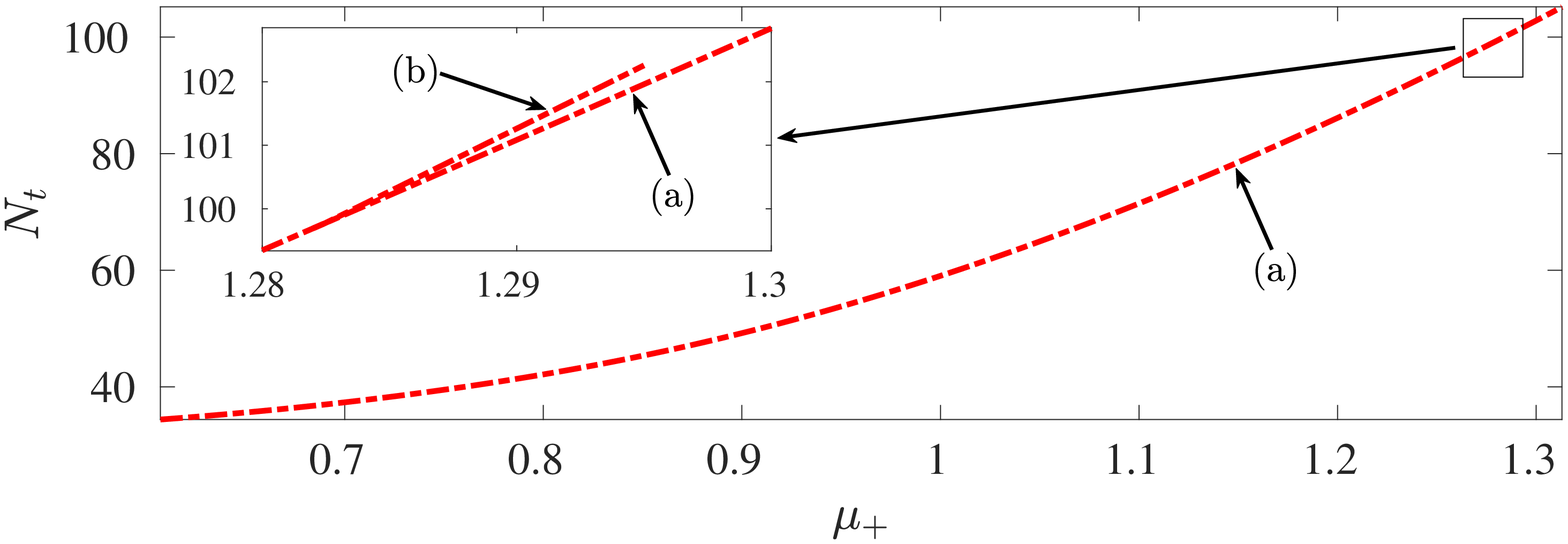}
\includegraphics[height=.17\textheight, angle =0]{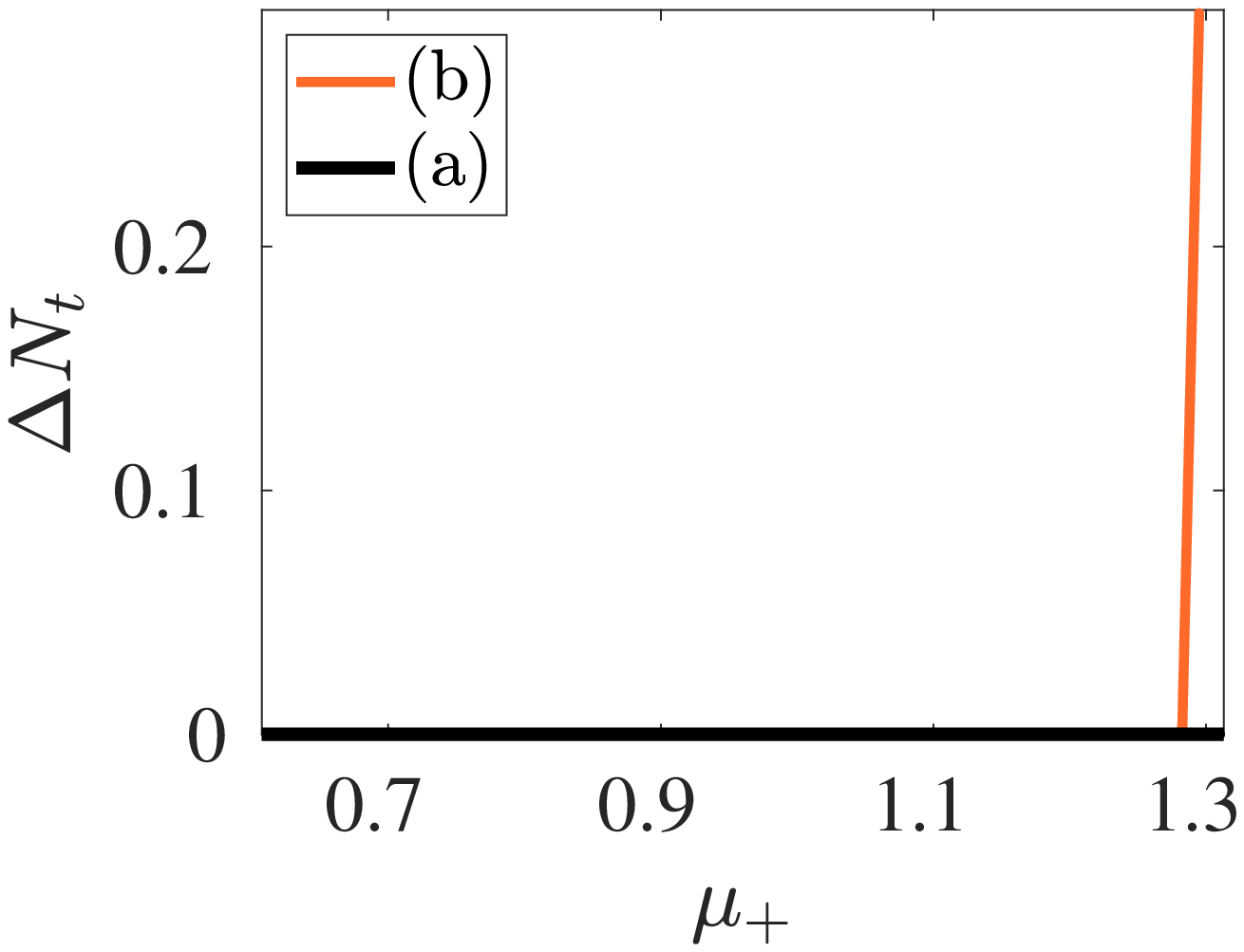}
\end{center}
\caption{Same as Fig.~\ref{fig0} but for the ring-dark-soliton-dipolar
(RDS-dipolar) branch. The densities and associated phases are shown for 
values of $\mu_{+}$ of $\mu_{+}=0.7$ (a) and $\mu_{+}=1.29$ (b), respectively
in the first and second rows. Note that the branch of panel (b) emerges at 
$\mu_{+}\approx 1.282$. The bottom left and right panels present the total 
number of atoms and total-number-of-atoms difference both as functions of 
$\mu_{+}$.
}
\label{fig6}
\end{figure}

\begin{figure}[htp]
\vskip -0.5cm
\begin{center}
\includegraphics[height=.25\textheight, angle =0]{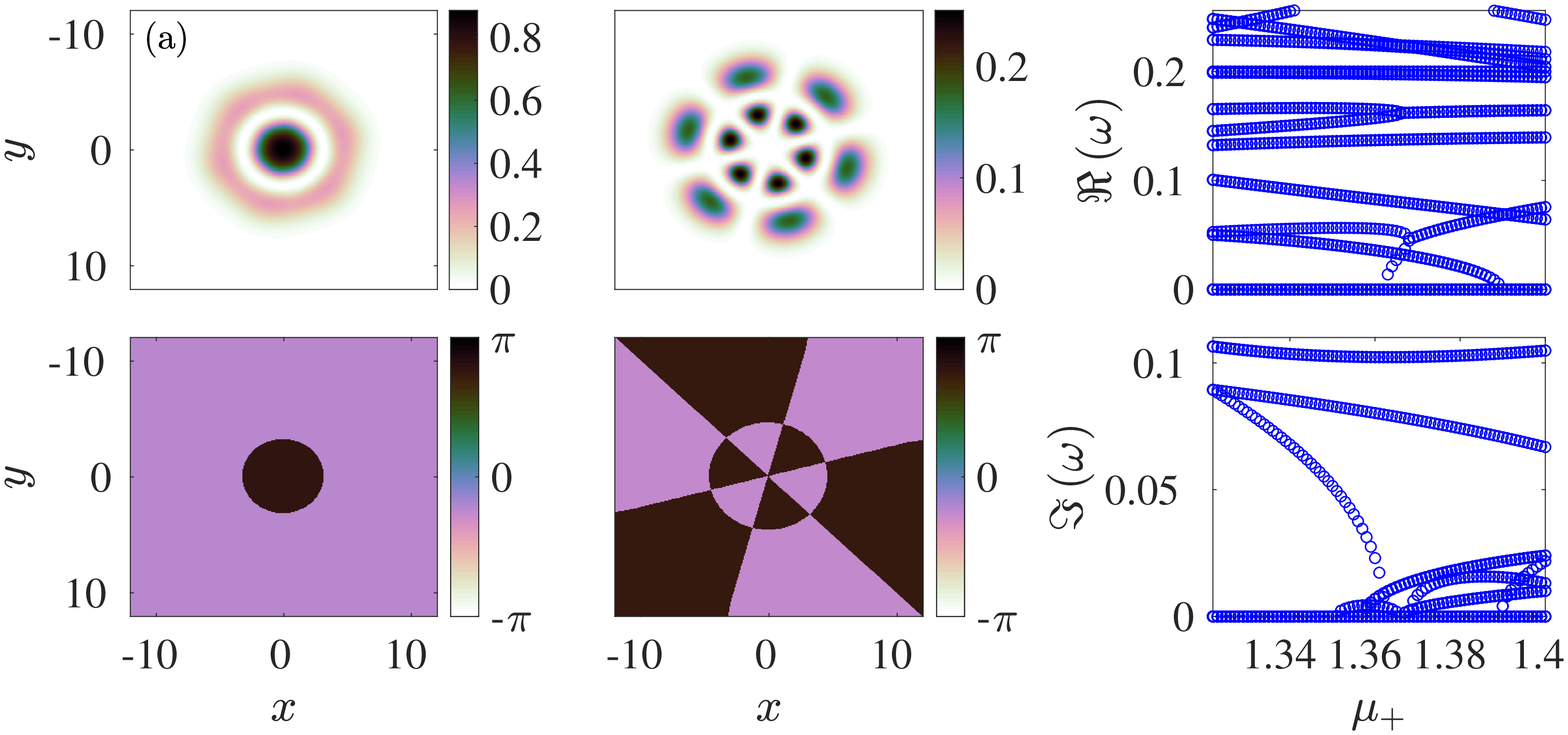}
\includegraphics[height=.25\textheight, angle =0]{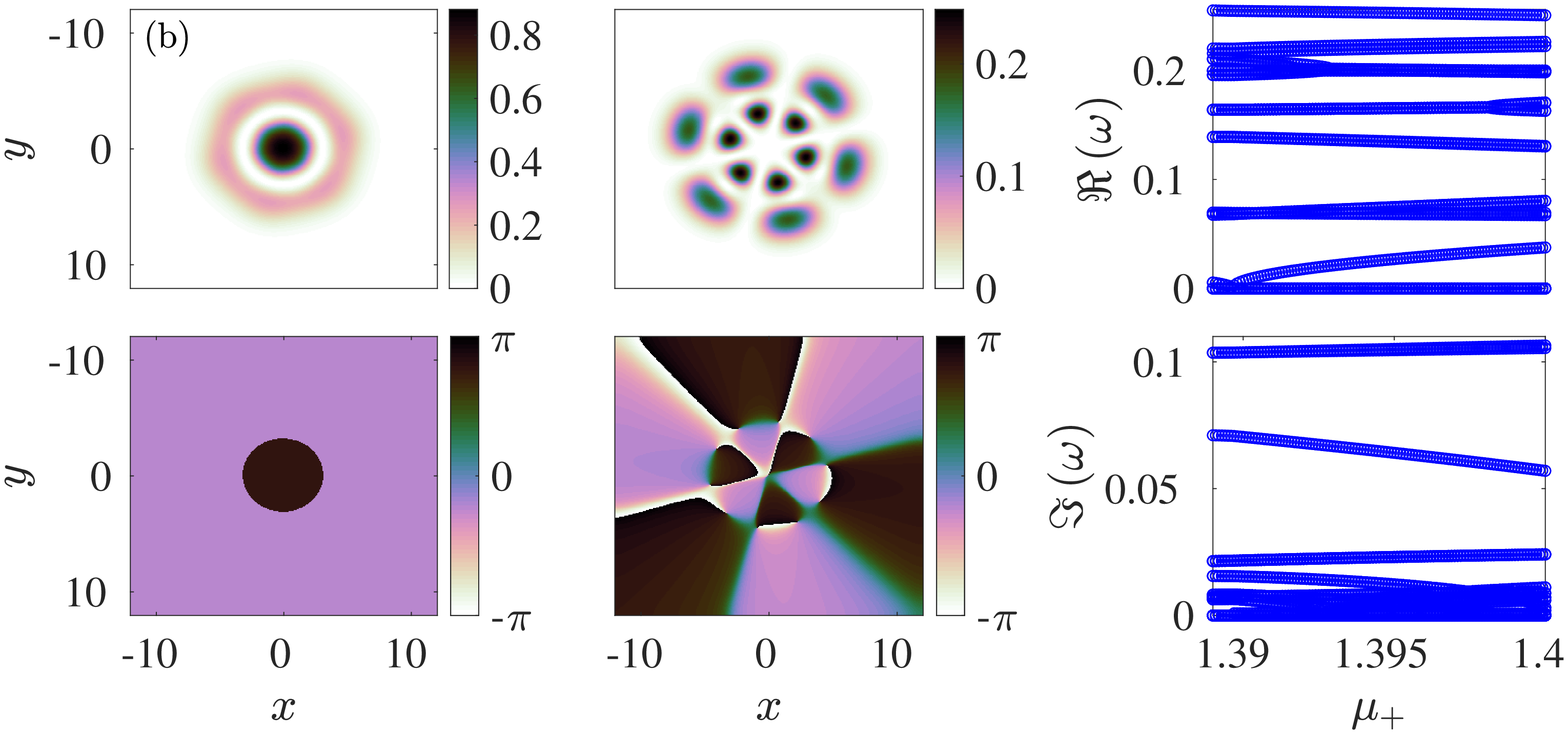}
\includegraphics[height=.17\textheight, angle =0]{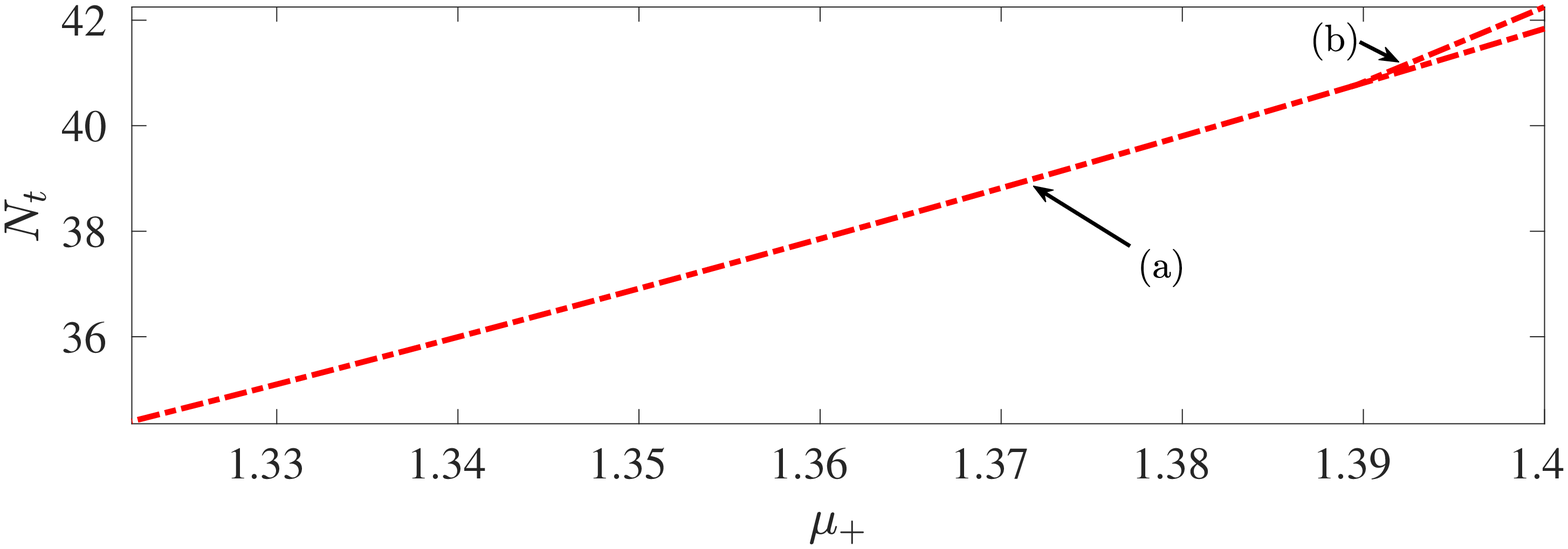}
\includegraphics[height=.17\textheight, angle =0]{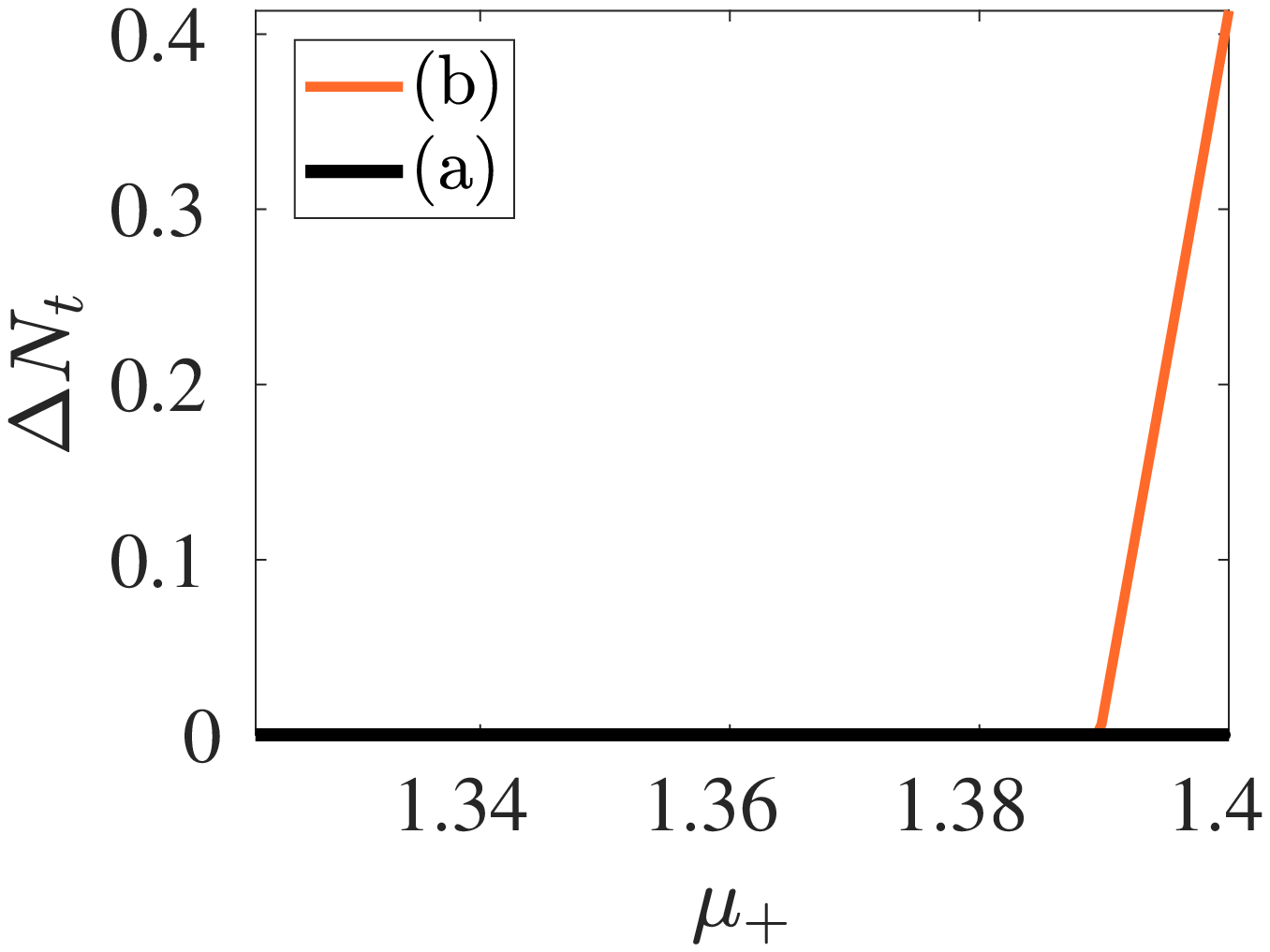}
\end{center}
\caption{Same as Fig.~\ref{fig0} but for the hexagonal type RDS-necklace 
branch. Similarly, the densities and associated phases are shown for a 
value of $\mu_{+}$ of $\mu_{+}=1.4$ for both branches. The branch (b) 
emerges at $\mu_{+}\approx 1.39$ via a pitchfork bifurcation. Again, 
the bottom left and right panels present the total number of atoms and 
total-number-of-atoms difference both as functions of $\mu_{+}$.
}
\label{fig7}
\end{figure}
The state of Fig.~\ref{fig6}(a) involves a symmetry-broken state where the first component 
features an elliptical type of a RDS, while the second component represents a dipolar 
($|1,0\rangle_{(\textrm{c})}$ in the notation of the Cartesian Hermite eigenstates) structure 
bearing a $\pi$ phase shift between the two matter wave blobs. This branch is highly unstable 
and in fact, classified as exponentially unstable over $\mu_{+}\approx [0.607,1.313]$ as is 
shown in the bottom left panel of Fig.~\ref{fig6} where the first component vanishes past 
$\mu_{+}\approx 1.313$. However, and close to the limit in $\mu_{+}$ where the first component 
vanishes, this branch undergoes a similar symmetry-breaking bifurcation as in the previous cases
in this work giving birth to the branch of Fig.~\ref{fig6}(b) (see also the bottom right panel 
of Fig.~\ref{fig6}). This daughter branch involves a series of density blobs in the first component
centered along the nodal line of the second component. The latter features a vortex quadrupole 
(this is hard to discern at the level of the density but more transparent at the level of the 
phase of this component) which is exponentially unstable as is also shown in the inset in the 
bottom left panel of Fig.~\ref{fig6}; its interval of existence over $\mu_{+}$ is rather 
narrow (the first component is non-vanishing for $\mu_{+}\approx[1.282,1.295]$). It should be noted
in passing that the branch of Fig.~\ref{fig6}(b) is a typical example
among the pitchfork bifurcations
that are expected to emerge from Fig.~\ref{fig6}(a) at values of $\mu_{+}\approx 0.67, 0.9$ 
and $\mu_{+}\approx 0.99$ where the first two will correspond to reverse pitchfork bifurcations 
(see, Figs.~\ref{fig8_set_1}(c) and~\ref{fig8_set_2}(g) next for an example of such a bifurcation).

On the other hand, and as per the branch of Fig.~\ref{fig7}(a), the first component features similarly 
a RDS whereas the second component involves a hexapolar double soliton necklace. The latter state 
(which also deforms the RDS into a pattern with hexagonal symmetry due to the nonlinear interactions)
was identified in the previous work of~\cite{egc_16} in the single-component case and classified as 
exponentially unstable. In the present two-component case, the bound mode of Fig.~\ref{fig7}(a) is 
exponentially unstable over the reported interval of $\mu_{+}\approx[1.3219,1.4]$ (see the bottom left
panel of Fig.~\ref{fig7}). The double solitonic necklace emerges at $\mu_{+}\approx 1.3219$; Subsequently, 
a real eigenvalue passes through the origin at $\mu_{+}\approx 1.39$ giving birth (via a pitchfork bifurcation)
to the branch of Fig.~\ref{fig7}(b) (see the bottom right panel of
Fig.~\ref{fig7}). This
branch, in turn, 
features a vortex necklace (notice the modification in the relevant
phase
profile from the purely real parent branch of Fig.~\ref{fig7}(a))
and is classified as exponentially unstable over the reported interval of 
$\mu_{+}\approx[1.39,1.4]$ as is shown in the bottom left panel of Fig.~\ref{fig7}. 
\begin{figure}[htp]
\vskip -0.5cm
\begin{center}
\includegraphics[height=.25\textheight, angle =0]{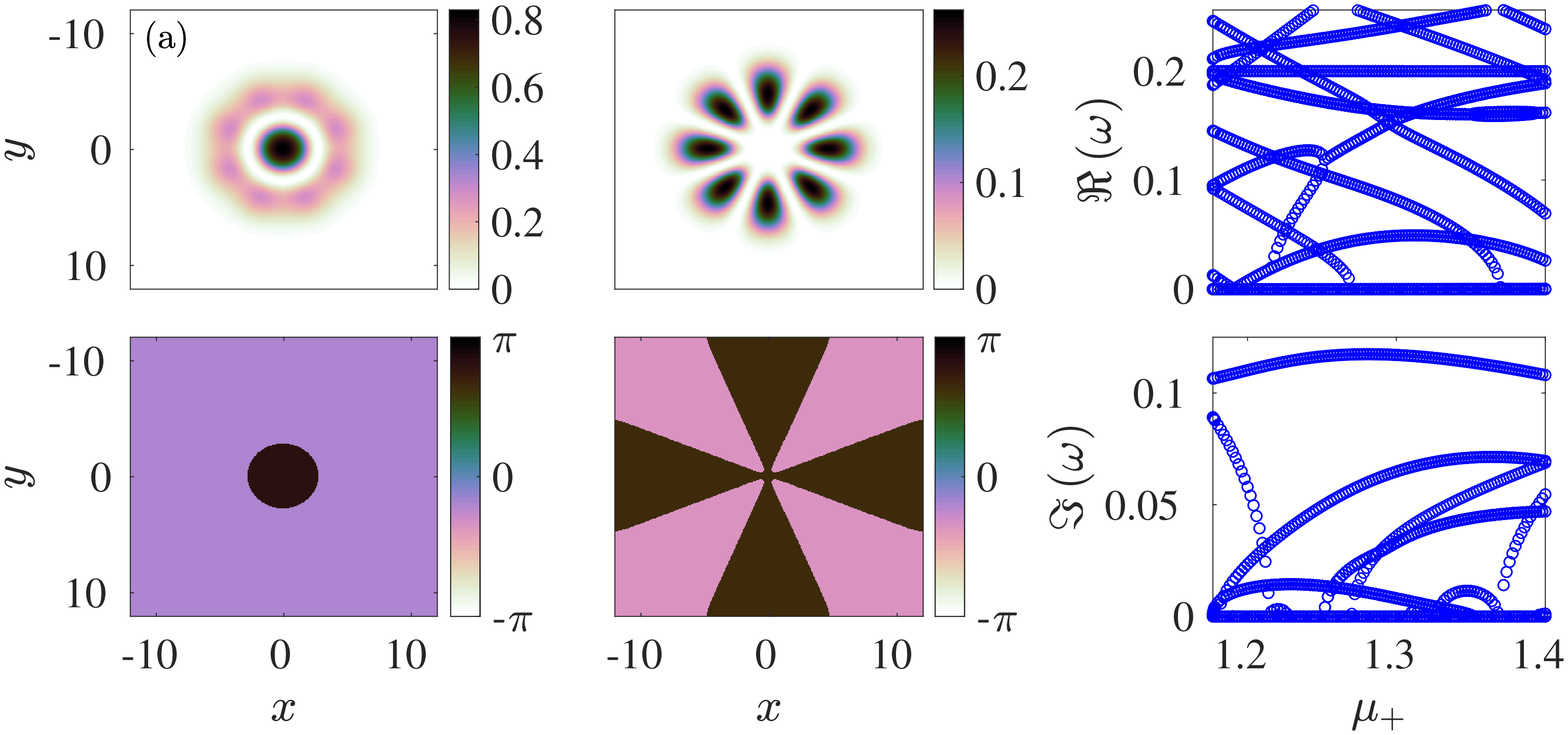}
\includegraphics[height=.25\textheight, angle =0]{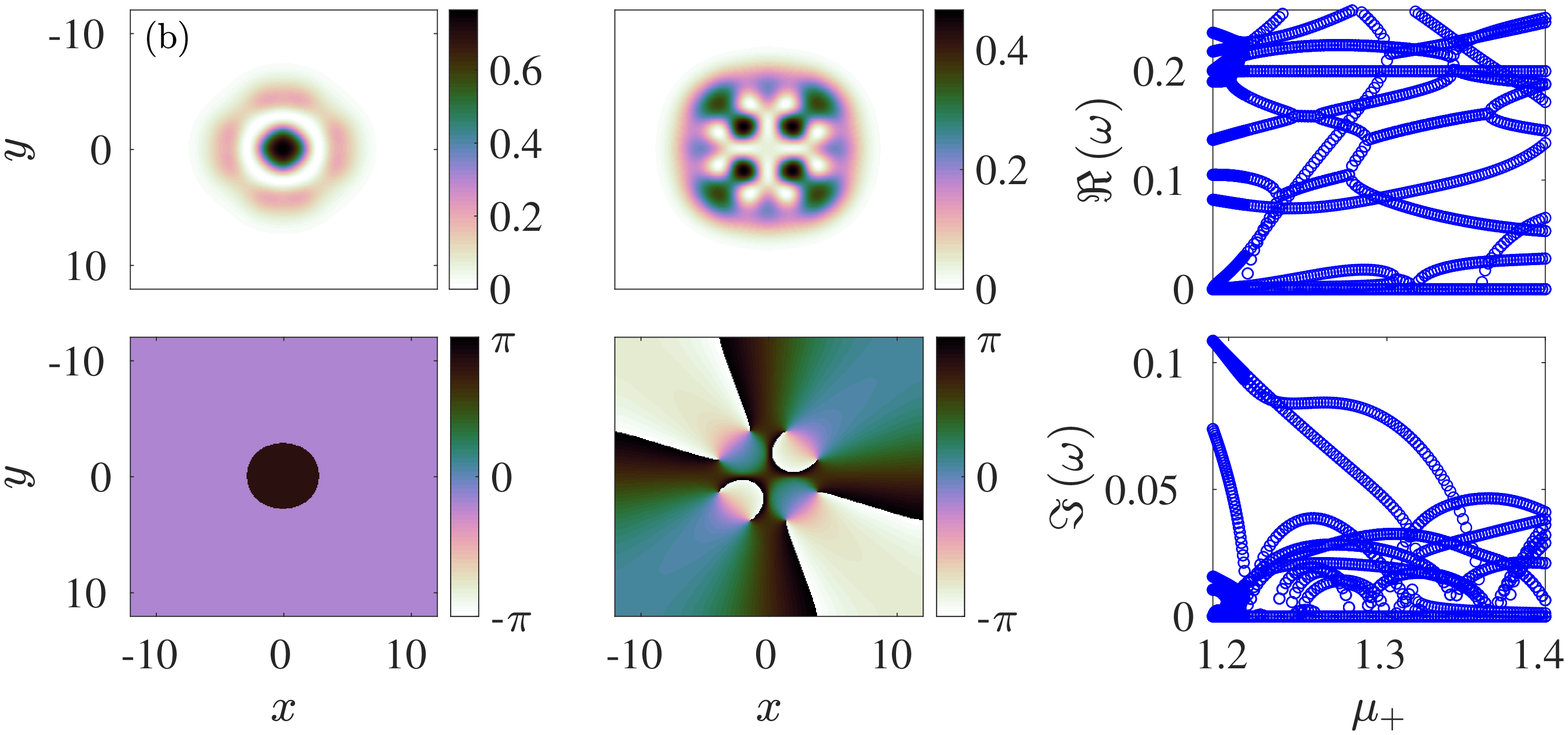}
\includegraphics[height=.25\textheight, angle =0]{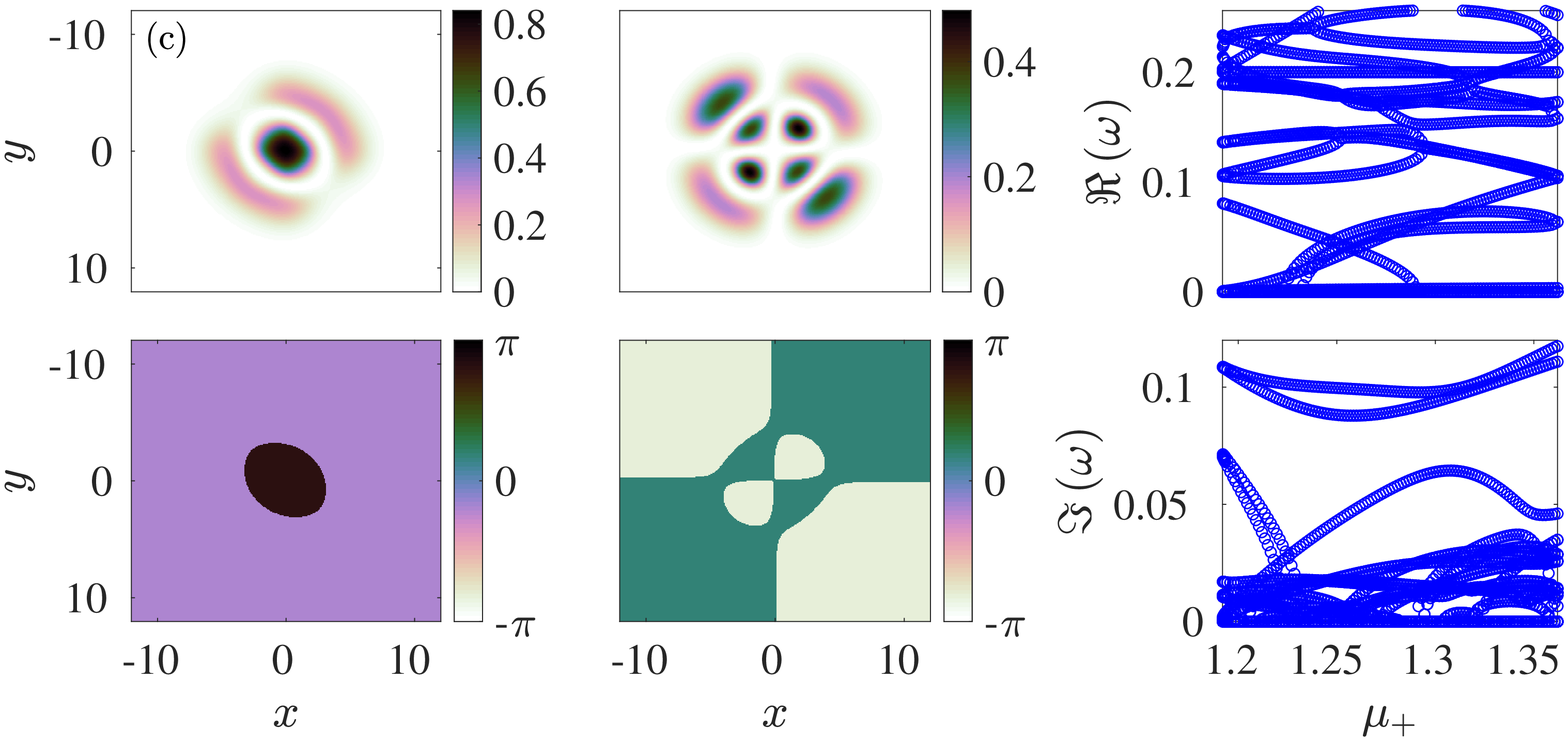}
\includegraphics[height=.25\textheight, angle =0]{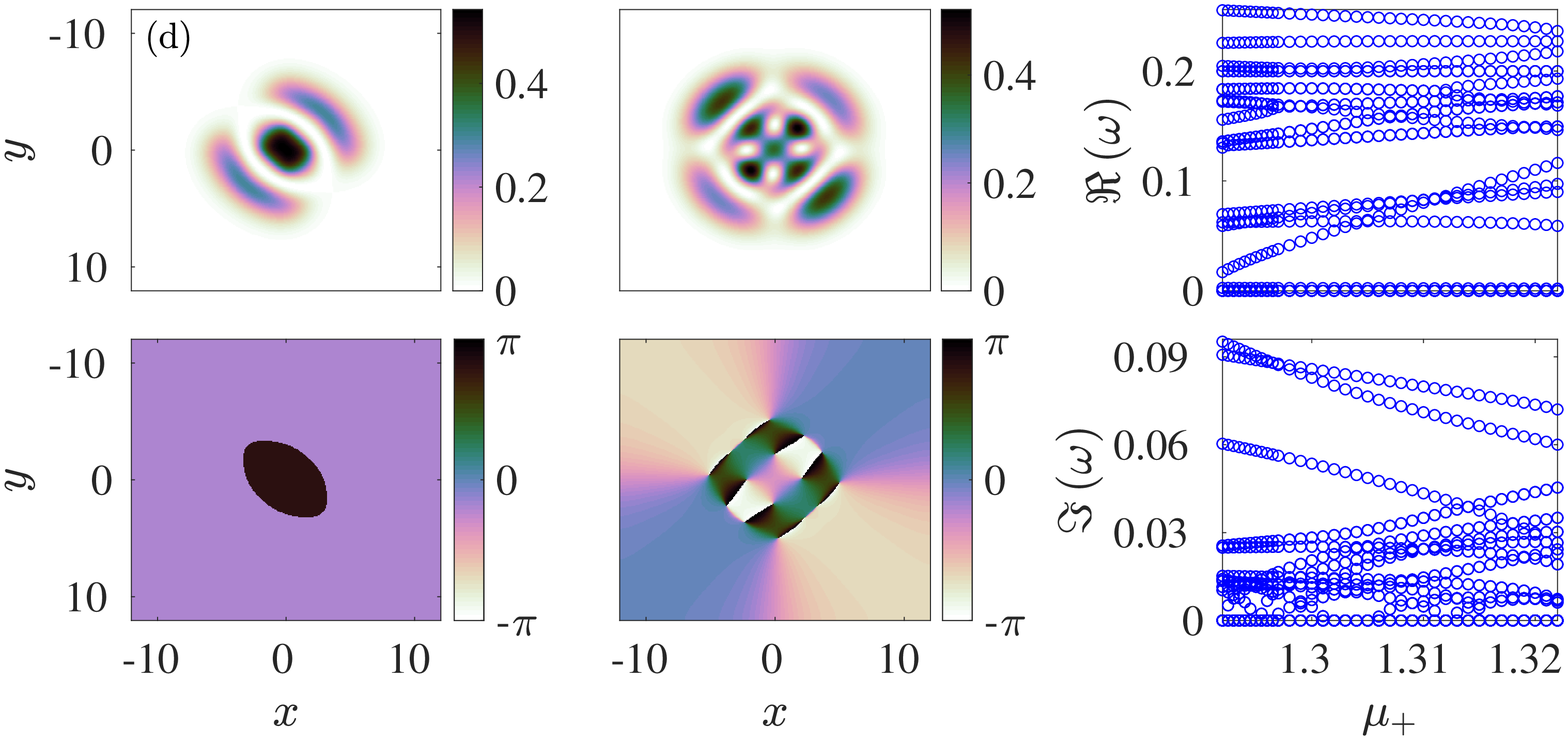}
\end{center}
\caption{
Same as Fig.~\ref{fig0} but for the RDS-necklace branch. All 
densities and phases are shown for a value  of $\mu_{+}=1.3$ 
except of (d) corresponding to $\mu_{+}\approx 1.32$. The branch of panel 
(a) emerges at $\mu_{+}\approx 1.177$. The branches (b) and (c) emanate 
from (a) at $\mu_{+}\approx 1.19$ and $\mu_{+}\approx 1.192$, respectively. 
The branch (c) merges with the branch of Fig.~\ref{fig8_set_2}(g) at 
$\mu_{+}\approx 1.362$ (see text for details). Finally, the branch of 
panel (d) emerges at $\mu_{+}\approx 1.291$ from (c).
}
\label{fig8_set_1}
\end{figure}

\begin{figure}[htp]
\vskip -0.5cm
\begin{center}
\includegraphics[height=.25\textheight, angle =0]{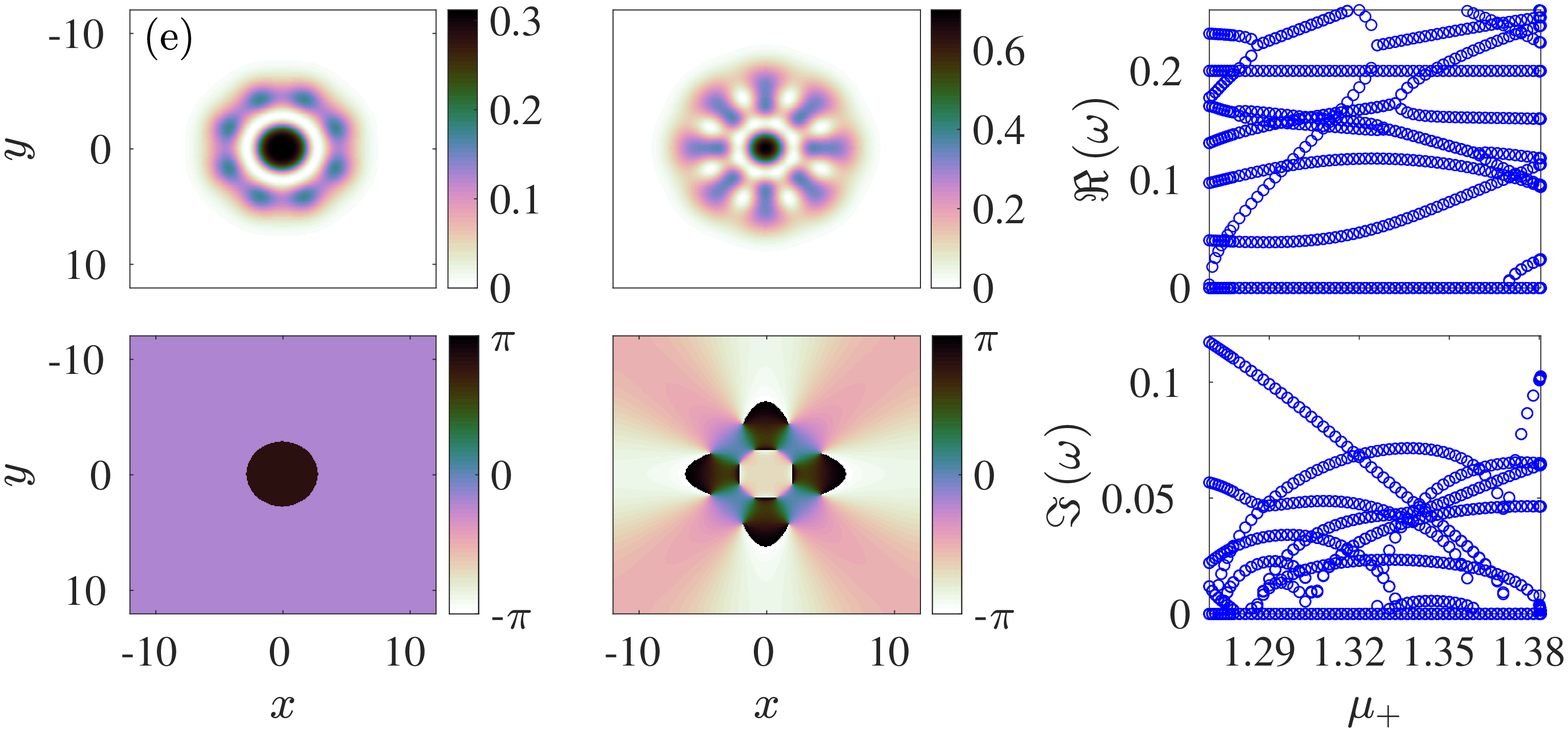}
\includegraphics[height=.25\textheight, angle =0]{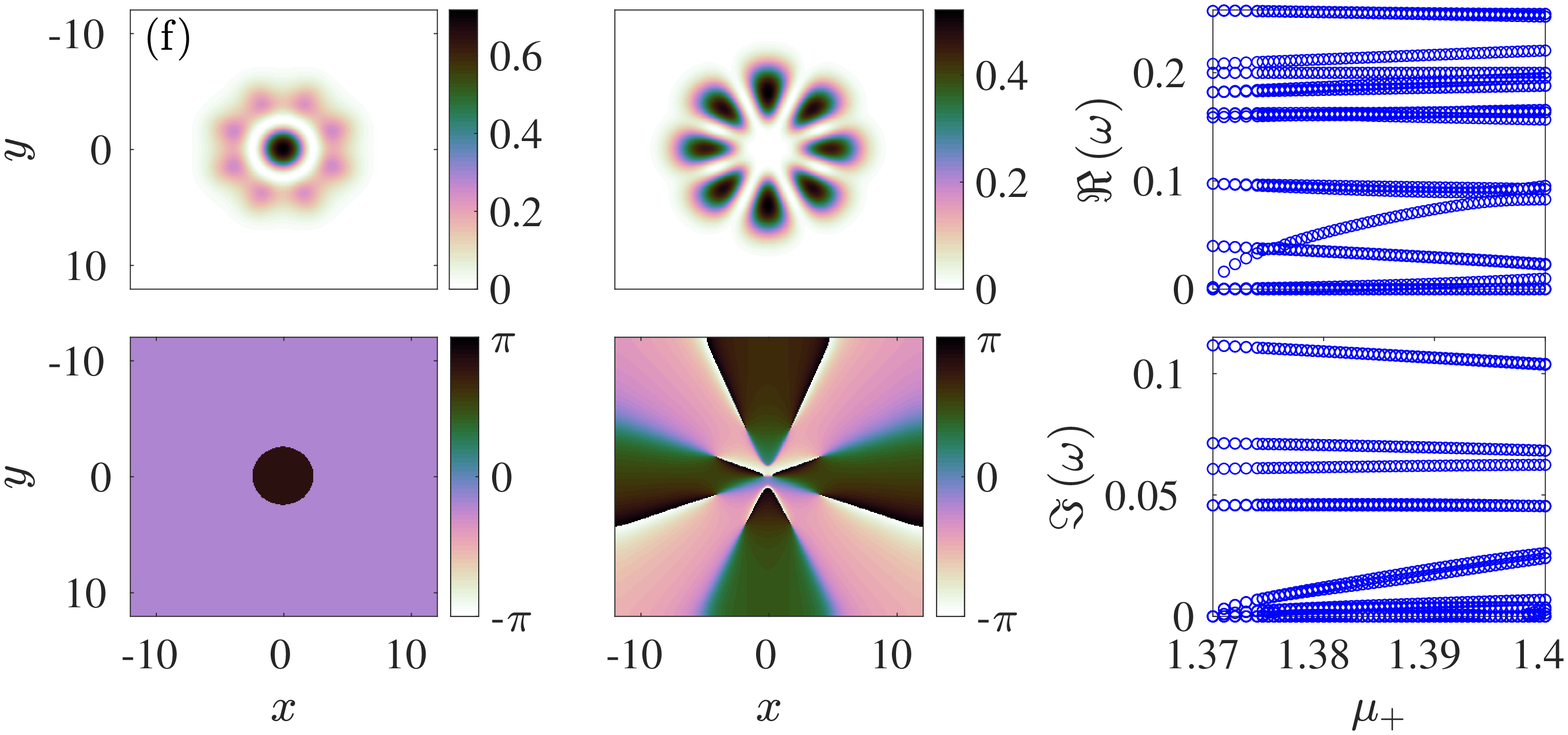}
\includegraphics[height=.25\textheight, angle =0]{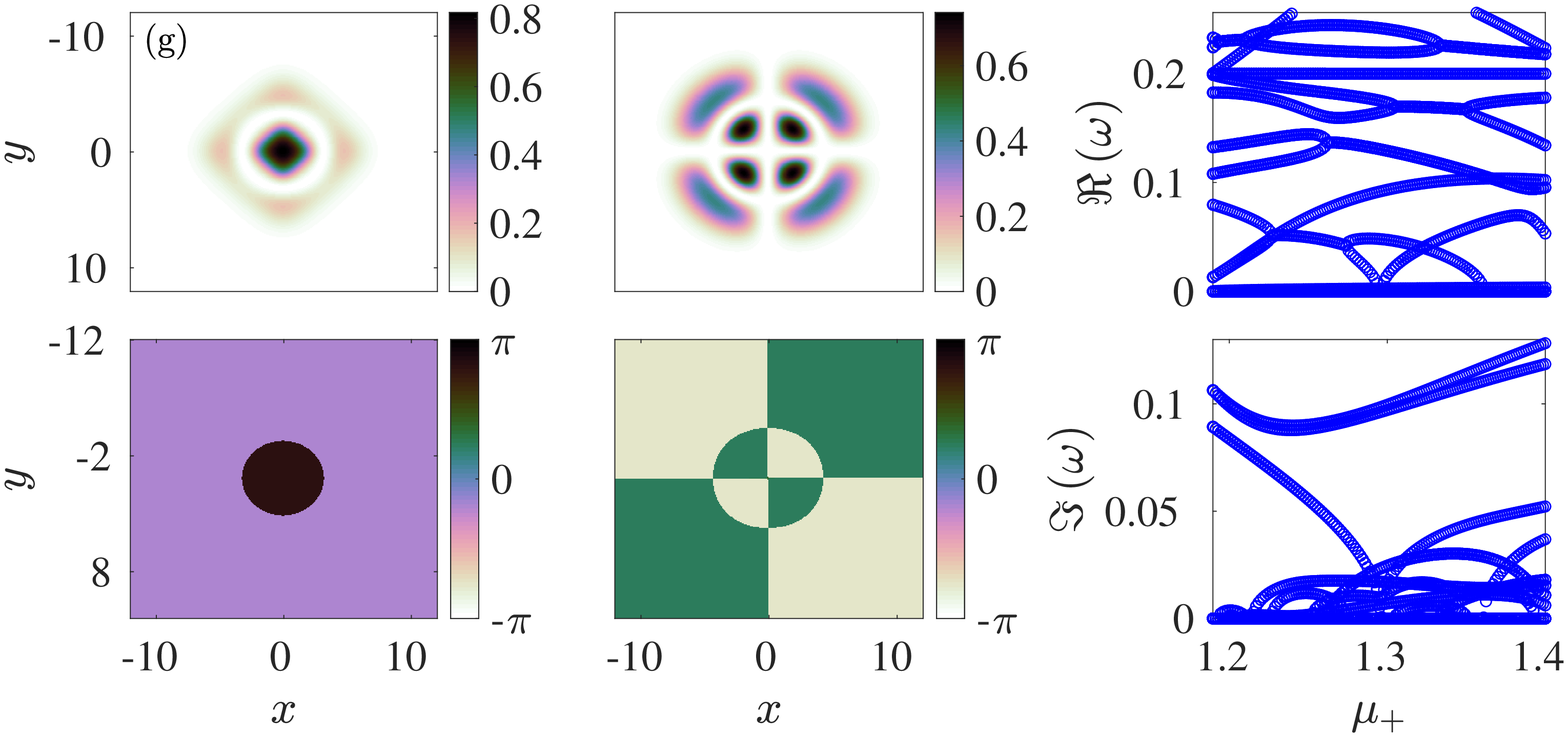}
\includegraphics[height=.25\textheight, angle =0]{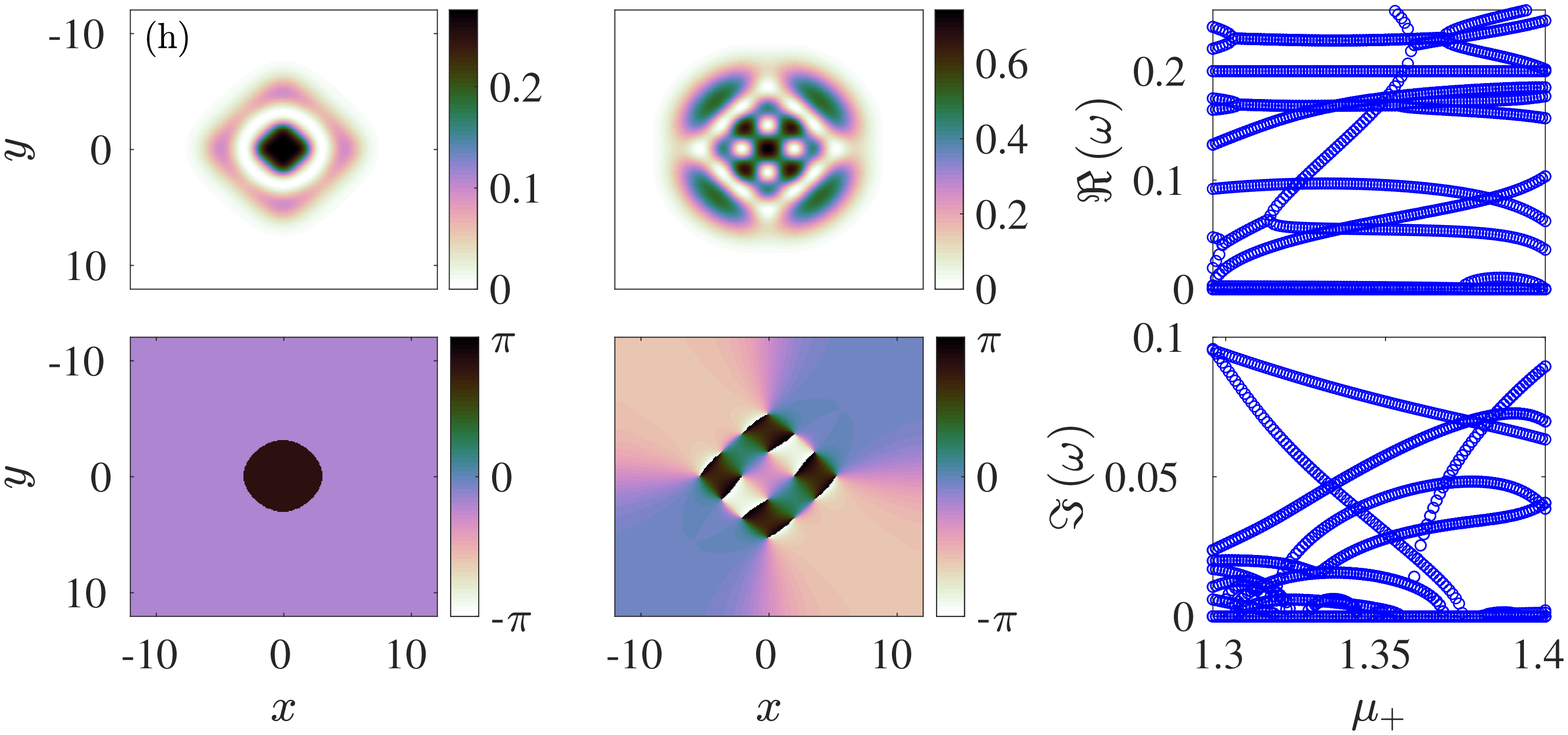}
\end{center}
\caption{
Continuation of Fig.~\ref{fig8_set_1}. Densities and phases are presented
for values of $\mu_{+}$ of $\mu_{+}=1.32$ (e), $\mu_{+}=1.4$ (f), $\mu_{+}=1.4$
(g), and $\mu_{+}=1.4$ (h), respectively. The branch of panel (e) emerges 
at $\mu_{+}\approx 1.27$ and vanishes at $\mu_{+}\approx 1.381$ whereas the
branch of (f) emerges at $\mu_{+}\approx 1.37$ and is shown up to $\mu_{+}=1.4$.
Finally, the branch of panel (g) emerges at $\mu_{+}\approx 1.1896$ (linear limit)
giving birth to branch (h) at $\mu_{+}\approx 1.296$. 
}
\label{fig8_set_2}
\end{figure}  
\begin{figure}[htp]
\vskip -0.5cm
\begin{center}
\includegraphics[height=.17\textheight, angle =0]{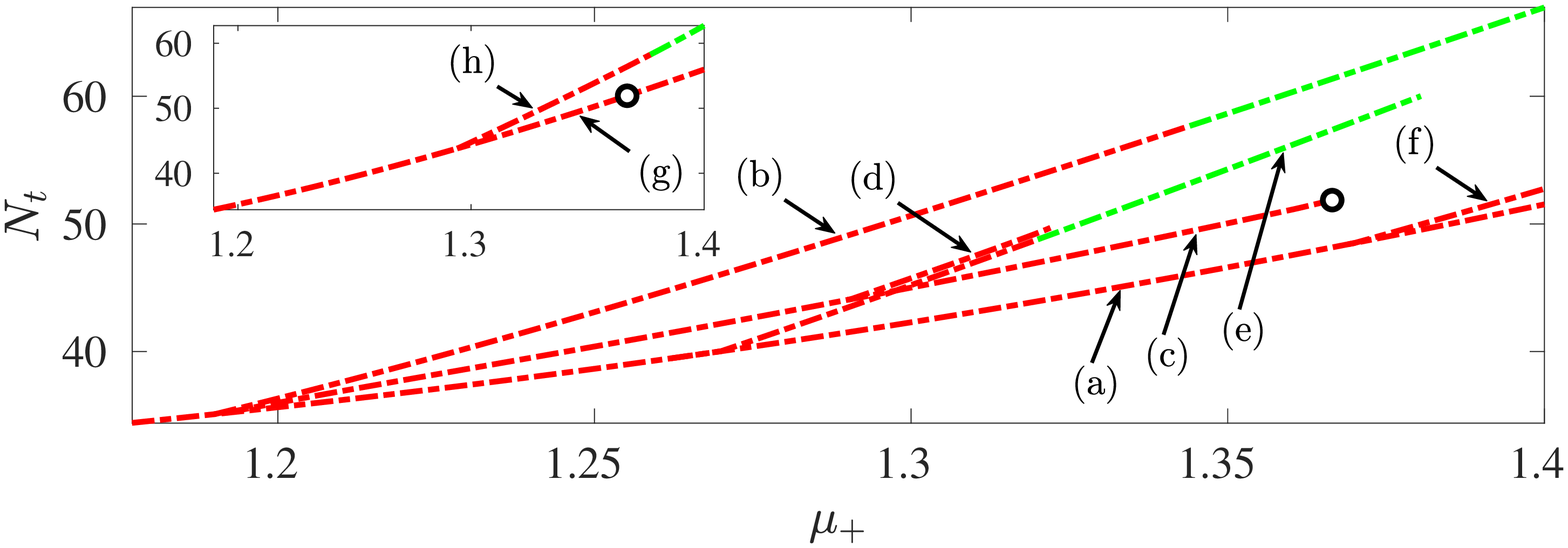}
\includegraphics[height=.17\textheight, angle =0]{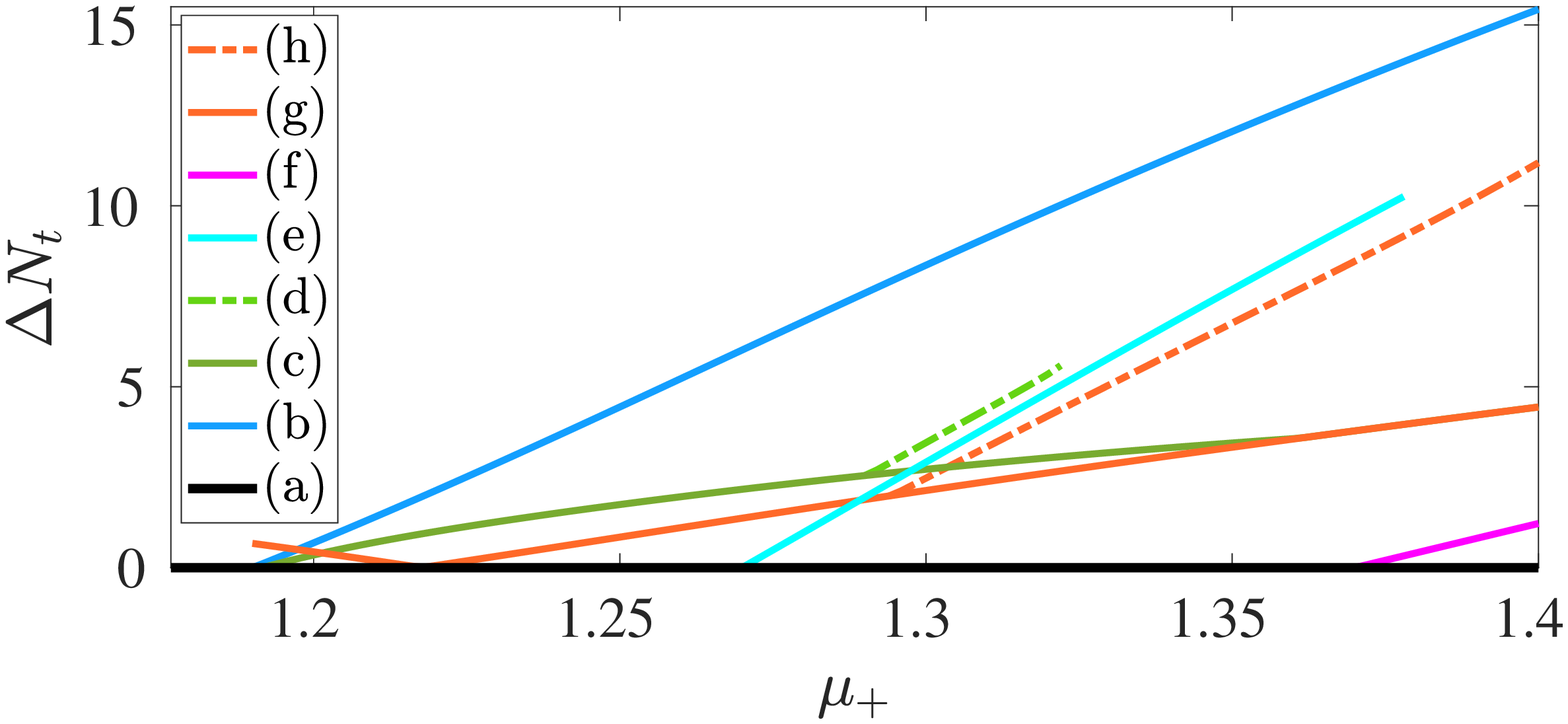}
\end{center}
\caption{
Continuation of Fig.~\ref{fig8_set_1} and also~\ref{fig8_set_2}. The total number of atoms 
and total number of atoms difference are shown in the top and 
bottom panels, respectively as functions of $\mu_{+}$. The black
open circles in the top panel are explained in the text.
}
\label{fig8_set_3}
\end{figure} 

We turn now our focus on the branches presented in Figs.~\ref{fig8_set_1},
\ref{fig8_set_2} and \ref{fig8_set_3}. Specifically, the 
branch of Fig.~\ref{fig8_set_1}(a) involves a RDS-necklace state whose first
component is a RDS of octagonal type and bears small ``blobs'' in its density. 
These blobs are complementary (due to inter-component repulsion) to the ones of 
the solitonic necklace of the second component. This bound mode emerges from the
linear limit at $\mu_{+}\approx 1.177$ and is classified (according to our spectral
stability analysis) as exponentially unstable over
$\mu_{+}=[1.177,1.4]$ (see also the
top panel of Fig.~\ref{fig8_set_3}). In addition, this branch gives birth to the 
structures of Fig.~\ref{fig8_set_1}(b) and~\ref{fig8_set_1}(c), respectively through
pitchfork bifurcations. In particular, the state of Fig.~\ref{fig8_set_1}(b) emerges 
at $\mu_{+}\approx 1.19$ and features a vortex necklace in the second component (see 
also the bottom panel of Fig.~\ref{fig8_set_3}). It can be discerned from the second 
component of Fig.~\ref{fig8_set_1}(b) that the blobs of the necklace of its parent 
branch [cf. Fig~\ref{fig8_set_1}(a)] re-arrange themselves due to the emergence of 
vorticity therein. Based on our spectral stability analysis, this state is classified 
as exponentially unstable over $\mu_{+}\approx [1.19,1.344]$ and oscillatorily unstable 
past $\mu_{+}\approx 1.344$ (see the top panel of Fig.~\ref{fig8_set_3}). Furthermore, 
the soliton necklace in the second component of the branch of Fig.~\ref{fig8_set_1}(a) 
undergoes a density re-arrangement in its blobs at $\mu_{+}\approx 1.192$ resulting in 
the bound state of Fig.~\ref{fig8_set_1}(c) whose first component is morphed into an elliptical
RDS state and is classified as exponentially unstable over $\mu_{+}\approx [1.192,1.362]$. 
Although we do not exhaustively discuss secondary bifurcations in this work, we report the 
emergence of the branch depicted in Fig.~\ref{fig8_set_1}(d) which emanates via a pitchfork 
bifurcation of its parent branch [cf. Fig.~\ref{fig8_set_1}(c)] at $\mu_{+}\approx 1.291$. 
This branch is exponentially unstable over the interval of $\mu_{+}\approx [1.291,1.322]$
(results past $\mu_{+}\approx 1.322$ are not shown) as is depicted in the top panel of Fig.~\ref{fig8_set_3}. 
Subsequently, the branch of Fig.~\ref{fig8_set_1}(a) undergoes two further pitchfork bifurcations
at $\mu_{+}\approx 1.27$ and $\mu_{+}\approx 1.37$, respectively. The first bifurcating state is
shown in Fig.~\ref{fig8_set_2}(e) which involves a vortex necklace of octagonal shape in the second
component. The latter state was identified in the single-component NLS equation in~\cite{egc_16} 
(and references therein) as a combination of a double ring configuration and a necklace. According
to our stability analysis results shown in the top panel of Fig.~\ref{fig8_set_3}, this branch is 
exponentially unstable for $\mu_{+}\approx [1.27,1.32]$ and oscillatorily unstable for $\mu_{+}\approx[1.32, 1.381]$
(note that the second component vanishes past $\mu_{+}\approx 1.381$). At the same time, the branch 
of Fig.~\ref{fig8_set_2}(f) features another vortex necklace in its second component. This branch is
classified as exponentially unstable over $\mu_{+}\approx [1.37,1.4]$ (results are not shown past 
$\mu_{+}=1.4$). Surprisingly, the branch of Fig.~\ref{fig8_set_1}(c) at $\mu_{+}\approx 1.362$ merges
with the branch of Fig.~\ref{fig8_set_2}(g) (see also the black open circles in the bifurcation diagram
in the top panel of Fig.~\ref{fig8_set_3}). This merging of
\ref{fig8_set_1}(c)
with~\ref{fig8_set_2}(g) is a canonical example of a {reverse} pitchfork
bifurcation where the branch of Fig.~\ref{fig8_set_2}(g) is the parent
branch.
This branch is classified as exponentially unstable. However, as $\mu_{+}$ increases and at 
$\mu_{+}\approx 1.296$, the branch of Fig.~\ref{fig8_set_2}(g) undergoes a pitchfork bifurcation giving birth 
to the state of Fig.~\ref{fig8_set_2}(h) bearing a cluster of twelve vortices in the outer ring and four 
charge-one vortices in arranged in a cross shape. Notice that adjacent vortices in the pattern bear opposite 
charges lending the overall pattern a vanishing net vorticity. This branch is classified as exponentially 
unstable except for a narrow interval of $\mu_{+}\approx[1.377,1.4]$ over which it appears (dominantly to be) 
oscillatorily unstable as is shown in the top panel of Fig.~\ref{fig8_set_3} (this branch is not shown for 
values of $\mu_{+}$ past $\mu_{+}\approx 1.4$). Our results on all the above branches are summarized in the 
top and bottom panels of Fig.~\ref{fig8_set_3} highlighting the dominant instability and bifurcations, respectively.
\begin{figure}[htp]
\vskip -0.5cm
\begin{center}
\includegraphics[height=.25\textheight, angle =0]{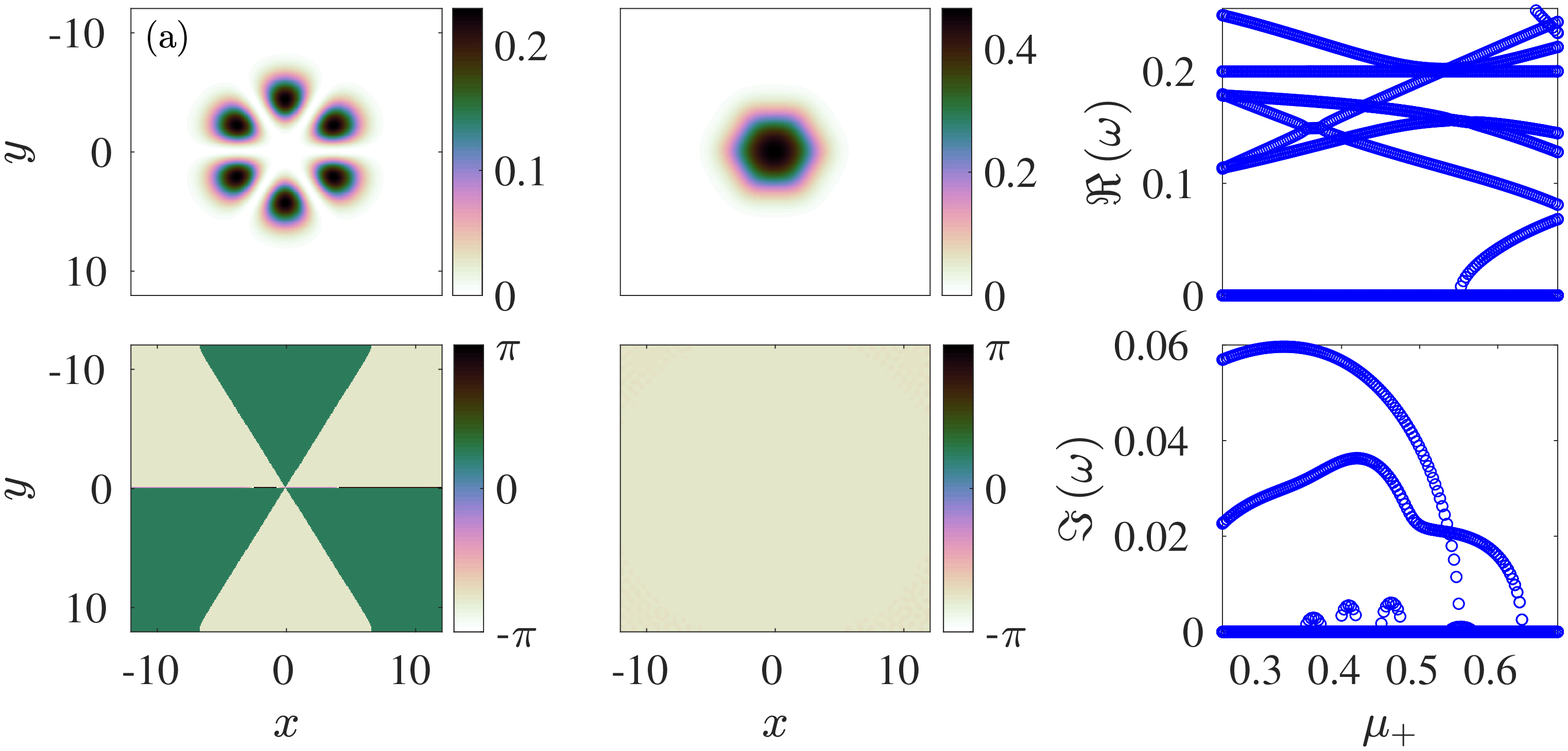}
\includegraphics[height=.25\textheight, angle =0]{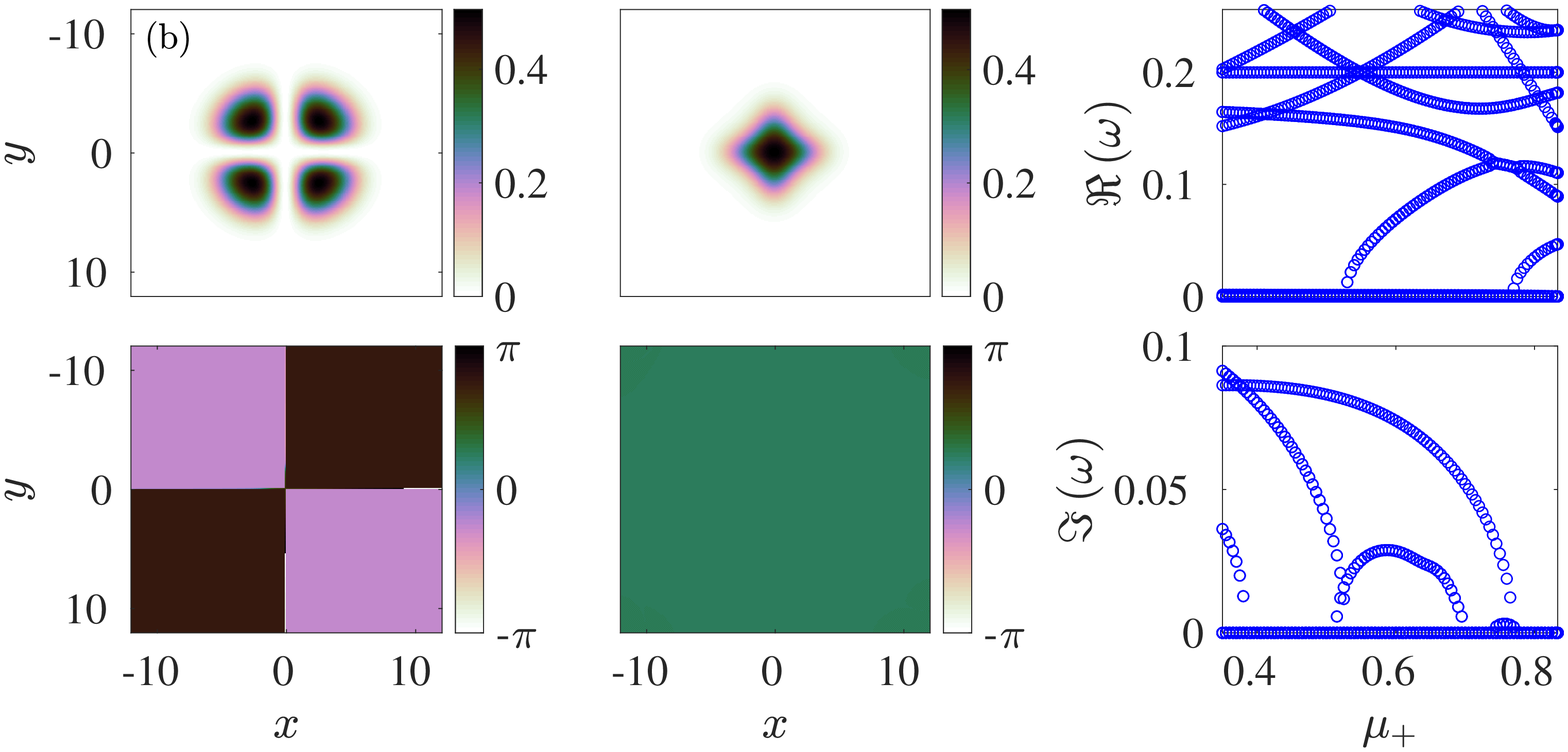}
\includegraphics[height=.25\textheight, angle =0]{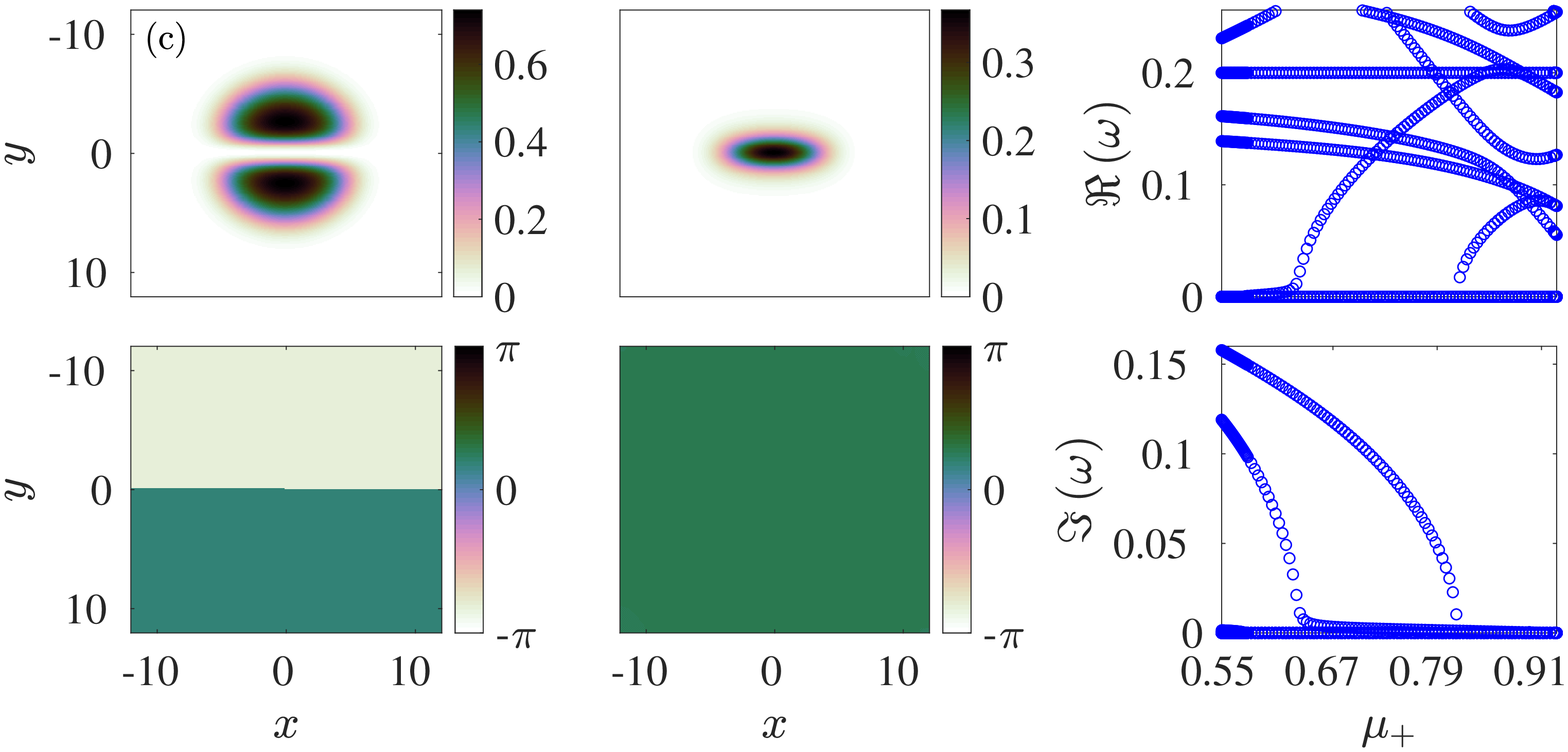}
\includegraphics[height=.25\textheight, angle =0]{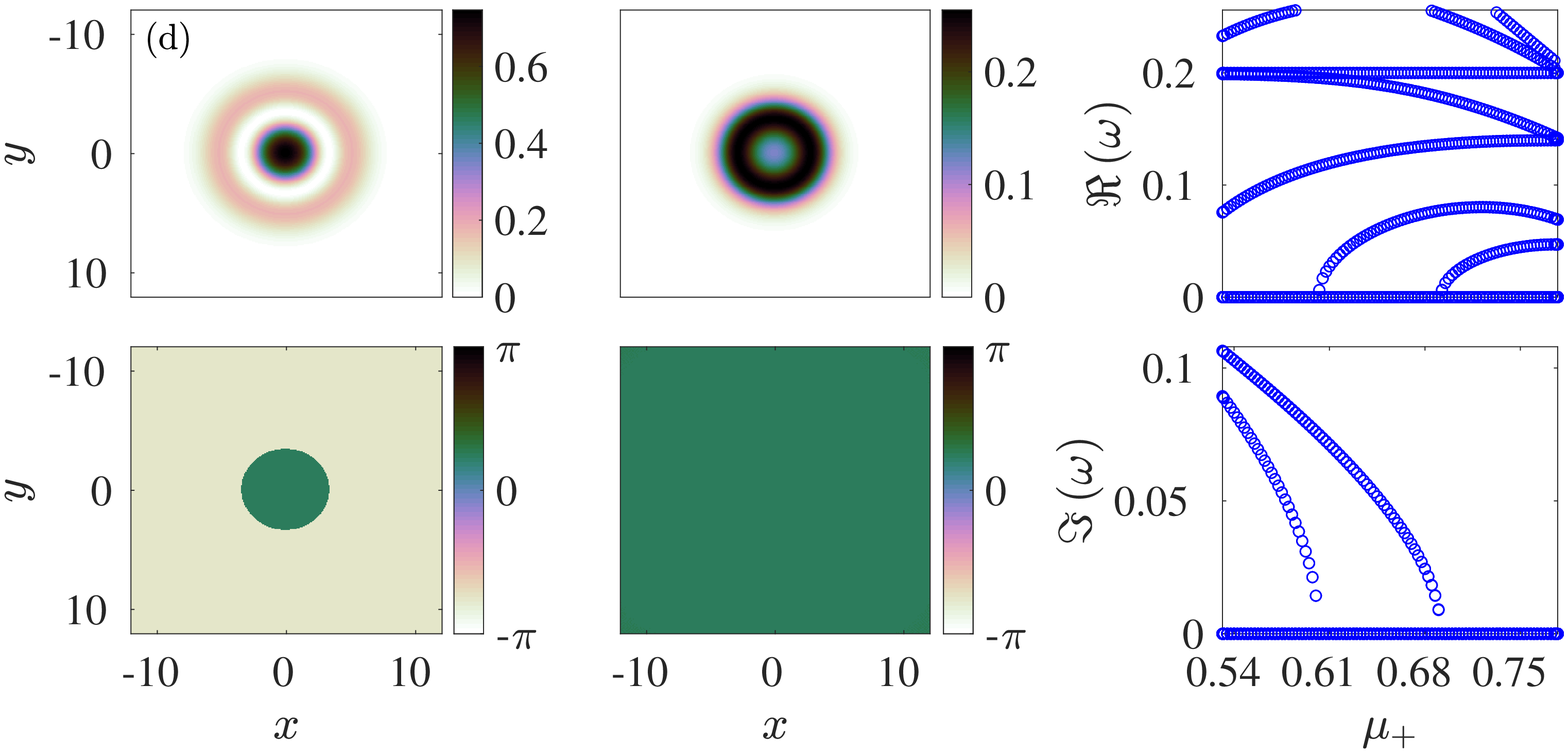}
\end{center}
\caption{Branches of solutions that are reversed versions of 
ones discussed previously in this work. Densities and phases 
as well as associated spectra are shown for $\mu_{+}=0.52$ (a)
(reversed version of Fig.~\ref{fig0}(a)), $\mu_{+}=0.6$ (b) 
(reversed version of Fig.~\ref{fig2}(a)), $\mu_{+}=0.68$ (c) 
(reversed version of Fig.~\ref{fig1}(a)), and $\mu_{+}=0.65$ (d) 
(reversed version of Fig.~\ref{fig3}(a)), respectively.
}
\label{fig9}
\end{figure}

\begin{figure}[htp]
\vskip -0.5cm
\begin{center}
\includegraphics[height=.25\textheight, angle =0]{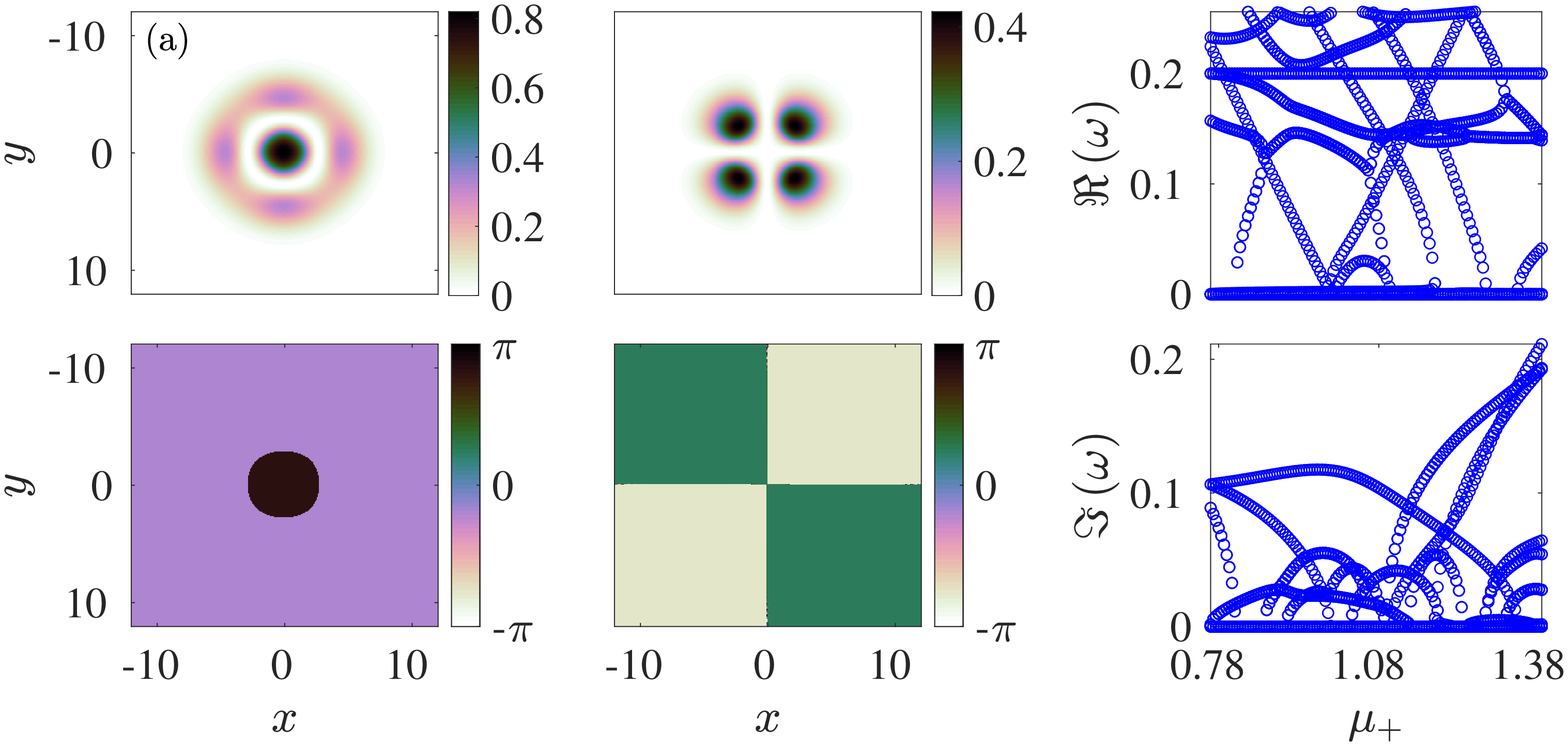}
\includegraphics[height=.25\textheight, angle =0]{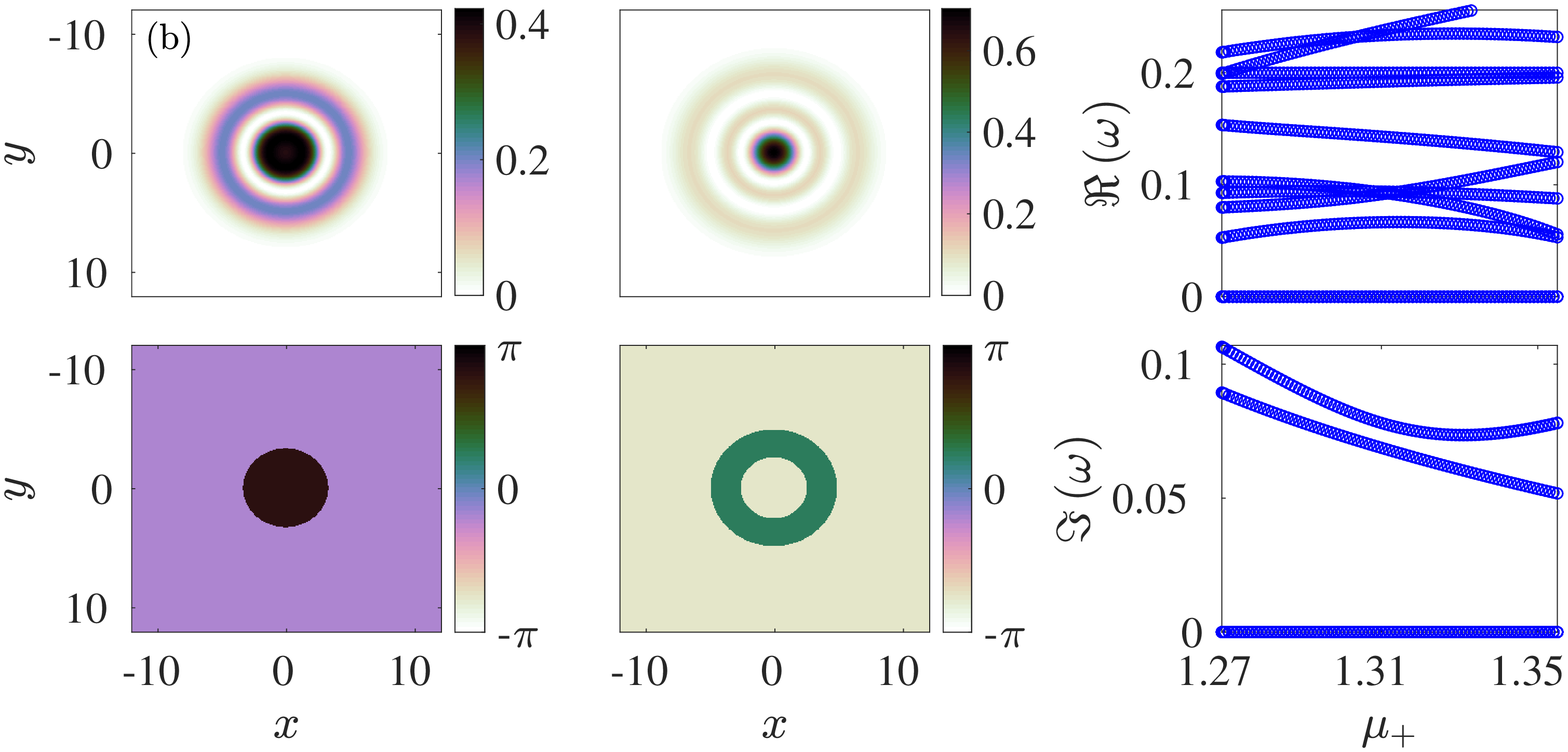}
\includegraphics[height=.25\textheight, angle =0]{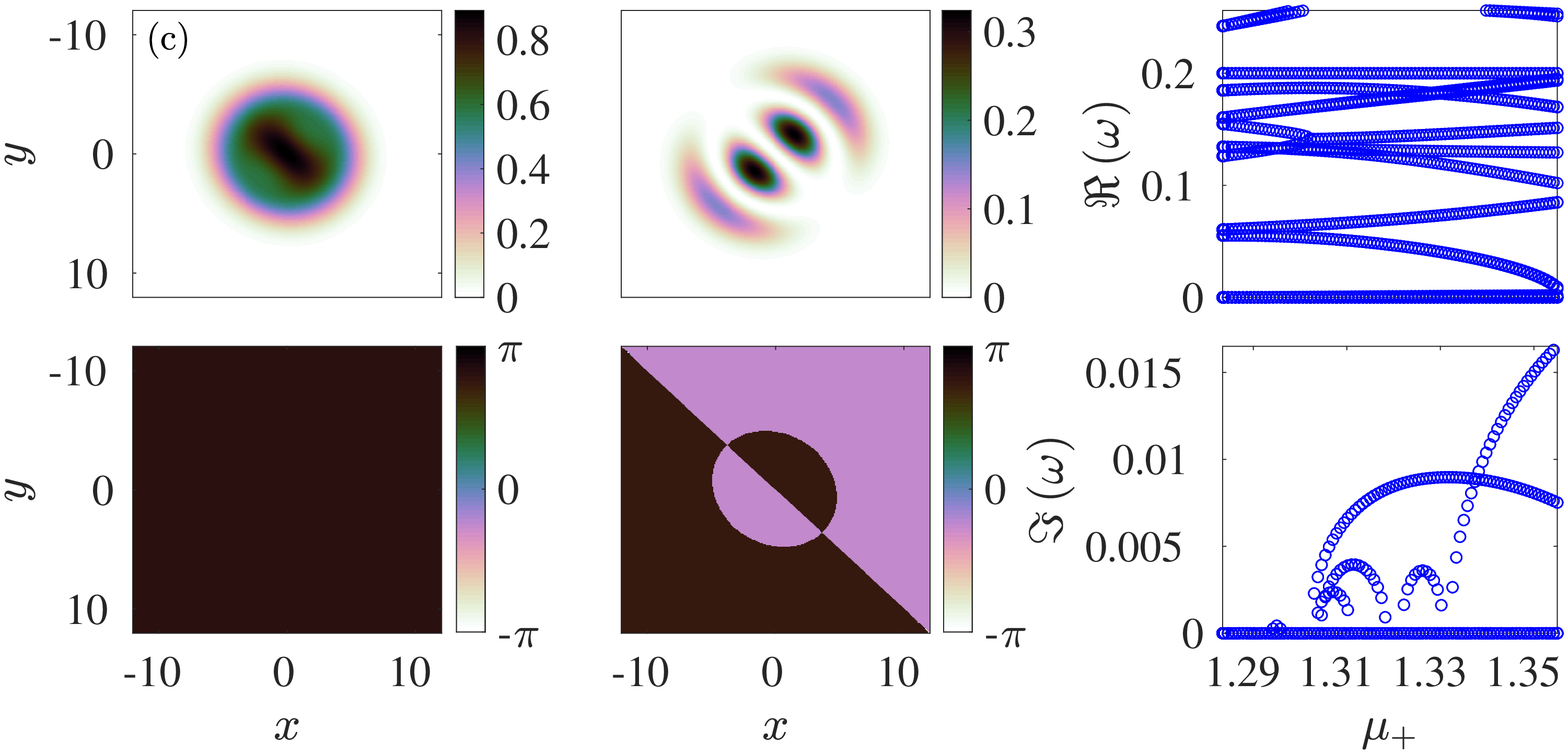}
\includegraphics[height=.25\textheight, angle =0]{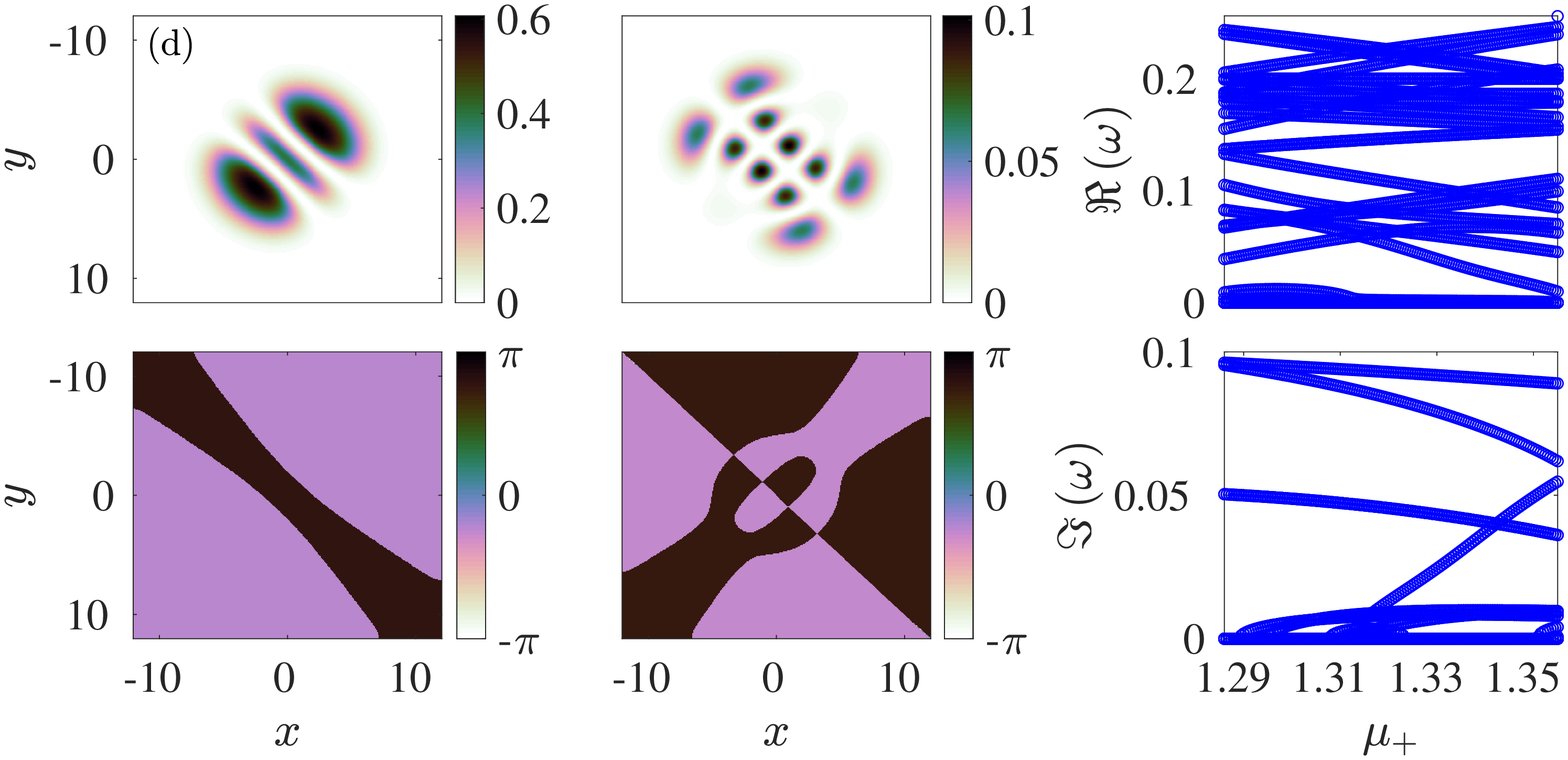}
\end{center}
\caption{
Same as Fig.~\ref{fig9}. Note that panel (a) depicted for $\mu_{+}=0.95$
is a reversed version of Fig.~\ref{fig4}(a). The branches of panels
(b)-(d) are all presented for a value of $\mu_{+}$ of $\mu_{+}=1.32$
at which the first component does not become the trivial solution.
}
\label{fig10}
\end{figure}

Finally, we discuss about the branches presented in Figs.~\ref{fig9} and~\ref{fig10} that the 
DCM identified. In Figs.~\ref{fig9}(a)-(d) and Fig.~\ref{fig10}(a), we observe a series of 
states that are variants of states that we have already encountered, albeit now in reverse 
i.e., with the second component playing the role of the first one and vice versa. Recall that 
in the case of the equal $g_{ij}$ (the so-called Manakov model integrable in one-dimension~\cite{abl3}), 
these states would be interchangeable, however the weak asymmetry (and corresponding immiscibility) 
distinguishes between these states and the earlier ones. Indeed, Fig.~\ref{fig9}(a) provides a 
reversed analogue of Fig.~\ref{fig0}(a) whereas the branch of Fig.~\ref{fig9}(b) can be similarly 
connected to Fig.~\ref{fig2}(a). These branches involve multipoles (a hexapole and a quadrupole) 
coupled to a fundamental state. The branch of Fig.~\ref{fig9}(c) presents a DB solitonic stripe, 
analogous to that of Fig.~\ref{fig1}(a). Also, the branch of Fig.~\ref{fig9}(d) is tantamount to 
Fig.~\ref{fig3}(a), namely the ring dark-bright soliton, whereas the branch of Fig.~\ref{fig10}(a) 
is strongly reminiscent of the branch of Fig.~\ref{fig4}(a) involving a RDS and the $|1,1\rangle_{(\textrm{c})}$ 
(in its Cartesian classification) state. In view of these similarities and the slight asymmetries of
the $g_{ij}$, the precise instability details of the states of Fig.~\ref{fig9}(a)-(d) and 
Fig.~\ref{fig10}(a) are somewhat different, yet qualitatively similar to the examples that we
have already studied above. It should be noted that we performed the continuation of all the above 
branches over $\mu_{+}$ and  stopped when one of the components was found to be below numerical
precision, i.e., when effectively the states become single component waveforms with the other 
component being trivial. Indicatively, the intervals of existence (and stability) of the states 
shown in Fig.~\ref{fig9} are $\mu_{+}\approx [0.2474,0.6769]$ (a) (stable over $\mu_{+}\approx[0.634,0.6769]$), 
$\mu_{+}\approx [0.542,0.9275]$ (b) (stable over $\mu_{+}\approx[0.84,0.9275]$), 
$\mu_{+}\approx [0.35,0.8335]$ (c) (stable over $\mu_{+}\approx[0.775,0.8335]$), 
$\mu_{+}\approx [0.531,0.778]$ (d) (stable over $\mu_{+}\approx [0.6925,0.778]$),
as well as the one of Fig.~\ref{fig10}(a) is $\mu_{+}\approx[0.764,1.385]$ (and is exponentially 
unstable over the relevant range in $\mu_{+}$). 

\begin{figure}[htp]
\vskip -0.5cm
\begin{center}
\includegraphics[height=.25\textheight, angle =0]{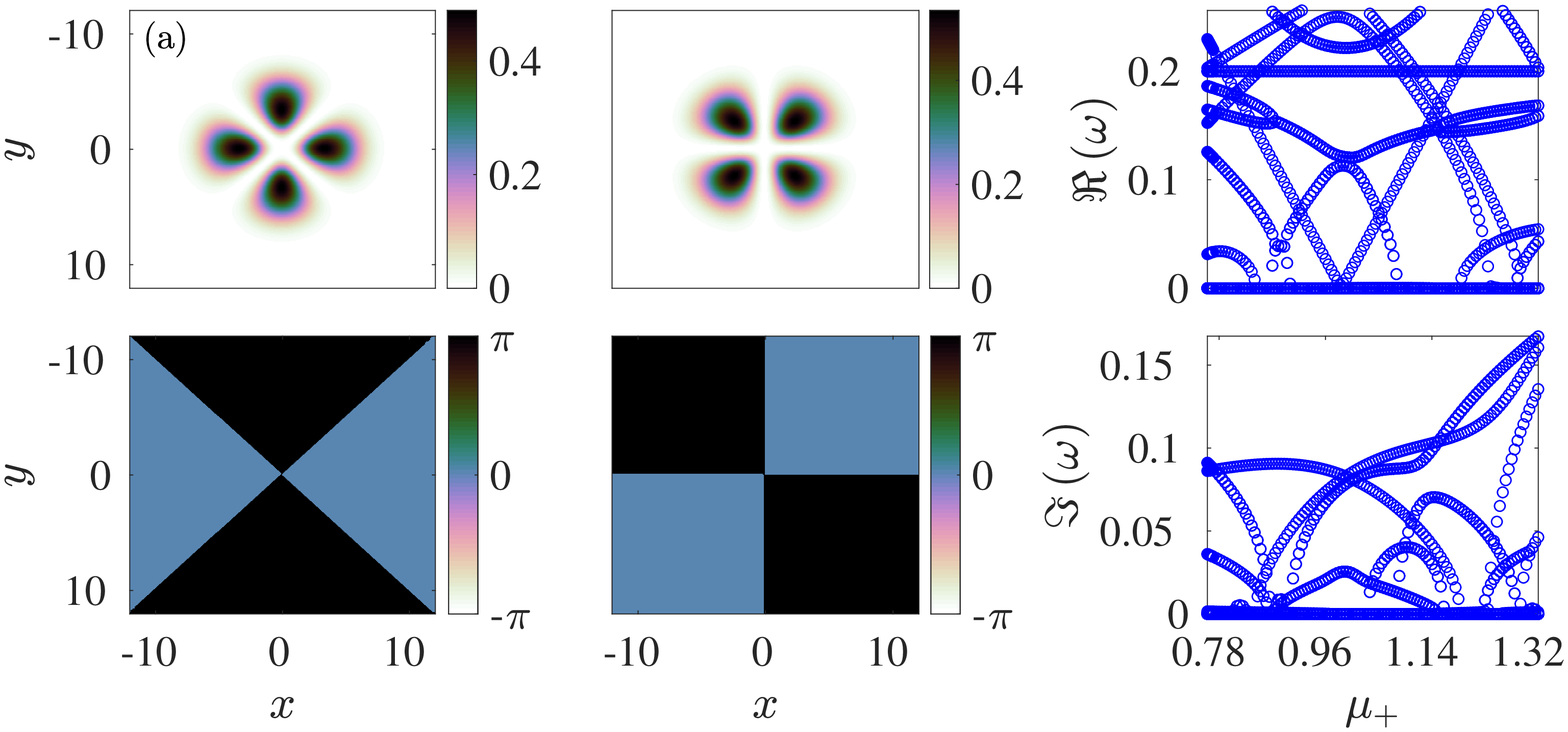}
\includegraphics[height=.25\textheight, angle =0]{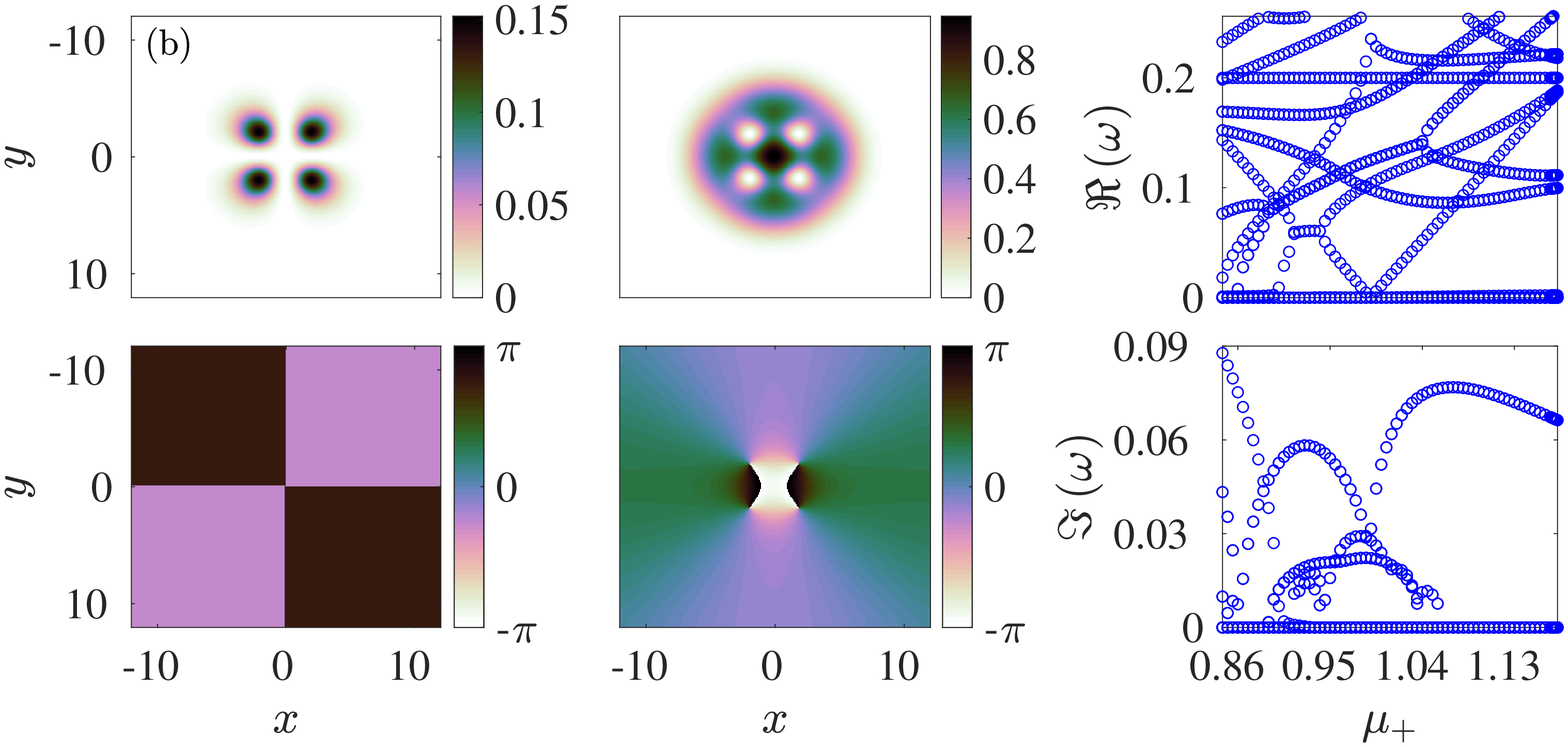}
\includegraphics[height=.17\textheight, angle =0]{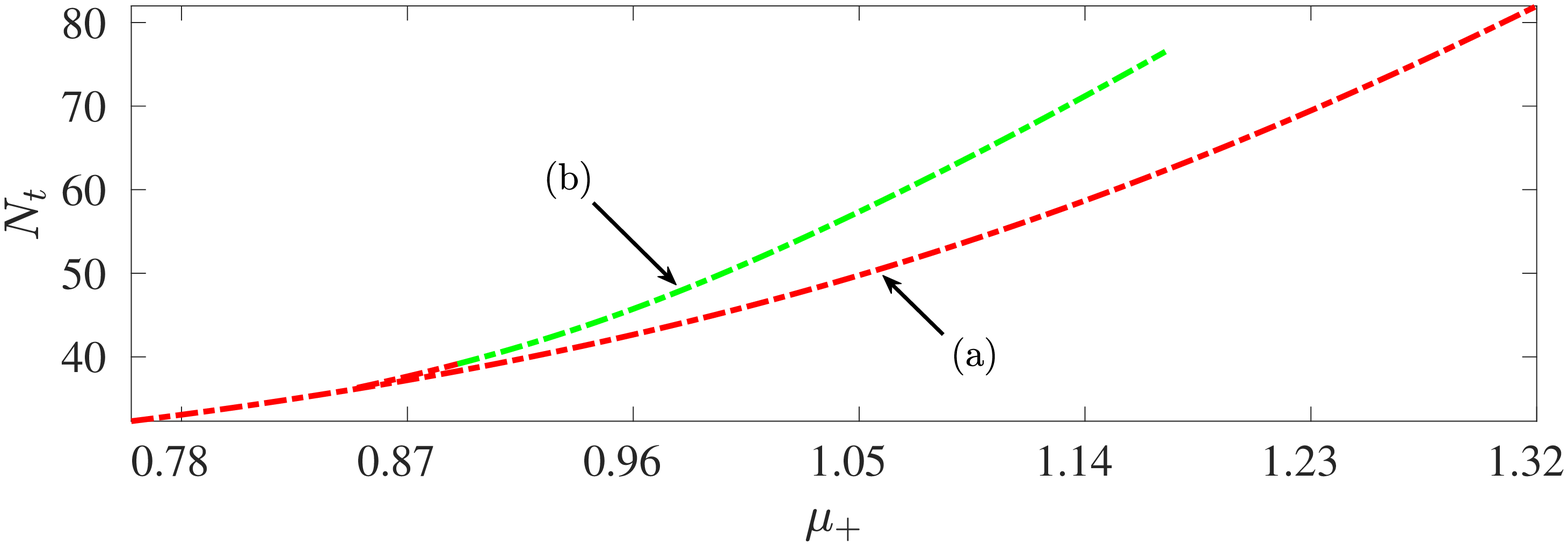}
\includegraphics[height=.17\textheight, angle =0]{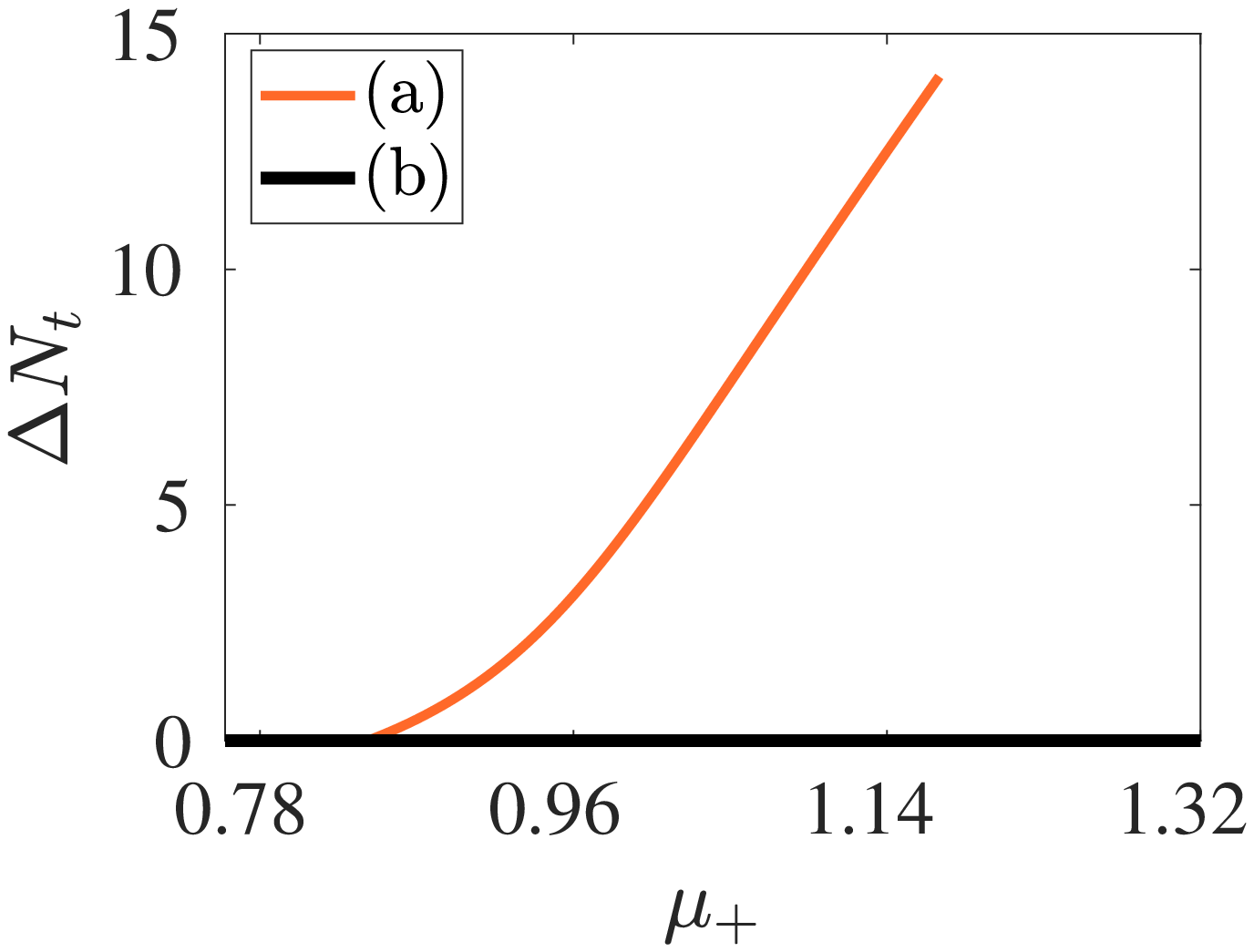}
\end{center}
\caption{
Same as Fig.~\ref{fig0} but for the rotated dipolar branch. Densities
and phases are shown in panels (a) and (b) for values of $\mu_{+}$ of
$\mu_{+}=1$ and $\mu_{+}=1.1$, respectively. The bound state involving 
the vortex quadrupole in the second component of panel (b) emerges at 
$\mu_{+}\approx 0.845$. Similarly, the bottom left and right panels present
the total number of atoms and number-of-atoms difference, respectively.
}
\label{fig11}
\end{figure}

As we advance to panels (b)-(d) of Fig.~\ref{fig10}, we encounter higher excited state combinations. 
For instance, in Fig.~\ref{fig10}(b), a double RDS in the second component is coupled to a single RDS 
in the first component (this bound state emerges at $\mu_{+}\approx 1.269$). It is important to recall, 
in line with the discussion of~\cite{egc_16}, that these ring states are eigenstates 
(with $l=0$ azimuthal index) of the associated Laguerre eigenmodes of the two-dimensional system. In 
particular, these are modes with $k=2$ in the second component and $k=1$ in the first component. 
Subsequently, in Fig.~\ref{fig10}(c), a fundamental state in the first component is coupled with the 
multipole $|2,1\rangle_{(\textrm{c})}+|0,3\rangle_{(\textrm{c})}$ in the notation of the Cartesian 
Hermite eigenstates resulting in a bound state  emerging at $\mu_{+}\approx 1.283$ as a stable branch 
until $\mu_{+}\approx 1.302$. The branch of Fig.~\ref{fig10}(d) emerges at $\mu_{+}\approx 1.286$ and 
features a higher order example of a Cartesian excited state. Here, a $|4,1 \rangle_{(\textrm{c})}$ 
state in the second component is coupled to a $|2,0 \rangle_{(\textrm{c})}$ one in the first component. 
This bound state features a multiplicity of exponentially unstable modes, in addition to a number of 
oscillatory instabilities in suitable parametric ranges. Finally, it should be noted that all the above
states with the exception of that of Fig.~\ref{fig10}(c) are very highly unstable, as can be seen from 
the right panels of Fig.~\ref{fig10}.
Yet, it is relevant to comment that some of these highly excited
states like the one depicted in Fig.~\ref{fig10}(c) may feature
intervals
of spectral stability and the highly accurate numerical techniques
used
herein can identify these states and their corresponding stability intervals.

\subsection{Further linear modes and DCM} \label{sec:extra_discussion}
The DCM was initialized at $\mu_+ = 0.492$, using as initial guess for Newton's method a Gaussian
in each component; this converged to the solution shown in Fig 15(c). At each subsequent continuation 
step, the solutions at the previous step were used as initial guesses. This strategy does not exploit 
our analytical knowledge of the problem, specifically our knowledge of eigenstates in the linear limit. 
Despite this disadvantage, the DCM identified a large number of solutions.

\begin{figure}[htp]
\vskip -0.5cm
\begin{center}
\includegraphics[height=.25\textheight, angle =0]{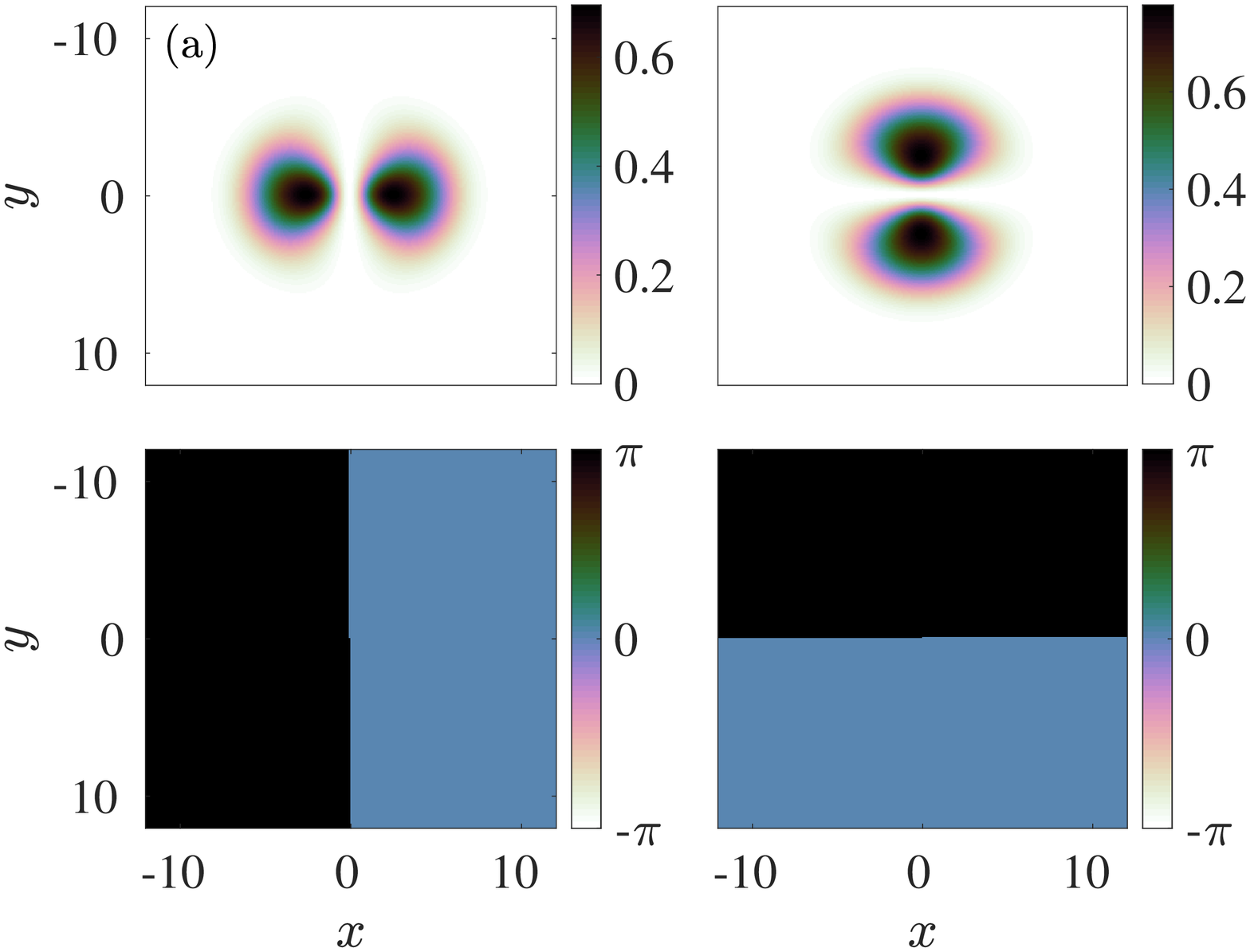}
\includegraphics[height=.25\textheight, angle =0]{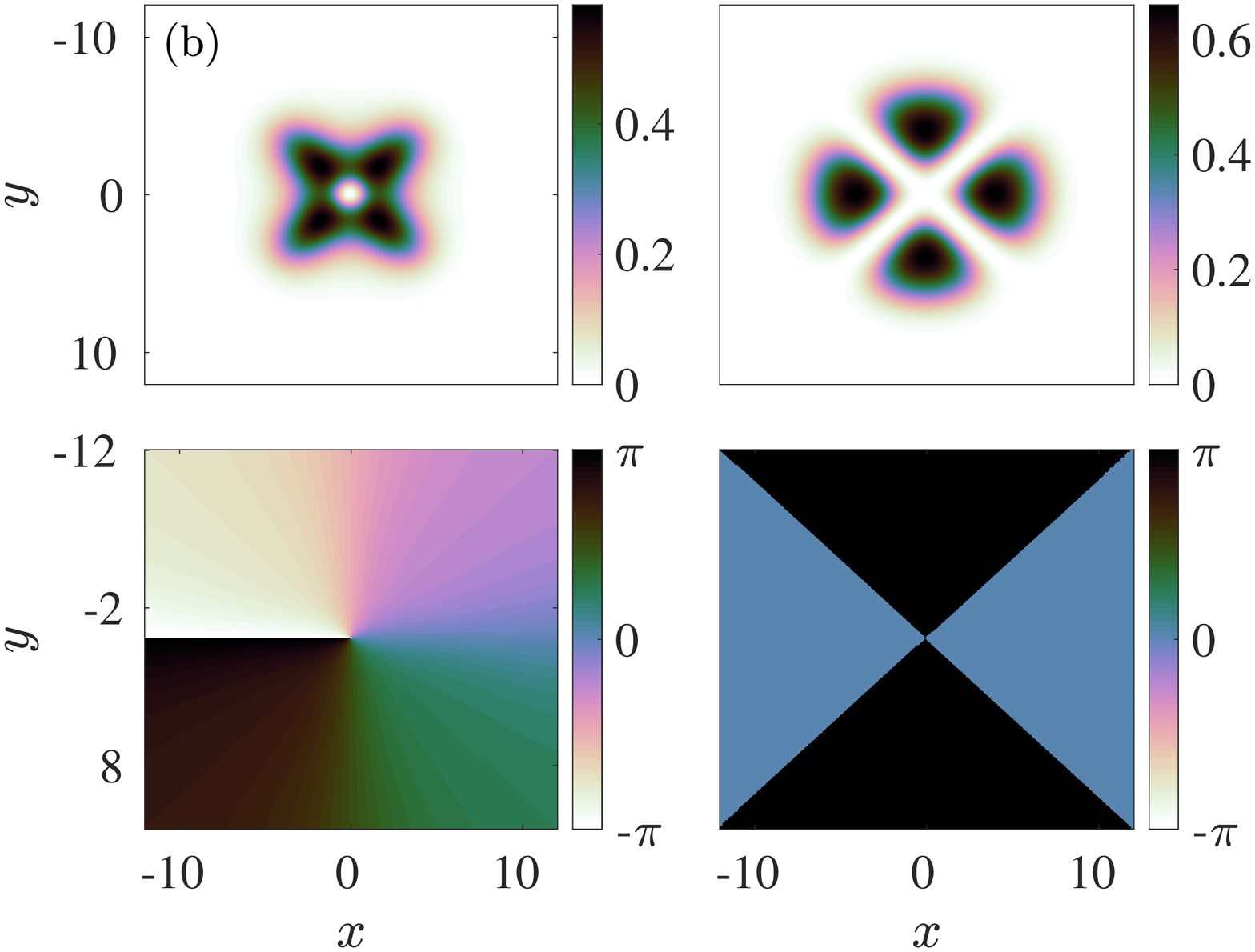}
\includegraphics[height=.25\textheight, angle =0]{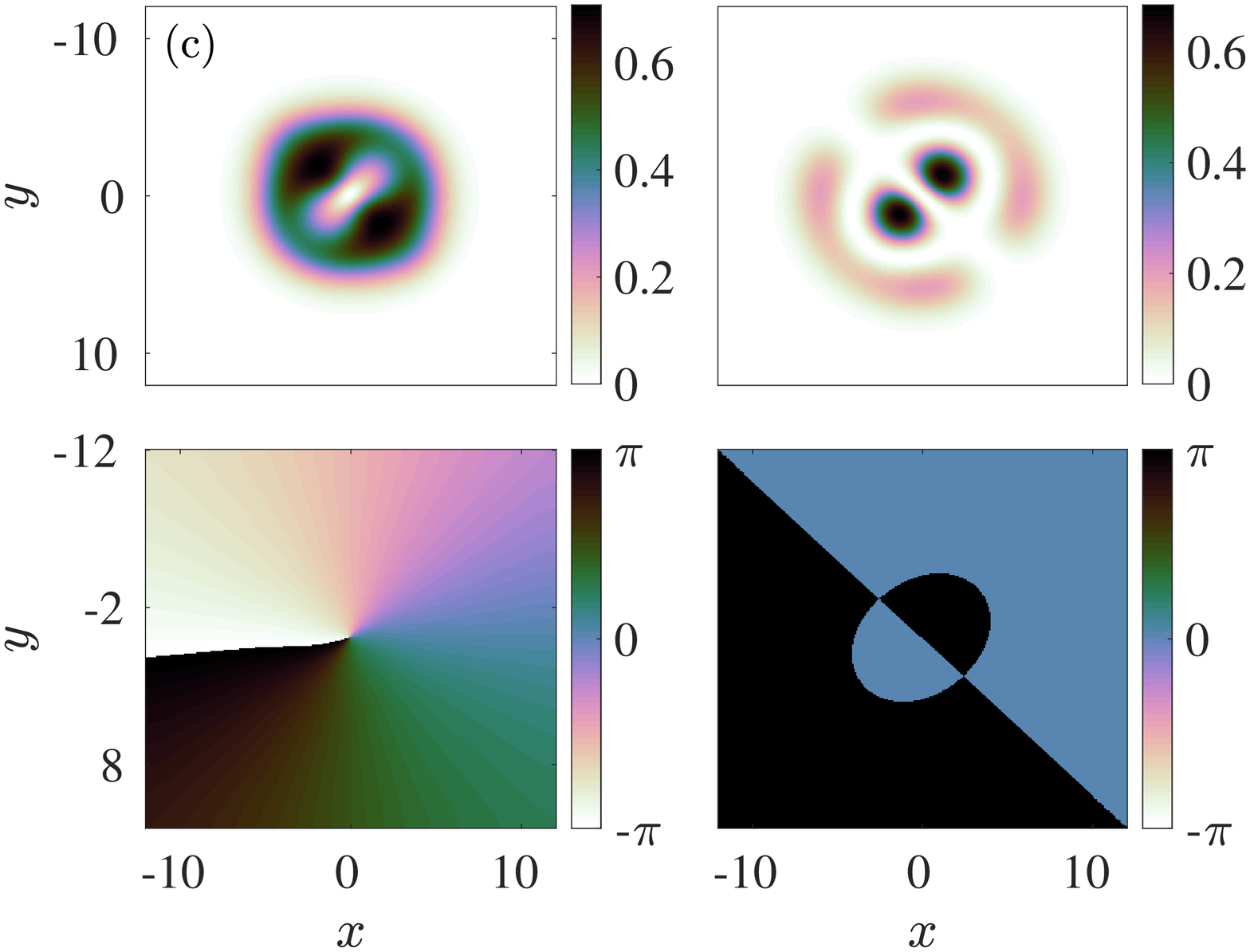}
\includegraphics[height=.25\textheight, angle =0]{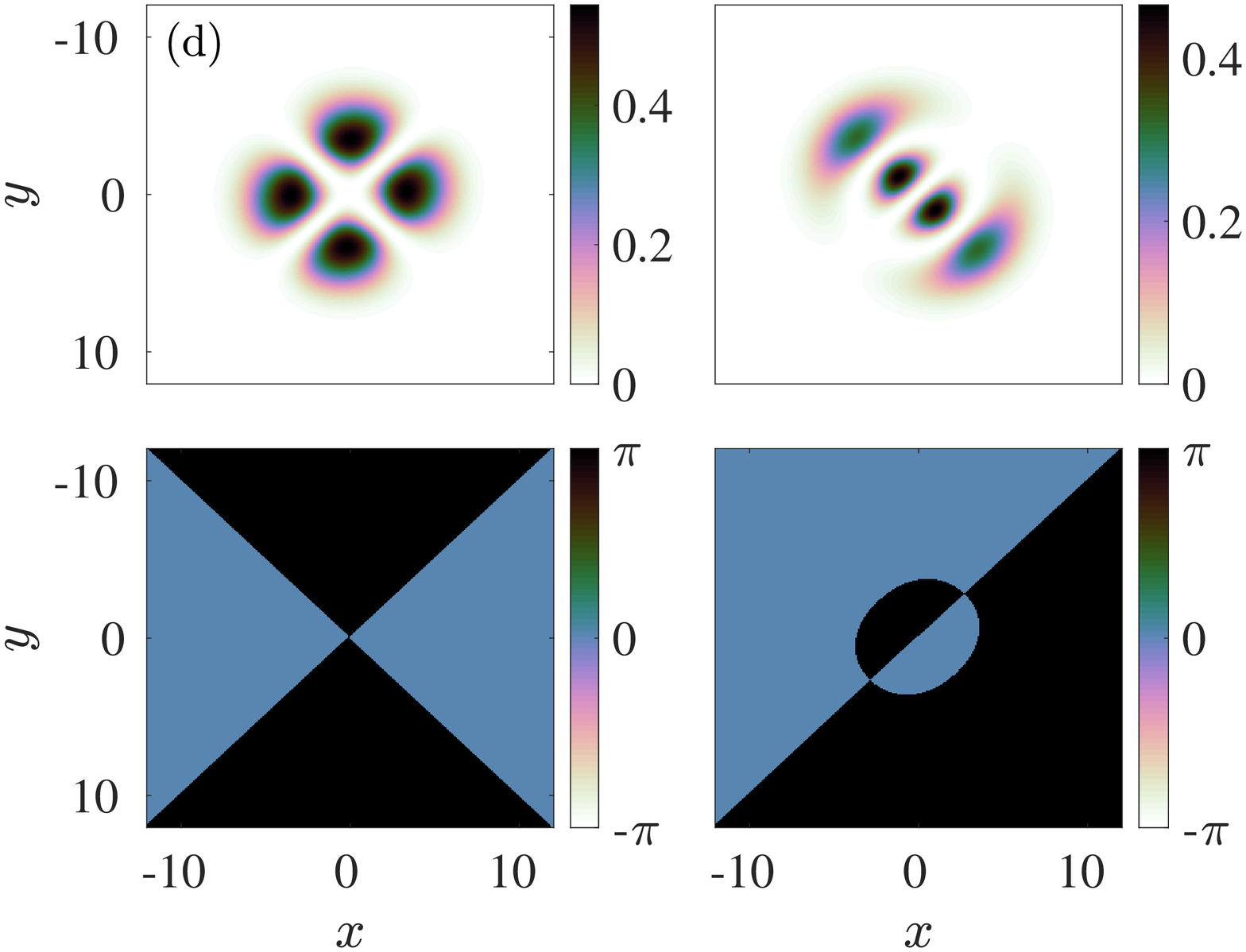}
\includegraphics[height=.25\textheight, angle =0]{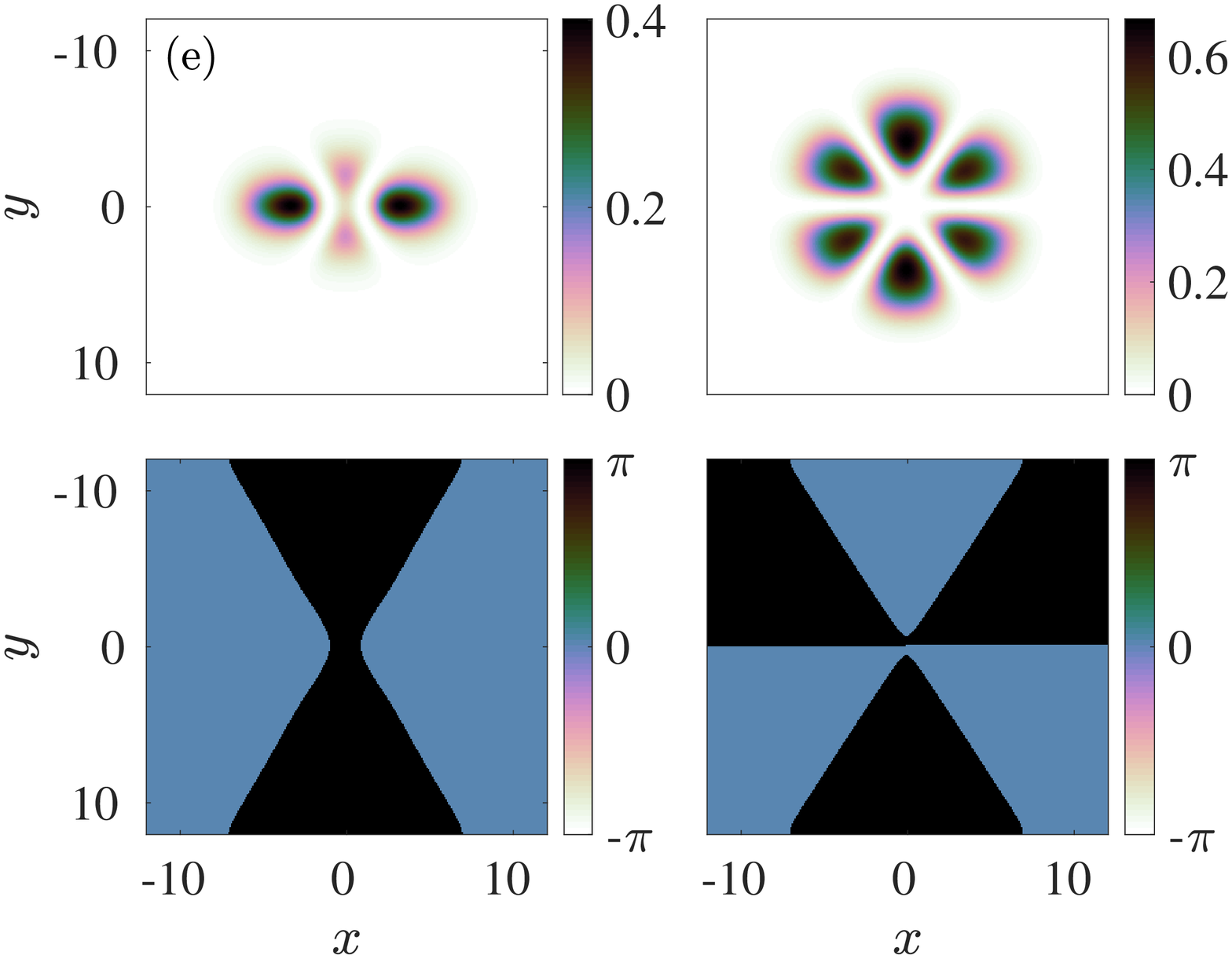}
\includegraphics[height=.25\textheight, angle =0]{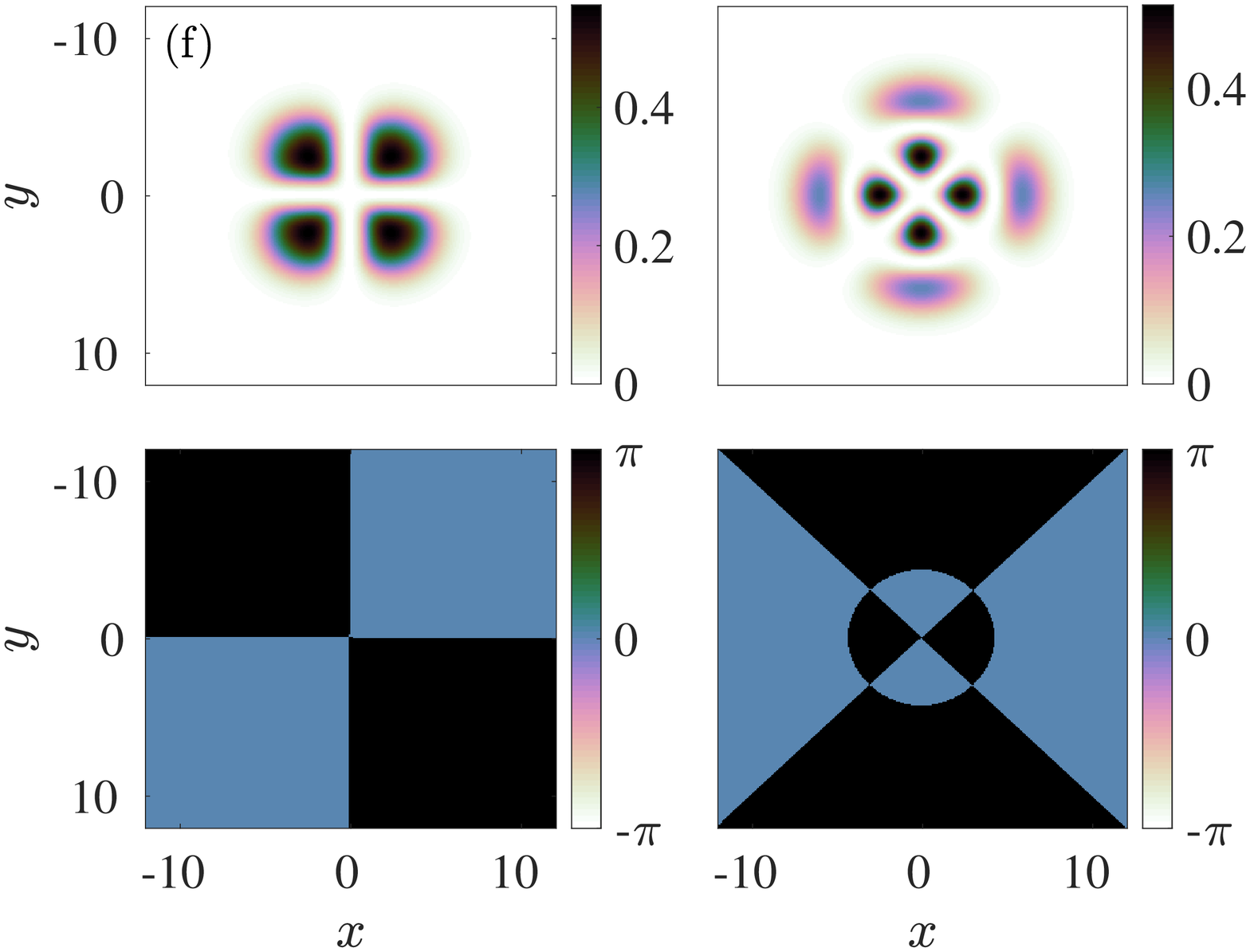}
\includegraphics[height=.25\textheight, angle =0]{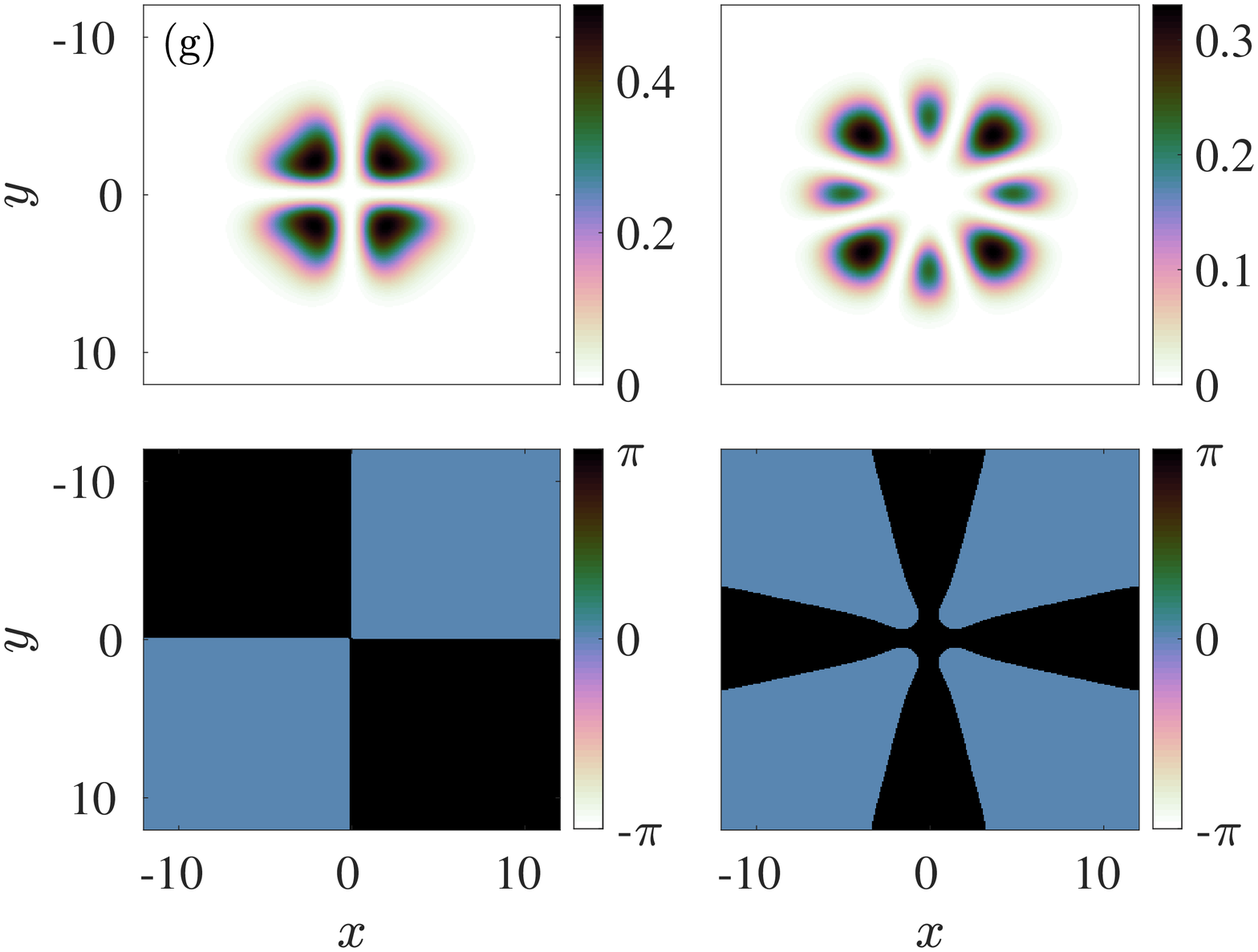}
\includegraphics[height=.25\textheight, angle =0]{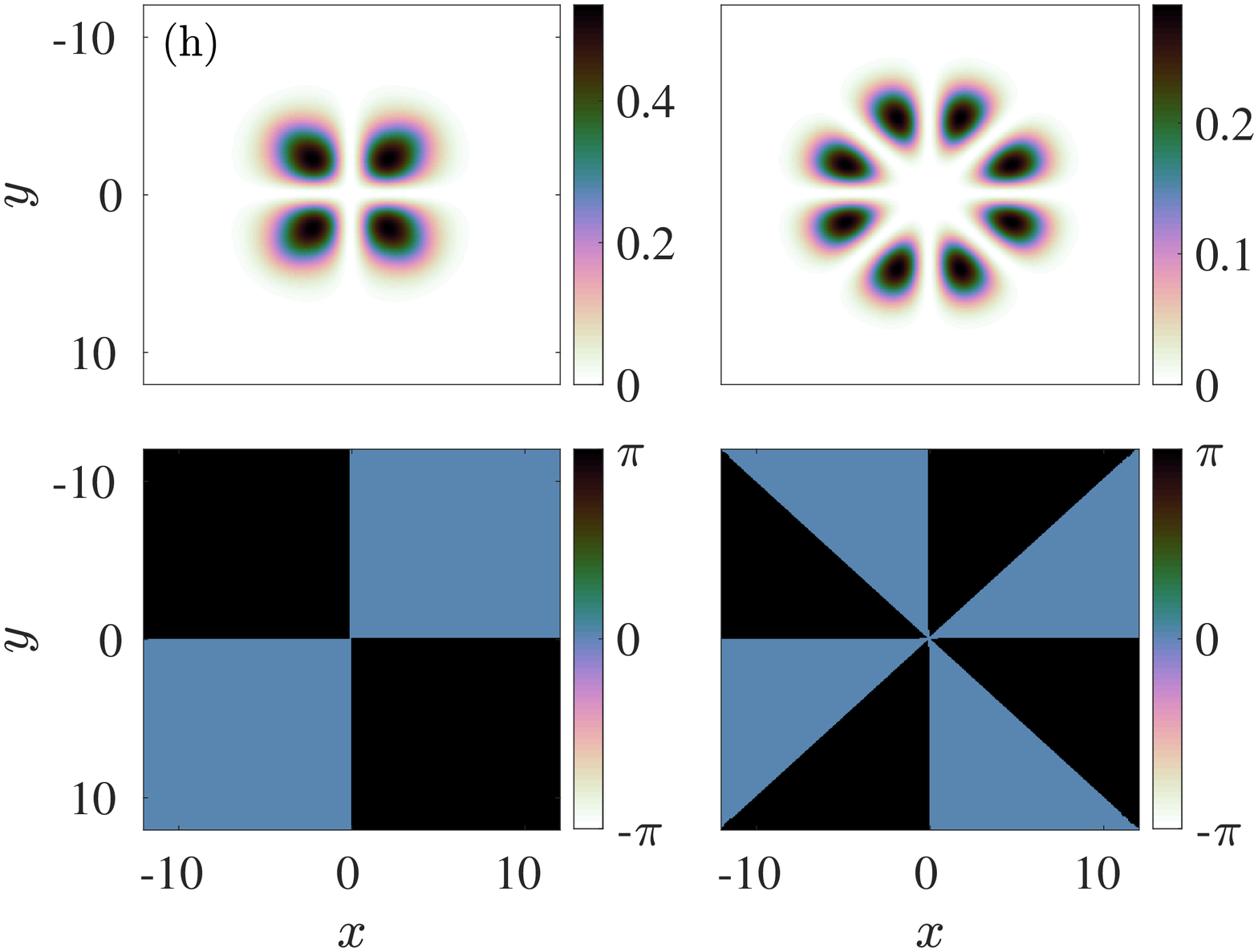}
\end{center}
\caption{
Other branches of solutions identified via continuation from the linear limit (see text
for details). Densities and phases are shown for $\mu_{+}=1$ (a) and $\mu_{+}=1.21$ (b), 
$\mu_{+}=1.26$ (c), $\mu_{+}=1.1$ (d), $\mu_{+}=1.232$ (e), $\mu_{+}=1.27$ (f), 
$\mu_{+}=1.3$ (g) and $\mu_{+}=1.3$ (h), respectively.
} 
\label{fig12}
\end{figure}

However, it did not identify all known branches (e.g.~ones that are present in the single-component 
NLS equation). This motivated an effort to identify further ones (in addition to the states obtained 
via DCM) using the physical understanding of the system and the linear limits. That is to say, we utilized
linear eigenstates in each of the components in either a Cartesian [cf.~Eq.~\eqref{extra1}] or a polar 
[cf.~Eqs.~\eqref{extra2} and~\eqref{extra3}] form and continued relevant combinations for suitable choices
of $\mu_+$ and $\mu_-$ to the nonlinear regime, i.e., for non-vanishing values of $N_+$ and $N_-$. To this
end, we briefly discuss a few examples (among several others) shown in Figs.~\ref{fig11} and~\ref{fig12} that 
could be used in addition to the results of the DCM method. We note in passing that we have found numerous
additional branches to the ones reported here (stemming from the linear limit). However, we refrain from 
discussing all of them here to avoid cluttering the manuscript with additional figures.

We begin our discussion with the quadrupolar complex of Fig.~\ref{fig11}(a). In particular, 
one can envision the $|1,1\rangle_{(\textrm{c})}$ state in the one component co-existing with 
its rotated version in the second component (due to the mutual repulsion between components 
and their immiscibility) as is shown in the figure. This bound state emerges at $\mu_{+}\approx 0.76$ 
and is classified as exponentially unstable over $\mu_{+}\approx [0.76,1.32]$ (we did not perform 
the continuation past $\mu_{+}=1.32$). From its associated spectrum, a cascade of pitchfork 
bifurcations can be discerned, the first of which takes place at $\mu_{+}\approx 0.845$. The branch
that emerges out of this mechanism is presented in Fig.~\ref{fig11}(b) featuring a vortex quadrupole 
in the second component (the latter state was identified in~\cite{egc_16} for the single-component 
NLS equation). The emerging branch amounts to a vortex bright quadrupole. It is exponentially unstable 
for $\mu_{+}\approx [0.845,0.89]$ and then oscillatorily unstable until $\mu_{+}\approx 1.172$ where 
the first component vanishes, thus reducing to the single-component case. The single component vortex 
quadrupole features an interval of oscillatory instability but is otherwise dynamically stable; see 
e.g.,~\cite{egc_16} and references therein. 

To offer an anthology of possible states that we have been able to identify on the basis of continuation
from the linear limit, we refer the reader to Fig.~\ref{fig12}. These (together with the branches shown 
in Fig.~\ref{fig11}) constitute 
examples of nonlinear eigenstates of the problem that we have continued
from their respective linear limit up to $\mu_{+}=1.32$. However, it is important to highlight that these 
states were {\it not} obtained via the DCM.

Although typical examples of the relevant  solutions are shown, we do not present their associated spectra: 
we simply present them as representative examples of the richness of additional available nonlinear patterns 
that a physical understanding of relevant limits can enable us to access. Nevertheless, we briefly comment on
intervals of stability (when applicable) and instability of these waveforms. Similarly as in Fig.~\ref{fig11}(a), 
one can construct a dipole state (in the form of $|1,0\rangle_{(\textrm{c})}$ in the first component) featuring 
its rotated version in the second component as is shown in Fig.~\ref{fig12}(a). This state emerges at $\mu_{+}\approx 0.681$
and is generally exponentially unstable except for the interval of $\mu_{+}\approx[0.9262,1.05]$ in which it 
appears to be oscillatorilly unstable. The bound state mode of Fig.~\ref{fig12}(b) features a single charge
vortex in the first component and the $|1,1\rangle_{(\textrm{c})}$ in the second component. This complex is 
classified as stable for $\mu_{+}\approx[1.095,1.097]$ (as it emerges from the linear limit in the second component) 
but then becomes oscillatorily unstable except of $\mu_{+}\approx[1.195,1.214]$. Furthermore, the solution of 
Fig.~\ref{fig12}(c) involves a deformed vortex in the first component and a $\Phi$ mode in the second component. 
This state bifurcates from the linear limit at $\mu_{+}\approx1.191$ and is classified as stable for 
$\mu_{+}\approx[1.191,1.198]$. For larger values of $\mu_{+}$ it becomes oscillatorily unstable up to $\mu_{+}\approx 1.212$ 
where it starts featuring a dominant exponentially unstable mode. Next, all solutions of panels (d)-(h) of Fig.~\ref{fig12} 
involve the $|1,1\rangle_{(\textrm{c})}$ state in the first component and the states in the second component bifurcate at 
$\mu_{+}\approx 0.9462$, $\mu_{+}\approx 1.089$, $\mu_{+}\approx 1.098$, $\mu_{+}\approx 1.22$, and $\mu_{+}\approx 1.224$, 
respectively. In particular, the solution of Fig.~\ref{fig12}(d) features a $|3,0\rangle_{(\textrm{c})}$ in the second component 
whereas the $|1,1\rangle_{(\textrm{c})}$ in Fig.~\ref{fig12}(e) traps a soliton necklace as this is the case with the second 
component of Fig.~\ref{fig12}(g) and (h). In Fig.~\ref{fig12}(f), the solution that is trapped in the second component can be
approximated in their linear limit by a linear combination of Cartesian eigenstates~\cite{egc_16} (see also the second component
of Fig.~\ref{fig8_set_2}(g)). The solutions of Figs.~(d)-(h) are all classified as exponentially unstable.


\section{Concluding remarks and future challenges} \label{sec:conclusion}
In the present work, we have shed some light on the wealth and complexity of the pattern formation 
that arises in the context of two-component atomic Bose-Einstein condensates, revealed by a
state-of-the-art numerical technique. Naturally, some of the
states that we explored have been previously considered in earlier studies such as~\cite{pola_pra_2012,stockhofe_jpb_2011,wenlong18} 
(among others) or represent two-component extensions of single-component ones. However, several of 
the states found have not previously been discussed in the literature, to the best of our knowledge.
In addition to identifying these states, we explored their bifurcation structure and gave a systematic 
mapping of the parametric regions (as a function of the chemical potential of the second component) 
where the states were potentially stable, exponentially or oscillatorily unstable.

Naturally, there are numerous directions of further work to pursue. A clear starting point is that of seeking 
a way for obtaining a complete ``cartography'' of the possible nonlinear states of the model. We saw 
that the DCM offers a rich repository of such states. Similarly, we argued that the linear limit and 
its theoretical understanding provides plenty more. Additionally bifurcation events from the existing 
states lead to even more. Yet, no toolbox at the moment appears to allow a systematic classification of 
the full span of the highly nonlinear states in this system. A summary of the current techniques and their
advantages and disadvantages will be a useful tool for further efforts. Moreover, to date we have focused 
on two-dimensional states, but it is particularly relevant to study three-dimensional configurations in 
both single but also multi-component settings. Another direction where recently solitonic pattern formation 
has been pursued is that of spinor BECs; see e.g.~\cite{Bersano2018} for a recent example of solitonic states. 
It is thus of particular interest to examine how spin-dependent interactions especially may modify the 
states stemming from a three-component analogue of the Manakov model. The three-component Manakov setting 
in this case represents effectively the spin-independent interaction, while the spin-dependent one is weak, 
but rather complex in that it involves nontrivial phase dynamics between the components~\cite{kawueda}. 
These are extensions worth pursuing from a physical perspective. 

From the numerical analysis/algorithmic perspective, we should comment that further extensions to the DCM 
that we are currently exploring would enable it to discover more solutions. First, using the linear eigenstates
as initial guesses at the appropriate value of $\mu_+$ would give it more branches to track, and therefore 
more initial guesses to use at each continuation step. Second, branch switching algorithms (as implemented 
in, e.g.,~AUTO~\cite{doedel1986}) could be applied to branches identified via deflation, combining their advantages. 
Third, deflation can be combined with nested iteration to greatly enhance its robustness~\cite{adler2017}. 
Fourth, the use of more robust nonlinear solvers (improved line searches or trust region techniques) would 
make the solution of deflated problems more successful on average. Together, these extensions could substantially
enhance the ability of deflation to reveal previously unknown solutions to nonlinear partial 
differential equations and, thus, represent a direction worth pursuing
in its own right.

\begin{acknowledgments}
EGC is indebted to Hans Johnston (UMass) for his endless support and guidance 
throughout this work as well as providing computing resources. He extends 
his deepest gratitude to Eric Polizzi (UMass) and Pavel Holoborodko (Advanpix) 
for substantial assistance regarding the eigenvalue computations using FEAST 
and the Multiprecision Computing Toolbox for MATLAB, respectively. This 
work is supported by EPSRC grant EP/R029423/1 (PEF), the EPSRC Centre For 
Doctoral Training in Industrially Focused Mathematical Modelling (EP/L015803/1) 
(NB).
This
material is based upon work supported by the US National Science
Foundation under Grants No. PHY-1602994 and DMS-1809074
(PGK). PGK also acknowledges support from the Leverhulme Trust via a
Visiting Fellowship and the Mathematical Institute of the University
of Oxford for its hospitality during part of this work.
\end{acknowledgments}

\appendix

\begin{section}{State-of-the-art eigenvalue computations using FEAST}
\label{feast}
In this appendix we briefly give more details of the eigenvalue computations associated with 
Eq.~\eqref{eig_prob}. 
Upon convergence of Newton's method (with $10^{-9}$ relative and absolute tolerances, respectively), 
the stability matrix $A$ of Eq.~\eqref{eig_prob} is computed. For the finite difference discretization 
employed, the matrix $A$ is a $357,604\times357,604$ sparse matrix with $2,856,048$ non-zero elements. 
To compute its eigenvalues, we initially used MATLAB's \verb|eigs| command which employs a Krylov-Schur 
method~\cite{mathworks_eigs} to compute a subset of the eigenvalues. Although MATLAB did not raise any 
warnings during the computation (i.e., \verb|flag=0|), in some of the
cases
the obtained spectrum contained spurious eigenvalues. 
This finding was further investigated by calculating 
\begin{equation}
\frac{\underset{1\leq i\leq n_{\textrm{max}}}{\max}\left\|A\textbf{W}_{\textrm{R}}^{(i)}-%
\rho_{i}\textbf{W}_{\textrm{R}}^{(i)}\right\|_{\ell_{1}}}{\left\|A\right\|_{\ell_{1}}},
\label{manhattan_norm}
\end{equation}
where $n_{\textrm{max}}$ is the number of eigenpairs computed, and $\textbf{W}_{\textrm{R}}^{(i)}$ 
is the $i$th right eigenvector associated with the eigenvalue $\rho_{i}$. For instance, Eq.~\eqref{manhattan_norm} 
evaluated for the solution 
of the branch shown in Fig.~\ref{fig1}(c) and for $\mu_{+}=1.32$ gives $\approx 2.21$ for
$100$ eigenpairs. One possible explanation about why this happens is given next~\cite{pavel}. 
MATLAB's \verb|eigs| computes first the LU decomposition (with full pivoting and scaling) 
of the matrix $A$, which is performed very accurately. However, the matrix $A$ itself is 
ill-conditioned (as we will see subsequently) and any slight change in subspace vectors in
the Krylov-Schur algorithm will lead to a large change in resulting eigenvectors~\cite{saad}. 
As a consequence, all the accuracy is lost by using \verb|eigs| without any warning raised.

To further investigate this issue, we used the Multiprecision Computing Toolbox ``Advanpix''~\cite{advanpix}
which implements an extended precision version of \verb|eigs| function called \verb|mpeigs|.
We performed the eigenvalue computation of the branch shown in Fig.~\ref{fig1}(c) with $34$
digits (over $121$ distinct values of $\mu_{+}$) which took approximately 3 months of computing time 
on an Intel(R) Xeon(R) CPU E5-2670 0 @ 2.60GHz processor with 64GB of RAM. The respective 
results are shown with red stars in the left and right panels of Fig.~\ref{fig_comp_feast_mpeigs} 
corresponding to the real and imaginary parts of $\omega$, respectively. We further checked
Eq.~\eqref{manhattan_norm} for this computation and for, e.g., $\mu_{+}=1.32$ we found that 
the residual of Eq.~\eqref{manhattan_norm} for $100$ eigenpairs $\left(\rho,\textbf{W}_{\textrm{R}}\right)$ 
was $\approx 1.57\times10^{-18}$. This computation clearly suggests that extended precision is 
capable of diminishing any small perturbations in subspace vectors in the Krylov-Schur algorithm. 

However, a new algorithm (motivated by contour integration and density-matrix representation in
quantum mechanics) for solving eigenvalue problems known as FEAST was introduced in~\cite{eric_1} 
(see also~\cite{tang_eric} for details). FEAST itself combines accuracy, efficiency and robustness 
while exhibiting natural parallelism at multiple levels. Recently, the algorithm was generalized
and applied to non-Hermitian matrices~\cite{kestyn_eric_tang} (and references therein). In our
present work, we used FEAST extensively to calculate the spectra of all branches shown with 
$10^{-8}$ relative tolerance (on the residuals) as the stopping criterion (e.g., for the branch
of Fig.~\ref{fig1}(c) and for $\mu_{+}=1.32$, Eq.~\eqref{manhattan_norm} gives $\approx 1.39\times 10^{-8}$
for $100$ eigenpairs). It should be noted that FEAST converges quite quickly ($3$ to $5$ iterations
were required in most of the cases we studied), thus providing us with a robust tool for solving 
large eigenvalue problems. Its execution time varies depending on the number of eigenvalues inside
a prescribed contour. Indicatively, the computation of the eigenvalues of the branch of Fig.~\ref{fig1}(c) 
took approximately $5$ hours.

\begin{figure}[pt]
\begin{center}
\includegraphics[height=.19\textheight, angle =0]{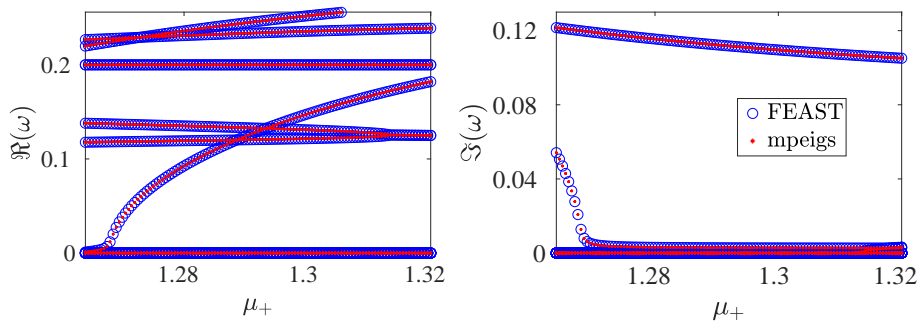}
\end{center}
\cprotect\caption{The eigenfrequencies of Eq.~\eqref{eig_prob} obtained by using 
the FEAST eigenvalue solver and \verb|mpeigs| are shown with blue open circles 
and red stars for comparison.}
\label{fig_comp_feast_mpeigs}
\end{figure}

To further demonstrate the robustness of the FEAST algorithm, we calculate the condition 
number of a simple eigenvalue $\rho$~\cite{saad} defined by
\begin{equation}
\kappa\left(\rho\right)=\frac{\left\|\textbf{W}_{\textrm{R}}\right\|_{\ell_{2}}%
\left\|\textbf{W}_{\textrm{L}}\right\|_{\ell_{2}}}%
{|\left(\textbf{W}_{\textrm{R}},\textbf{W}_{\textrm{L}}\right)|},
\label{cond_number}
\end{equation}
where $\textbf{W}_{\textrm{L}}$ is the left eigenvector associated with the eigenvalue 
$\rho$. Direct computation of Eq.~\eqref{cond_number} using the left and right eigenvectors
computed in FEAST (inside an elliptical contour) for $\mu_{+}=1.32$, we get a value of
$\underset{1\leq i\leq n_{\textrm{max}}}{\max}\kappa\left(\rho_{i}\right)\approx 3.37\times 10^{7}$,
again for $100$ eigenpairs. The spectra of the entire branch were computed using FEAST and 
are shown in Fig.~\ref{fig_comp_feast_mpeigs} with blue open circles. The agreement 
of the spectra obtained using \verb|mpeigs| and FEAST is clearly evident. FEAST has clearly 
demonstrated its robustness and accuracy (see also the discussion in~\cite{kestyn_eric_tang}) 
and is a powerful tool for studying the spectra of very large ill-conditioned matrices.
\end{section}

\end{document}